\documentclass[twocolumn,english,aps,superscriptaddress,floatfix,longbibliography]{revtex4-2}
\usepackage[latin9]{inputenc}
\usepackage{color}
\usepackage{babel}
\usepackage{verbatim}
\usepackage{textcomp}
\usepackage{amsmath}
\usepackage{amssymb}
\usepackage{graphicx}
\usepackage{esint}
\usepackage[unicode=true,pdfusetitle,
 bookmarks=true,bookmarksnumbered=false,bookmarksopen=false,
 breaklinks=true,pdfborder={0 0 0},pdfborderstyle={},backref=false,colorlinks=true]
 {hyperref}
\hypersetup{ linkcolor=blue,citecolor=blue,urlcolor=blue}

\makeatletter

\usepackage{color}
\usepackage{babel}

\usepackage{color}
\usepackage{dsfont}   
\usepackage{braket}

\newcommand{\rucl}{$\alpha$-RuCl$_3$}
\newcommand{\sumprime}{{\vphantom{\sum}}'}
\newcommand{\normord}[1]{\text{\bf :}{#1}\text{\bf :}}
\DeclareMathOperator{\Tr}{Tr}

\definecolor{orange}{rgb}{1,0.5,0}

\makeatother

\begin{document}
\title{Heisenberg-Kitaev model in a magnetic field: $1/S$ expansion}
\author{Pedro M. C\^onsoli}
\email{pedro.consoli@usp.br}
\affiliation{Instituto de F\'isica de S\~ao Carlos, Universidade de S\~ao Paulo, C.P. 369, S\~ao Carlos, SP, 13560-970, Brazil}
\affiliation{Institut f\"ur Theoretische Physik and W\"urzburg-Dresden Cluster of Excellence ct.qmat, Technische Universit\"at Dresden, 01062 Dresden, Germany}
\author{Lukas Janssen}
\affiliation{Institut f\"ur Theoretische Physik and W\"urzburg-Dresden Cluster of Excellence ct.qmat, Technische Universit\"at Dresden, 01062 Dresden, Germany}
\author{Matthias Vojta}
\affiliation{Institut f\"ur Theoretische Physik and W\"urzburg-Dresden Cluster of Excellence ct.qmat, Technische Universit\"at Dresden, 01062 Dresden, Germany}
\author{Eric C. Andrade}
\affiliation{Instituto de F\'isica de S\~ao Carlos, Universidade de S\~ao Paulo, C.P. 369, S\~ao Carlos, SP, 13560-970, Brazil}


\begin{abstract}
The exact solution of Kitaev's spin-$1/2$ honeycomb spin-liquid model has sparked an intense search for Mott insulators hosting bond-dependent Kitaev interactions, of which Na$_{2}$IrO$_{3}$ and $\alpha$-RuCl$_{3}$ are prime examples. Subsequently, it has been proposed that also spin-$1$ and spin-$3/2$ analogs of Kitaev interactions may occur in materials with strong spin-orbit coupling. As a minimal model to describe these Kitaev materials, we study the Heisenberg-Kitaev Hamiltonian in a consistent $1/S$ expansion, with $S$ being the spin size. We present a comprehensive study of this model in the presence of an external magnetic field applied along two different directions, {[}001{]} and {[}111{]}, for which an intricate classical phase diagram has been reported. In both settings, we employ spin-wave theory in a number of ordered phases to compute phase boundaries at the next-to-leading order in $1/S$, and show that quantum corrections substantially modify the classical phase diagram. More broadly, our work presents a consistent route to investigate the leading quantum corrections in spin models that break spin-rotational symmetry.
\end{abstract}
\date{\today}
\maketitle


\section{Introduction}

The combined effects of strong electron-electron interaction and spin-orbit coupling has stimulated the search for unconventional phases of matter in transition metal oxides with partially filled 4$d$ and 5$d$ shells \citep{witczak-krempa14,rau16,trebst17,winter17,janssen19,takagi19}. As originally demonstrated by Jackeli and Khaliullin \citep{jackeli09}, the effective spin model for these Mott insulators in edge-sharing octahedral geometries contains, in general, bond-dependent Ising-like exchange interactions, which lie at the heart of Kitaev's honeycomb model \citep{kitaev06}. The $S=1/2$ Kitaev model on tricoordinated lattices is exactly solvable by mapping it onto a model of free Majorana fermions coupled to Z$_{2}$ gauge fields, showing a gapless spin-liquid ground state \citep{kitaev06}. Interestingly, this spin liquid becomes a non-Abelian topological spin liquid upon applying a small magnetic field \citep{kitaev06,jiang11}. At intermediate field strengths, and depending on the orientation of the field, recent numerical studies have uncovered the existence of a further, presumably gapless, field-induced spin-liquid phase \citep{jiang18,liang18,nasu18,zhu18,gohlke18,jiang19,hickey19,patel19,zou20} between this low-field topological spin liquid and the high-field polarized phase.

On the experimental side, it is now well established that Kitaev-type interactions are relevant for the honeycomb iridates \citep{choi12,singh12} and for \rucl$\,$ \citep{plumb14,sears15,banerjee16}, in which the Ir$^{4+}$ and Ru$^{3+}$ ions form effective $j=1/2$ local moments. Nevertheless, the realization of quantum spin liquids in the strong spin-orbit coupling regime has remained a challenge because more realistic models for these compounds include additional interactions that tend to drive different kinds of long-range magnetic order \citep{chaloupka10,kimchi11,chaloupka13,rau14,ioannis15}. In \rucl, the long-range magnetic order can be suppressed by applying an in-plane magnetic field \citep{johnson15,sears17,wolter17,kelley18b,kasahara18a,janssen19}. Remarkably, for tilted field directions, an approximately half-integer quantized thermal Hall conductance~\citep{kasahara18b} has been found, indicative of a gapped topological spin liquid with chiral Majorana edge mode \citep{aviv18,ye18,gao19}. Signatures of a new phase, intermediate between the low-field ordered and high-field polarized phases, have also been obtained for in-plane magnetic fields from magnetocaloric effect \citep{balz2019} and magnetostriction measurements \citep{gass2020}.

On the theoretical side, there is growing evidence that much of the rich physics of $S=1/2$ Kitaev model is also present for larger values of $S$. A new class of spin-$1$ Kitaev materials was recently proposed \citep{stavropoulos19}, with a number of specific materials presented as candidates, e.g., the layered antimonates $A_{3}$Ni$_{2}$SbO$_{6}$ ($A= \text{Li}, \text{Na}$)~\citep{zvereva15}. The $S=1$ Kitaev model is not exactly solvable, although it shares many of the properties of its $S=1/2$ version \citep{baskaran08,koga18,ioannis18,dong19,lee20b}, including the behavior in the presence of a magnetic field \citep{zhu20,khait20,hickey20}. Furthermore, different Cr-based compounds have recently been proposed as candidates for $S=3/2$ Kitaev systems~\citep{stavropoulos19,lee20,xu20}. More generally, higher-$S$ effective spin-orbital models with bond-dependent interactions have also been discussed~\citep{nussinov15,natori16,natori18}.

The nearest-neighbor Heisenberg-Kitaev model \citep{chaloupka10} has emerged as a minimal model to describe the various Kitaev materials, with further symmetry-allowed interactions being important in some of them. Remarkably, the Heisenberg-Kitaev model displays highly nontrivial behavior already in the classical limit, $S\to\infty$. While the spin liquid phases shrink to isolated points in the phase diagram, with high ground-state degeneracy, the physics in applied magnetic fields is extremely rich due to the non-Heisenberg interactions, and there is a plethora of field-induced phases with complex magnetic ordering \citep{janssen16,janssen17,price17,chern20}. A systematic study of this physics away from the classical limit, i.e., for different spin sizes $S$, is lacking.

In this paper, we therefore study the nearest-neighbor Heisenberg-Kitaev model in an external magnetic field using an expansion in $1/S$. Our primary focus is the stability of the ordered phases, following the work of Ref.~\citep{janssen16}, for the two field directions: $\left[001\right]$ and $\left[111\right]$ in the cubic spin-space basis~\citep{janssen19}. Specifically, we analyze the model by applying spin-wave theory both to the ordered \citep{chern20,cookmeyer18} and the high-field polarized phases \citep{janssen16,janssen17,mcclarty18,darshan18}. Importantly, the noncollinearity of the canted ordered states requires nonlinear spin-wave theory for a consistent $1/S$ expansion \citep{zhitomirsky98,coletta12}, which, for small values of $S$, introduces sizable modifications to the classical phase diagrams obtained in Ref.~\citep{janssen16}.

Finally, we remark that our work goes beyond the investigation of Kitaev materials. We present a well-defined $1/S$ expansion \citep{zhitomirsky98,coletta12,rau18} that can be applied to any generic spin model lacking SU(2) symmetry. It thus stands as an accessible analytical formalism beyond linear spin-wave theory to complement numerical methods, e.g., exact diagonalization or density matrix renormalization group, which are typically used to study complex magnetic systems, but restricted to small clusters.

The remainder of this paper is organized as follows. In Sec~\ref{sec:model}, we describe our model and develop a theoretical framework to consistently account for next-to-leading order terms in $1/S$. Our method is then applied to the cases of $\textbf{h}\parallel\left[001\right]$ and $[111]$ in Secs.~\ref{sec:h001} and \ref{sec:h111}, respectively, where we also show phase diagrams and magnetization curves for specific values of $S$. We conclude in Sec.~\ref{sec:conclusion}. Details of our calculations, a number of analytical results and spin-wave spectra for the several phases studied in our work are given in the appendices.


\section{Model and spin-wave theory} \label{sec:model}

As a minimal model to describe the physics of Kitaev materials, we consider the nearest-neighbor Heisenberg-Kitaev (HK) Hamiltonian \citep{chaloupka10,chaloupka13}
\begin{equation}
\mathcal{H}=J\sum_{\left\langle ij\right\rangle}\textbf{S}_{i}\cdot\textbf{S}_{j}+K\sum_{\left\langle ij\right\rangle _{\gamma}}S_{i}^{\gamma}S_{j}^{\gamma}-\textbf{h}\cdot\sum_{i}\textbf{S}_{i},\label{eq:hk}
\end{equation}
where $\gamma\in\left\{ x,y,z\right\}$ labels the three different links on the honeycomb lattice. For convenience, we absorb all constants that appear in the effective moment $g\mu_{\mathrm{B}}\textbf{S}$ of each pseudospin into the field $\textbf{h}:=g\mu_{B}\mu_{0}\textbf{H}$. In addition, from now on we shall parametrize the HK couplings as $J=A\cos\varphi$ and $K=2A\sin\varphi$, where $A>0$ is an overall energy scale \citep{chaloupka13}.

Because the Kitaev term breaks spin-rotational symmetry, the response of the system acquires a strong dependence on the direction of the external field $\textbf{h}$. Here, we give all field directions in the cubic spin basis $\left\{\hat{\textbf{x}},\hat{\textbf{y}},\hat{\textbf{z}}\right\}$ and label them in the form $\left[xyz\right]$, so that $\textbf{h}\parallel\left[xyz\right]$ reads $\textbf{h}\propto x\hat{\textbf{x}}+y\hat{\textbf{y}}+z\hat{\textbf{z}}$. In \rucl, the cubic axes $\hat{\textbf{x}}$, $\hat{\textbf{y}}$ and $\hat{\textbf{z}}$ point along nearest-neighbor Ru-Cl bonds. Therefore, the $\left[111\right]$ direction (often referred to as $c^{*}$ axis) is perpendicular to the honeycomb plane, whereas the in-plane crystallographic $a$ and $b$ axes in the monoclinic notation are along the $\left[11\bar{2}\right]$ and $\left[\bar{1}10\right]$ directions, respectively~\citep{janssen19}. Hence, $\textbf{h}\parallel\left[001\right]$ describes a configuration in which the magnetic field lies along an intermediate direction in the $ac^{*}$ plane.

At zero field, the HK model realizes four different ordered states as a function of the interaction parameter $\varphi$: Besides the usual ferromagnetic and N\'eel antiferromagnetic states near the Heisenberg limits $\varphi = \pi$ and $0$, respectively, stripy and zigzag states are stabilized for increasing Kitaev interactions~\citep{chaloupka13}. In the classical limit, formally corresponding to $S\to\infty$, the Kitaev spin-liquid phases near $\varphi = \pm \pi/2$ shrink to isolated points in the phase diagram, which are characterized by extensive classical ground-state degeneracies~\citep{baskaran08, price13}.

The ordered moments in the N\'eel, stripy, and zigzag phases point along the cubic spin-space axes at zero field~\citep{janssen19}. Consequently, for a field along the $[001]$ direction, the spins in these phases can always align perpendicular to an infinitesimal field and cant homogeneously towards the magnetic field axis at small finite fields, until a continuous phase transition towards the polarized state is reached at some critical field strength.

\begin{figure}
\centering
\includegraphics[width=8.5cm]{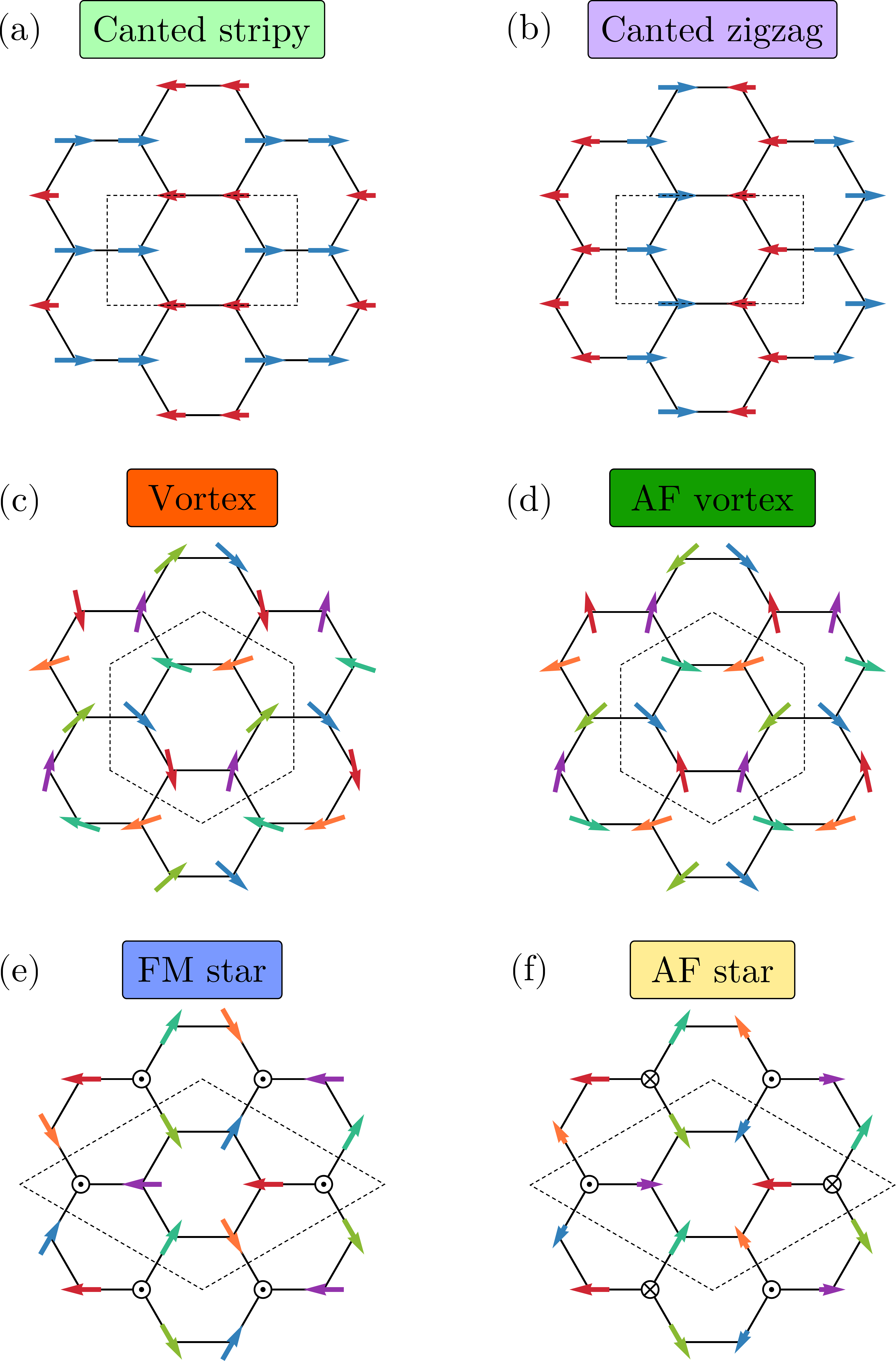}
\caption{Spin configurations of ordered phases of the HK model in a $\left[111\right]$ field, projected onto a plane perpendicular to $[111]$. The respective magnetic unit cells are shown in dashed lines. Unequal lengths of the projected spins in the canted zigzag, canted stripy and AF star configurations reflect the occurrence of nonuniform canting.
\label{fig:projphases}}
\end{figure}

This situation changes dramatically for a field along the $[111]$ direction. In this case, the stripy and zigzag states cannot align perpendicular to this axis, prohibiting a homogeneous canting towards the magnetic field axis. The inhomogeneously canted stripy and zigzag states therefore compete with other states that allow an energetically more efficient canting mechanism, potentially leading to metamagnetic transitions between different ordered phases at intermediate field strengths. In fact, for this field configuration, the classical analysis of Ref.~\citep{janssen16} found six novel field-induced phases in addition to the canted versions of four zero-field phases. Two of these field-induced phases have unit cells consisting of at least 18 sites (or may be even incommensurate~\citep{price17}) and cover only a very small region of the phase diagram. One might therefore speculate that these two phases may be destabilized upon the inclusion of quantum fluctuations for small values of $S$. Representative spin configurations of the other four field-induced phases, dubbed vortex, antiferromagnetic (AF) vortex, ferromagnetic (FM) star, and AF star in Ref.~\citep{janssen16}, together with those of the canted stripy and canted zigzag states, are shown in Fig.~\ref{fig:projphases}. These four field-induced phases have magnetic unit cells of six and eight sites, and span a comparatively large parameter region in the phase diagram. Their fate at small values of $S$ therefore represents an important open problem, which we address in this work.


\subsection{Classical reference states}

The starting point for our spin-wave analysis is the parametrizations of the classical phases that arise from the Hamiltonian, Eq.~\eqref{eq:hk}, for a given field direction. On general grounds, each phase is characterized by a magnetic unit cell composed of $N_{\mathrm{s}}$ spins, so that a particular parametrization specifies a total of $N_{\mathrm{s}}$ pairs of angles. By
labeling the different sites in the magnetic unit cell with the subindex $\mu\in\left\{ 1,\ldots,N_{\mathrm{s}}\right\}$, we then attribute to each spin an azimuthal and a polar angle, denoted here by $\phi_{\mu}$ and $\theta_{\mu}$, respectively, with the polar angles measured with respect to the field direction~\citep{janssen16}. To fix the parametrization angles $\left\{ \boldsymbol{\phi},\boldsymbol{\theta}\right\} \equiv\left\{ \phi_{1},\ldots,\phi_{N_{\mathrm{s}}},\theta_{1},\ldots,\theta_{N_{\mathrm{s}}}\right\}$, we minimize the classical ground-state energy of Eq.~\eqref{eq:hk}.

In this work, we focus on the four field-induced phases displayed in Figs.~\ref{fig:projphases}(c)--\ref{fig:projphases}(f), in addition to the high-field polarized phase (not shown) and the canted versions of the stripy, zigzag [Figs.~\ref{fig:projphases}(a) and \ref{fig:projphases}(b)] and N\'eel (not shown) phases. In fact, as we shall see below, quantum fluctuations typically tend to destabilize states with large magnetic unit cells in favor of small-unit-cell states. For the purposes of this work, we hence make the simplifying assumption that the two additional large-unit-cell phases found in Ref.~\citep{janssen16}, which cover only a very small region of the phase diagram, are entirely destabilized by quantum fluctuations at the small values of $S$ we are interested in.


\subsection{Linear spin-wave theory} \label{subsec:LSWT}

In order to set up the spin-wave theory, for a given value of the interaction parameter $\varphi$, we rotate the spin coordinate system so that the transformed Hamiltonian bears a ferromagnetic ground state. This involves a set of $N_{\mathrm{s}}$ rotations which map the laboratory $\left\{ \hat{\textbf{x}},\hat{\textbf{y}},\hat{\textbf{z}}\right\}$ basis onto local $\left\{\hat{\textbf{e}}_{\mu1},\hat{\textbf{e}}_{\mu2},\hat{\textbf{e}}_{\mu3}\right\}$ bases which have $\hat{\textbf{e}}_{\mu3}$ pointing along the classical spin direction in magnetic sublattice $\mu$.
In this basis, we then employ the Holstein-Primakoff transformation \citep{holstein1940}
\begin{equation}
\begin{cases}
S_{i\mu}^{3}=S-a_{i\mu}^{\dagger}a_{i\mu},\\
S_{i\mu}^{-}=a_{i\mu}^{\dagger}\sqrt{2S-a_{i\mu}^{\dagger}a_{i\mu}},\\
S_{i\mu}^{+}=\sqrt{2S-a_{i\mu}^{\dagger}a_{i\mu}}\;a_{i\mu},
\end{cases}\label{eq:Holstein-Primakoff}
\end{equation}
where $a_{i\mu}^{\dagger}$ $\left(a_{i\mu}\right)$ is a bosonic creation (annihilation) operator. The additional subindex $i$ runs from $1$ to $N_{\mathrm{c}}$, the number of magnetic unit cells.

By expanding the spin ladder operators in powers of $a_{i\mu}^{\dagger}a_{i\mu}/2S$, one can then rewrite the Hamiltonian as a power series in $1/\sqrt{S}$,
\begin{equation}
\mathcal{H}=\sum_{n=0}^{\infty}S^{2-\frac{n}{2}}\mathcal{H}_{n}\;,
\label{eq:H series}
\end{equation}
where each term is labeled according to its order $n$ in bosonic operators.

In the linear spin-wave (LSW) regime, interactions between magnons are neglected, so that only the terms up to order $n=2$ in Eq.~\eqref{eq:H series} are retained. As the expansion is performed around a configuration that minimizes $\mathcal{H}_{0}$, the linear term $\mathcal{H}_{1}$ vanishes, and we end up with a simple quadratic Hamiltonian. After applying a Fourier transform, one finds
\begin{equation}
\mathcal{H}_{\mathrm{LSW}} = S^{2}E_{\mathrm{gs},0}+\frac{S}{2}\sum_{\textbf{k}}\left(\alpha_{\textbf{k}}^{\dagger}\mathbb{M}_{\textbf{k}}\alpha_{\textbf{k}}-\Tr\mathbb{A}_{\textbf{k}}\right).
\label{eq:H_LSW pre-Bogoliubov}
\end{equation}
Here, $S^{2}E_{\mathrm{gs},0} \equiv S^{2}\mathcal{H}_{0}$ is the classical ground-state energy and $\alpha_{\textbf{k}}^{\dagger}=\left(a_{\textbf{k}1}^{\dagger},\ldots,a_{\textbf{k}N_{\mathrm{s}}}^{\dagger},a_{-\textbf{k}1},\ldots,a_{-\textbf{k}N_{\mathrm{s}}}\right)$. Moreover, $\mathbb{M}_{\textbf{k}}$ is a $2N_{\mathrm{s}}\times2N_{\mathrm{s}}$ matrix that can generically be written in terms of two $N_{\mathrm{s}}\times N_{\mathrm{s}}$ submatrices, $\mathbb{A}_{\textbf{k}}$ and $\mathbb{B}_{\textbf{k}}$, as
\begin{equation}
\mathbb{M}_{\textbf{k}} =
\begin{pmatrix}
\mathbb{A}_{\textbf{k}} & \mathbb{B}_{\textbf{k}} \\
\mathbb{B}_{\textbf{k}}^{\dagger} & \mathbb{A}_{-\textbf{k}}^{\text{T}}
\end{pmatrix}.
\label{eq:Mk matrix}
\end{equation}
After a Bogoliubov transformation \citep{blaizot1986}, (see Appendix \ref{sec:app bog transf} for details) we obtain
\begin{equation}
\mathcal{H}_{\mathrm{LSW}} = S^{2} E_{\mathrm{gs},0} + S E_{\mathrm{gs},1}+S\sum_{\textbf{k}\mu}\epsilon_{\textbf{\text{k}\ensuremath{\mu}}}b_{\textbf{k}\mu}^{\dagger}b_{\textbf{k}\mu},\label{eq:H_LSW post-Bogoliubov}
\end{equation}
where $b_{\textbf{k}\mu}^{\dagger}$ $\left(b_{\textbf{k}\mu}\right)$ creates (annihilates) a magnon with momentum $\textbf{k}$ and energy $\epsilon_{\textbf{k}\mu}$, $\mu$ labels the $N_{\mathrm{s}}$ magnon bands, and
\begin{equation}
E_{\mathrm{gs},1}=\frac{1}{2}\sum_{\textbf{k}}\left(\sum_{\mu}\epsilon_{\textbf{k}\mu}-\Tr\mathbb{A}_{\textbf{k}}\right)
\label{eq:epsilon_q}
\end{equation}
is the next-to-leading order (NLO) contribution in $1/S$ to the ground-state energy,
\begin{align}
E_{\mathrm{gs}}\left(\varphi,h,\frac{1}{S}\right) = S^{2} \sum_{n=0}^{\infty} \left( \frac{1}{S} \right)^n E_{\mathrm{gs},n} \left(\varphi,h\right).
\label{eq:ground-state energy}
\end{align}
In the absence of a magnetic field, the term $\Tr\mathbb{A}_{\textbf{k}}$ equals $S E_{\mathrm{gs},0}$, such that it combines with the leading term $S^{2}E_{\mathrm{gs},0}$ into $S\left(S+1\right) E_{\mathrm{gs},0}$. We emphasize, however, that this does not happen for $h\ne0$.

In Appendix \ref{sec:app SW spectra}, we present the LSW spectra of several of the ordered phases considered here for both $\textbf{h} \parallel \left[001\right]$ and $\textbf{h} \parallel \left[111\right]$.


\subsection{Quantum corrections to the magnetization} \label{subsec:NLO mag}

The theory presented in Sec.~\ref{subsec:LSWT} provides the means to calculate the NLO contribution in $1/S$ to the $T=0$ magnetization per site,
\begin{equation}
m_{h} =-\frac{1}{N}\frac{\partial E_{\mathrm{gs}}}{\partial h}
=-\frac{S^{2}}{N} \frac{\partial}{\partial h} \left[E_{\mathrm{gs},0} + \frac{E_{\mathrm{gs},1}}{S}+\mathcal{O}\!\left(\frac{1}{S^{2}}\right)\right],
\label{eq:mh -dEgsdh}
\end{equation}
where $N=N_\mathrm{s} N_\mathrm{c}$ denotes the total number of sites. With Eq.~\eqref{eq:mh -dEgsdh} at hands, let us consider a few results for $\textbf{h}\parallel\left[001\right]$. As mentioned before, the classical ground state in this setting is characterized by spins canting uniformly toward the $\left[001\right]$ direction from $h=0$ up to the classical critical field, $h_{\mathrm{c}0}$. At this point, all spins become parallel to $\textbf{h}$ and the classical ordered phase gives way to a fully polarized high-field phase. Consequently, the magnetization increases linearly with the field at leading order in $1/S$, reaching its saturation at $h_{\mathrm{c}0}$.

\begin{figure}
\centering
\includegraphics[width=8.5cm]{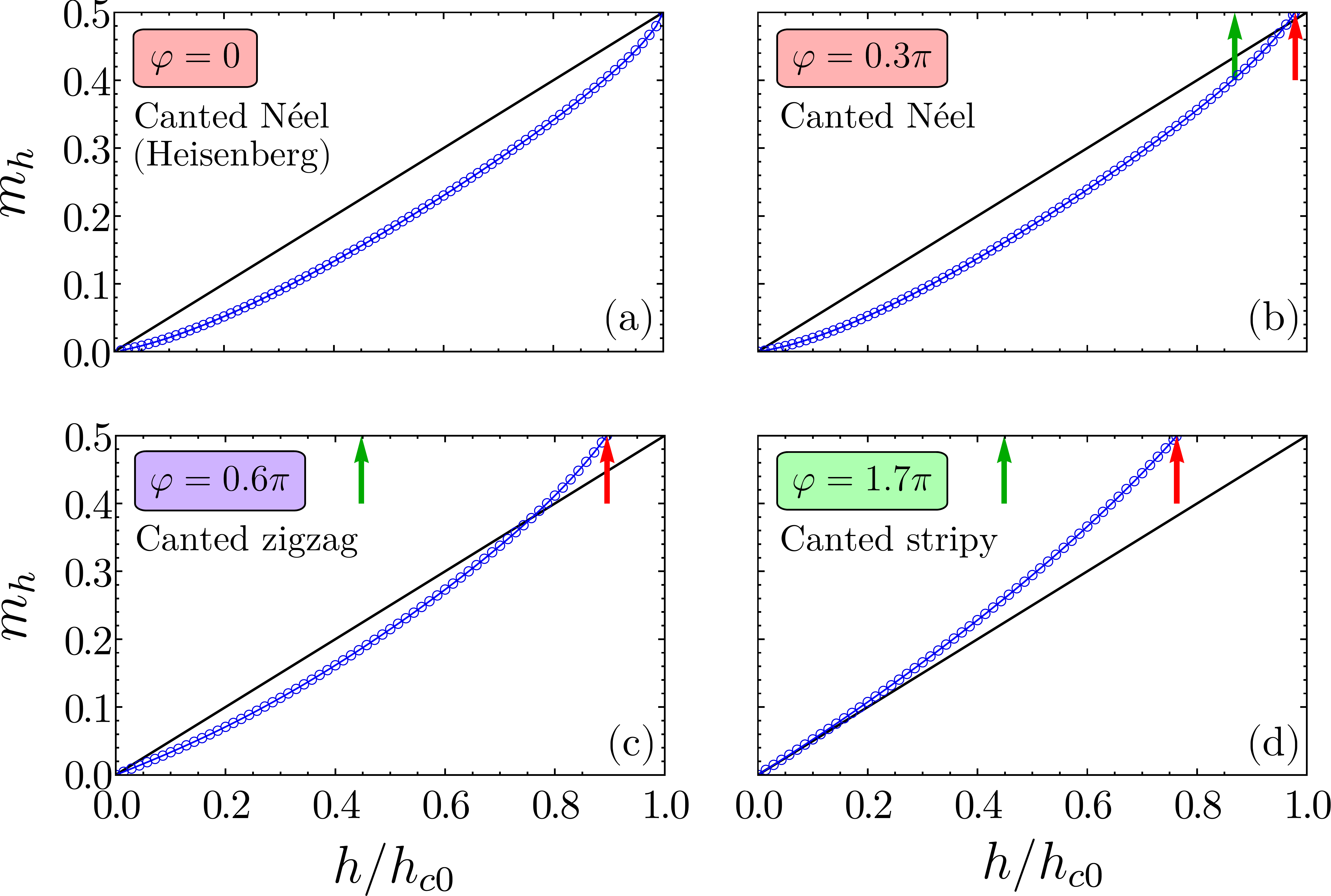}
\caption{Magnetization $m_h$ as a function of field $h$ in the HK model with $J=A\cos\varphi$ and $K=2A\sin\varphi$ in a magnetic field $\textbf{h}\parallel\left[001\right]$, at leading (black) and next-to-leading (blue) order in $1/S$ for $S=1/2$ and different values of $\varphi$. To aid the comparison, the horizontal axes have been rescaled by the respective classical critical fields $h_{\mathrm{c}0}$. Red arrows highlight an unphysical saturation of the magnetization curve, suggesting that, except in the Heisenberg limit (a), phase transitions occur below the classical critical field in (b)--(d). Green arrows indicate the positions of the corrected critical fields according to Secs.~\ref{subsec:1/hc expansion - 2nd order} and \ref{subsec:HF theory}.
\label{fig:mag curves 001}}
\end{figure}

However, such a simple picture changes shape as soon as quantum fluctuations are taken into account. While the SU(2) symmetric point generically exhibits a decrease in $m_{h}$ in the canted N\'eel phase~\citep{zhitomirsky98}, see Fig.~\ref{fig:mag curves 001}(a), a markedly different behavior emerges upon considering $K\neq0$, see Figs.~\ref{fig:mag curves 001}(b)--\ref{fig:mag curves 001}(b)(d). For sufficiently high fields, the $1/S$ correction to $m_{h}$ becomes positive, causing the NLO curves to cross their classical counterparts and saturate below $h_{\mathrm{c}0}$. Yet, because the polarized state is not an eigenstate of the full HK Hamiltonian, Eq.~\eqref{eq:hk}, quantum fluctuations take place even for $h\ge h_{\mathrm{c}0}$ and prevent the magnetization from saturating at any finite field in the high-field phase~\citep{janssen16,janssen17}. Hence, the portions of the NLO magnetization curves right below $h_{\mathrm{c}0}$ for $K \neq 0$ are guaranteed to be unphysical. Although we have presented results for $S=1/2$ and $\mathbf h \parallel [001]$ in Fig.~\ref{fig:mag curves 001}, such an inconsistency applies for all finite values of $S$ and also for other field directions.

We interpret these results as evidence for a reduction of the critical field $h_{\mathrm{c}}$ upon the inclusion of quantum corrections, for the presented values of $\varphi$. Below, we address the question of how this correction to the critical field can be computed in a systematic expansion in $1/S$.


\subsection{Quantum corrections to the direction of magnetic moments} \label{subsec:Cubic terms}

In LSW theory, the angles $\left\{ \boldsymbol{\phi},\boldsymbol{\theta}\right\}$ that parametrize the directions of the spins in the ordered phases are determined via the minimization of the classical Hamiltonian~$\mathcal{H}_{0}\left(\boldsymbol{\phi},\boldsymbol{\theta}\right)$. Consequently, the linear term $\mathcal{H}_{1} \left(\boldsymbol{\phi},\boldsymbol{\theta}\right)$ vanishes. Nevertheless, in dealing with noncollinear magnetic orders such as the canted phases discussed in Sec.~\ref{subsec:NLO mag}, additional single-boson contributions stem from the cubic term, $\mathcal H_3$, and lead to a renormalization of the parametrization angles, $\left\{ \boldsymbol{\phi}, \boldsymbol{\theta}\right\} \to\left\{\boldsymbol{\tilde{\phi}}, \boldsymbol{\tilde{\theta}}\right\}$, which affects physical observables already at NLO order in $1/S$ \citep{zhitomirsky98,coletta12}. In the following, we provide an outline of this procedure and connect it to the results presented in Sec.~\ref{subsec:NLO mag}.

We consider the effects of $\mathcal{H}_{3}$ in our calculations at the mean-field level \citep{zhitomirsky98}. We begin by writing $\mathcal{H}_{3}$ in normal order with respect to the Bogoliubov quasiparticles $b_{\textbf{k}\mu}^{\dagger}$ and $b_{\textbf{k}\mu}$,
\begin{equation}
\mathcal{H}_{3}=\normord{\mathcal{H}_{3}}+\mathcal{H}_{3}^{\left(1\right)},
\end{equation}
such that in $\normord{\mathcal{H}_{3}}$ all creation operators $b_{\textbf{k}\mu}^{\dagger}$ are placed to the left of annihilation operators $b_{\textbf{k}\mu}$.
Since $\normord{\mathcal{H}_{3}}$ only yields corrections beyond NLO in $1/S$ \citep{chubukov94,chernyshev09a,winter17a,rau18}, it will not be considered here, so that we are left with the single-boson term, $\mathcal{H}_{3}^{\left(1\right)}$. The new parametrization angles, $\boldsymbol{\tilde{\phi}}$ and $\boldsymbol{\tilde{\theta}}$, are then determined by rendering the complete linear term zero,
\begin{equation}
S^{3/2}\mathcal{H}_{1}\left(\boldsymbol{\tilde{\phi}},\boldsymbol{\tilde{\theta}}\right)+S^{1/2}\mathcal{H}_{3}^{\left(1\right)}\left(\boldsymbol{\tilde{\phi}},\boldsymbol{\tilde{\theta}}\right)=0.\label{eq:new minimum}
\end{equation}

In the spirit of Eq.~\eqref{eq:H series}, one can expand the new angles around their classical values in a power series in $1/S$,
\begin{align}
\tilde{\phi}_{\mu} & = \sum_{n=0}^{\infty}\left(\frac{1}{S}\right)^{n}\tilde{\phi}_{\mu n}
\equiv \phi_\mu + \frac{1}{S} \delta\phi_\mu + \mathcal O\!\left(\frac{1}{S^2}\right), \label{eq:1/S expansion ph} \\
\tilde{\theta}_{\mu} & =\sum_{n=0}^{\infty}\left(\frac{1}{S}\right)^{n}\tilde{\theta}_{\mu n}
\equiv \theta_\mu + \frac{1}{S} \delta\theta_\mu + \mathcal O\!\left(\frac{1}{S^2}\right), \label{eq:1/S expansion th}
\end{align}
where $\tilde{\phi}_{\mu0}\equiv\phi_{\mu}$, $\tilde{\theta}_{\mu0}\equiv\theta_{\mu}$, $\tilde{\phi}_{\mu1}\equiv\delta\phi_{\mu}$, $\tilde{\theta}_{\mu1}\equiv\delta\theta_{\mu}$, and $\mu = 1, \dots, N_{\mathrm{s}}$. After expanding Eq.~\eqref{eq:new minimum} up to order $S^{1/2}$, we encounter a system of linear equations that can be solved for $\delta\phi_{\mu}$ and $\delta\theta_{\mu}$. Their precise expressions, together with a detailed derivation of the linear system for the HK Hamiltonian, are given in Appendix \ref{sec:app cubic terms}.

With the values of $\left\{\delta\phi_{\mu},\delta\theta_{\mu}\right\}$, we can compute the magnetization curves from the relation
\begin{align}
m_{h} & =\frac{1}{N}\sum_{i\mu}\frac{\textbf{h}}{h}\cdot\left\langle \textbf{S}_{i\mu}\right\rangle =S\sum_{\mu}\cos\theta_{\mu}\nonumber \\
 & -\sum_{\mu}\left(\sin\theta_{\mu}\delta\theta_{\mu}+\frac{\cos\theta_{\mu}}{N}\sum_{\text{\textbf{k}}}\left\langle a_{\text{\textbf{k}}\mu}^{\dagger}a_{\text{\textbf{k}}\mu}\right\rangle \right)+\mathcal{O}\!\left(\frac{1}{S}\right),\label{eq:mh <>}
\end{align}
where the expectation values are calculated with respect to the vacuum of the Bogoliubov quasiparticles. Although Eqs.~\eqref{eq:mh -dEgsdh} and \eqref{eq:mh <>} are derived from different definitions and even require different levels of calculation within spin-wave theory, they must produce identical results, as both consistently include all contributions up to NLO order in $1/S$ \citep{zhitomirsky98,coletta12}. We have explicitly checked for different values of $\varphi$ that Eqs.~\eqref{eq:mh -dEgsdh} and \eqref{eq:mh <>} indeed lead to the same magnetization curves. This nontrivial crosscheck corroborates the calculations presented below which involve the angle corrections.

Notably, Eq.~\eqref{eq:mh <>} provides a new way to interpret the plots in Fig.~\ref{fig:mag curves 001}: While the second term inside the parentheses always leads to a reduction in the magnetization, the first term can be either positive or negative, depending on the sign of $\delta\theta_{\mu}$. Therefore, an increase in the magnetization can be understood as a consequence of a decrease in the canting angles $\left(\delta\theta_{\mu}<0\right)$, which expresses a tendency for premature alignment of the spins along the direction of the magnetic field. This supports our claim that the critical field is reduced to a value $h_{\mathrm{c}}\le h_{\mathrm{c}0}$ upon the inclusion of quantum corrections for the presented values of $\varphi$.

We emphasize that, even though the corrections to the magnetization computed in Eqs.~\eqref{eq:mh -dEgsdh} and \eqref{eq:mh <>} are equivalent, the calculation of the angle corrections required in the latter approach turns out to be essential for the determination of the quantum corrections to the critical field, as we discuss now.


\subsection{Quantum corrections to second-order transition lines: Ordered side} \label{subsec:1/hc expansion - 2nd order}

We can now turn to the goal of constructing a consistent $1/S$ expansion for the critical field $h_{\mathrm{c}}$, which will ultimately enable us to investigate the effect of quantum fluctuations on the phase diagram for arbitrary values of $S$. To start with, we must address the question of how to consistently define the critical field. While this is simply a matter of energy level crossings for first-order phase transitions, the answer is not at all obvious in the case of continuous phase transitions. Thus, let us focus on the latter case for a moment. If we were to base ourselves solely on properties of the ordered phases and on the results for $\varphi=0$, any of the following, apparently equivalent, defining conditions would seem to fit: (i) the saturation of the magnetization; (ii) the vanishing of quantum fluctuations; (iii) $\cos\tilde{\theta}_{\mu}(h_{\mathrm{c}}) = 1$ for all spins in the unit cell, $\mu = 1,\dots,N_{\mathrm{s}}$. However, our discussion in Sec.~\ref{subsec:NLO mag} allows us to rule out the first two immediately, since neither of these properties characterize the polarized phase in the presence of the Kitaev term.

Hence, we move on to the last criterion, which is most intimately connected to a semiclassical picture. In terms of the notation introduced in Sec.~\ref{subsec:Cubic terms}, the condition $\cos\tilde{\theta}_{\mu}(h_{\mathrm{c}}) = 1$ can be written as
\begin{equation}
1-\frac{1}{S}\tan\theta_{\mu}(h_{\mathrm{c}})\, \delta\theta_{\mu}(h_{\mathrm{c}})+\mathcal{O}\!\left(\frac{1}{S^{2}}\right) = \frac{1}{\cos\theta_{\mu}(h_{\mathrm{c}})}\,. \label{eq:cos(thtilde) = 1}
\end{equation}
However, we can simplify Eq.~\eqref{eq:cos(thtilde) = 1} by noting that all ordered phases which undergo continuous field-induced phase transitions in this study entail uniform canting at the classical level, and are thus governed by the equation $\cos\theta_{\mu}(h) \equiv \cos\theta(h) = h/h_{\mathrm{c}0}$ for all $\mu$. With this, we arrive at
\begin{equation}
\frac{1/h_{\mathrm{c}}}{1/h_{\mathrm{c}0}}=1-\frac{1}{S}\tan\theta(h_{\mathrm{c}0})\,\delta\theta(h_{\mathrm{c}0})+\mathcal{O}\!\left(\frac{1}{S^{2}}\right),\label{eq:1/hc exp cos(thtilde)}
\end{equation}
which gives a consistent $1/S$ expansion not for $h_{\mathrm{c}}$, but for $1/h_{\mathrm{c}}$, provided that the products $\tan\theta_{\mu}(h)\,\delta\theta_{\mu}(h)$ are analytic at $h_{\mathrm{c}0}$ and converge to the same value for all $\mu$ as $h\to h_{\mathrm{c}0}^{-}$.

At a first glance, it might seem that Eq.~\eqref{eq:1/hc exp cos(thtilde)} implies that the NLO contribution to $1/h_{\mathrm{c}}$ is zero, since $\tan\theta(h_{\mathrm{c}0}) = 0$. This is indeed what happens for a pure Heisenberg interactions. Nevertheless, as proven analytically for $\textbf{h}\parallel\left[001\right]$ in Appendix \ref{sec:app cubic terms}, $\delta\theta(h_{\mathrm{c}0})$ actually diverges upon the inclusion of the smallest Kitaev exchange. In fact, it does so in a way that, except at the Kitaev points $\varphi = \pm \pi/2$, the product $\tan\theta(h_{\mathrm{c}0})\,\delta\theta(h_{\mathrm{c}0})$ is always unique and finite, thus meeting the requirements for the validity of Eq.~\eqref{eq:1/hc exp cos(thtilde)}.

Another observation here is that Eq.~\eqref{eq:1/hc exp cos(thtilde)} follows directly from the condition $\cos\tilde{\theta}_{\mu}(h_{\mathrm{c}}) = 1$, without the need to postulate the existence of a $1/S$ expansion for any specific function of $h_{\mathrm{c}}$. This way, $1/h_{\mathrm{c}}$ emerges as a natural quantity to be considered in this framework. In general, there is of course a one-to-one correspondence between the expansions of $1/h_{\mathrm{c}}$ and $h_{\mathrm{c}}$, which can be used to deduce the coefficients of one expansion from those of the other.
However, as in any asymptotic series, when explicitly evaluating the truncated series at finite values of $S$, the numerical values obtained depend on whether one considers the inverse of the expansion of $1/h_{\mathrm{c}}$ or the expansion of $h_{\mathrm{c}}$ itself.
In fact, as we shall see below, the results obtained by evaluating the expansion of $1/h_{\mathrm{c}}$ for small values of $S$ turn out to be more consistent with the physical expectation. When computing explicit corrections to the critical field, we therefore evaluate Eq.~\eqref{eq:1/hc exp cos(thtilde)} directly, without further solving for $h_{\mathrm{c}}$.

Finally, we emphasize that, even after assuming that the classical magnetic order is characterized by uniform canting, our formalism allows the corrections to the canting angle to vary between different magnetic sublattices at fields below the classical critical field, $h<h_{\mathrm{c}0}$. Such a distinction will prove to be important later on, when we deal with a particular manifestation of quantum order-by-disorder (Sec.~\ref{subsec:neel ObD}).


\subsection{Quantum corrections to second-order transition lines: Disordered side} \label{subsec:HF theory}

As an alternative to the procedure described in Sec.~\ref{subsec:1/hc expansion - 2nd order}, one can construct a consistent $1/S$ expansion for $1/h_{\mathrm{c}}$ by applying spin-wave theory to the high-field polarized phase. The occurrence of a continuous transition to a symmetry-broken ordered phase is then signaled by the closure of the magnon excitation gap, which expresses the condensation of magnons in the system. Parenthetically, we note that the transition between the high-field phase and a topological Z$_2$ spin liquid would involve the closure of a vison gap instead, but this is beyond the realm of a $1/S$ expansion.

While the classical phase boundaries are obtained from LSW theory~\cite{janssen16}, NLO contributions generally require one to consider both cubic and quartic terms of the spin-wave Hamiltonian, Eq.~\eqref{eq:H series}. As the classical reference state in the polarized phase is collinear, the cubic part of the spin-wave Hamiltonian is identically zero (see Appendix \ref{sec:app NLSWT polarized} for further details), so that we can focus solely on the quartic terms.

Once more, we begin by writing $\mathcal{H}_{4}$ in normal order,
\begin{equation}
\mathcal{H}_{4}=\normord{\mathcal{H}_{4}}+\normord{\mathcal{H}_{4}^{\left(2\right)}}+\mathcal{H}_{4}^{\left(0\right)}.
\end{equation}
Here, $\normord{\mathcal{H}_{4}^{\left(2\right)}}$ and $\mathcal{H}_{4}^{\left(0\right)}$ represent the (also normal-ordered) quadratic and zero-order contributions which result as a byproduct of the bosonic commutation relations.
Since $\mathcal{H}_{4}^{\left(0\right)}$ consists of a momentum-independent shift in the ground-state energy and $\normord{\mathcal{H}_{4}}$ describes magnon decay processes, which only yield corrections beyond NLO in $1/S$ \citep{chubukov94,chernyshev09a,winter17a,rau18}, they can both be neglected. At NLO, the $1/S$ expansion is therefore equivalent to a Hartree-Fock approximation in this phase. The quantum corrections to the magnon spectrum thus follow entirely from
\begin{equation}
\mathcal{H}_{4}^{\left(2\right)}=\frac{1}{2}\sum_{\textbf{k}}\beta_{\textbf{k}}^{\dagger}\Sigma_{\textbf{k}}\beta_{\textbf{k}},\label{eq:H4(2)}
\end{equation}
which differs from $\normord{\mathcal{H}_{4}^{\left(2\right)}}$ by momentum-independent terms. Further details on the calculation of the static self-energy, $\Sigma_{\textbf{k}}$, and explicit results for $\textbf{h}\parallel\left[001\right]$ are given in Appendix \ref{sec:app NLSWT polarized}. After adding Eq.~\eqref{eq:H4(2)} to $\mathcal{H}_{2}$, we arrive at
\begin{equation}
\mathcal{H}_{2}+\mathcal{H}_{4}^{\left(2\right)}=\frac{1}{2}\sum_{\textbf{k}}\beta_{\textbf{k}}^{\dagger}\left(S\sigma_{3}\Omega_{\textbf{k}}+\Sigma_{\textbf{k}}\right)\beta_{\textbf{k}},\label{eq:H2 + H4(2)}
\end{equation}
where $\Omega_{\textbf{k}}=\text{diag}\left(\epsilon_{\textbf{k}1},\epsilon_{\textbf{k}2},\epsilon_{-\textbf{k}1},\epsilon_{-\textbf{k}2}\right)$ and $\sigma_{3}=\text{diag}\left(\mathds{1}_{N_{\mathrm{s}}},-\mathds{1}_{N_{\mathrm{s}}}\right)$ is a $2N_{\mathrm{s}}\times2N_{\mathrm{s}}$ generalization of the diagonal Pauli matrix.

The corrected spectrum, $E_{\textbf{k}\mu}$, is then determined by applying nondegenerate perturbation theory to Eq.~\eqref{eq:H2 + H4(2)}. Because we have expressed the perturbation in terms of the bosons which diagonalize the (unperturbed) LSW Hamiltonian, the result is simply
\begin{equation}
E_{\textbf{k}\mu}=S\epsilon_{\textbf{k}\mu}+\Sigma_{\textbf{k}}^{\mu\mu}.\label{eq:NLO spectrum}
\end{equation}
Note that only the diagonal elements of $\Sigma_{\textbf{k}}$ enter the spectrum. Together with the fact that $\Sigma_{\textbf{k}}$ is Hermitian, this guarantees that $E_{\textbf{k}\mu}$ is real. For explicit results in the case of $\textbf{h}\parallel\left[001\right]$, see Appendix \ref{sec:app NLSWT polarized}.

With this, one can use Eq.~\eqref{eq:NLO spectrum} to read off the first two terms in the $1/S$ expansion of the spin-wave gap
\begin{equation}
\Delta\!\left(\varphi,\frac{1}{h},\frac{1}{S}\right)=S\sum_{n=0}^{\infty}\left(\frac{1}{S}\right)^{n}\Delta_{n}\!\left(\varphi,\frac{1}{h}\right).\label{eq:Delta 1/S expansion}
\end{equation}
By attributing the index $\mu=1$ to the lower band of the spectrum and denoting the instability wave vector, i.e., the wave vector at which the gap closes at leading order, by $\textbf{Q}=\textbf{Q}\left(\varphi\right)$, we find that $\Delta_{0}\equiv\epsilon_{\textbf{Q}1}$ and $\Delta_{1}\equiv\Sigma_{\textbf{Q}}^{11}$ for $h$ above, but not too far from, the classical critical field~$h_\mathrm{c0}$.

\begin{figure}
\centering
$\vcenter{\hbox{\includegraphics[width=4.2cm]{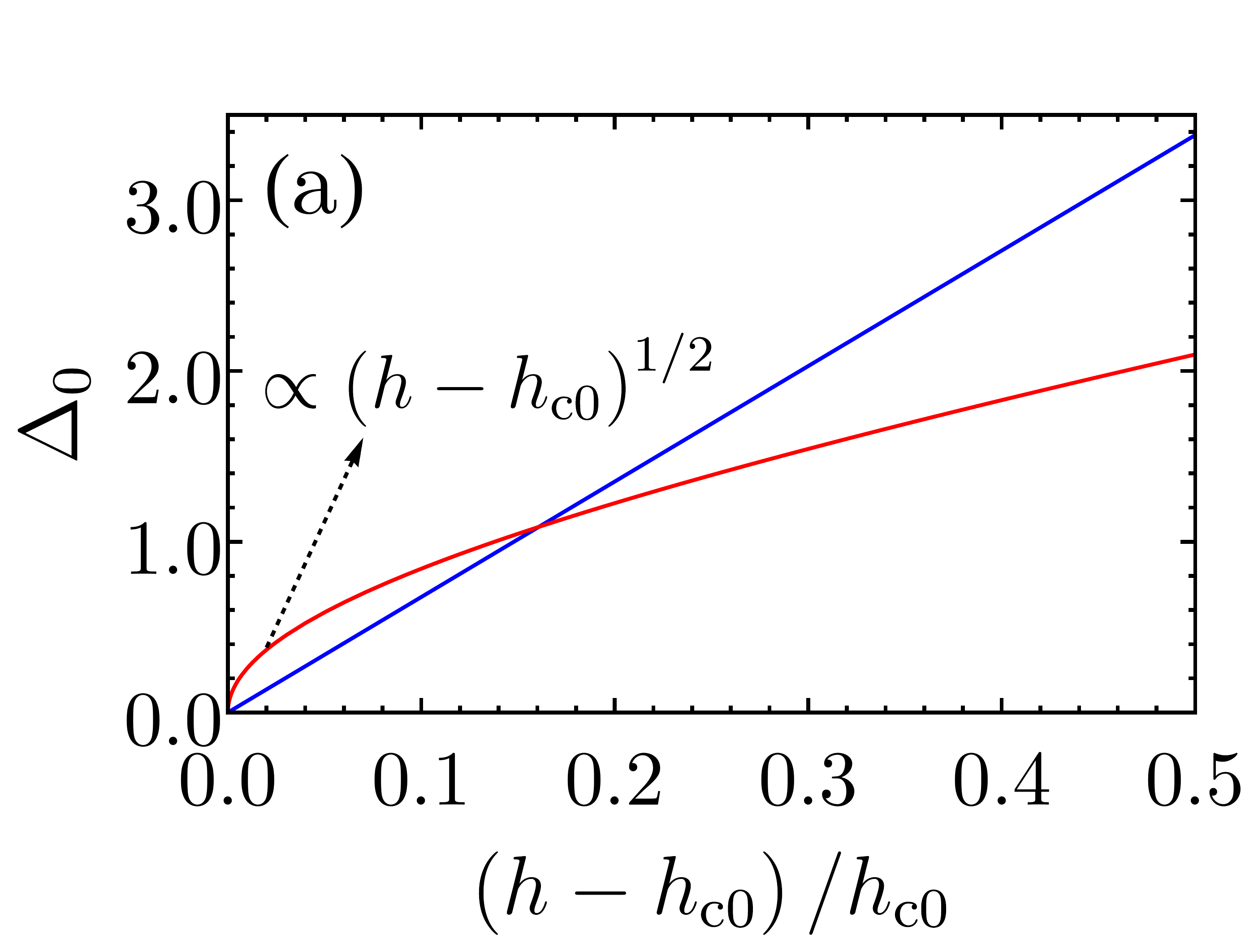}}}$
$\vcenter{\hbox{\includegraphics[width=4.2cm]{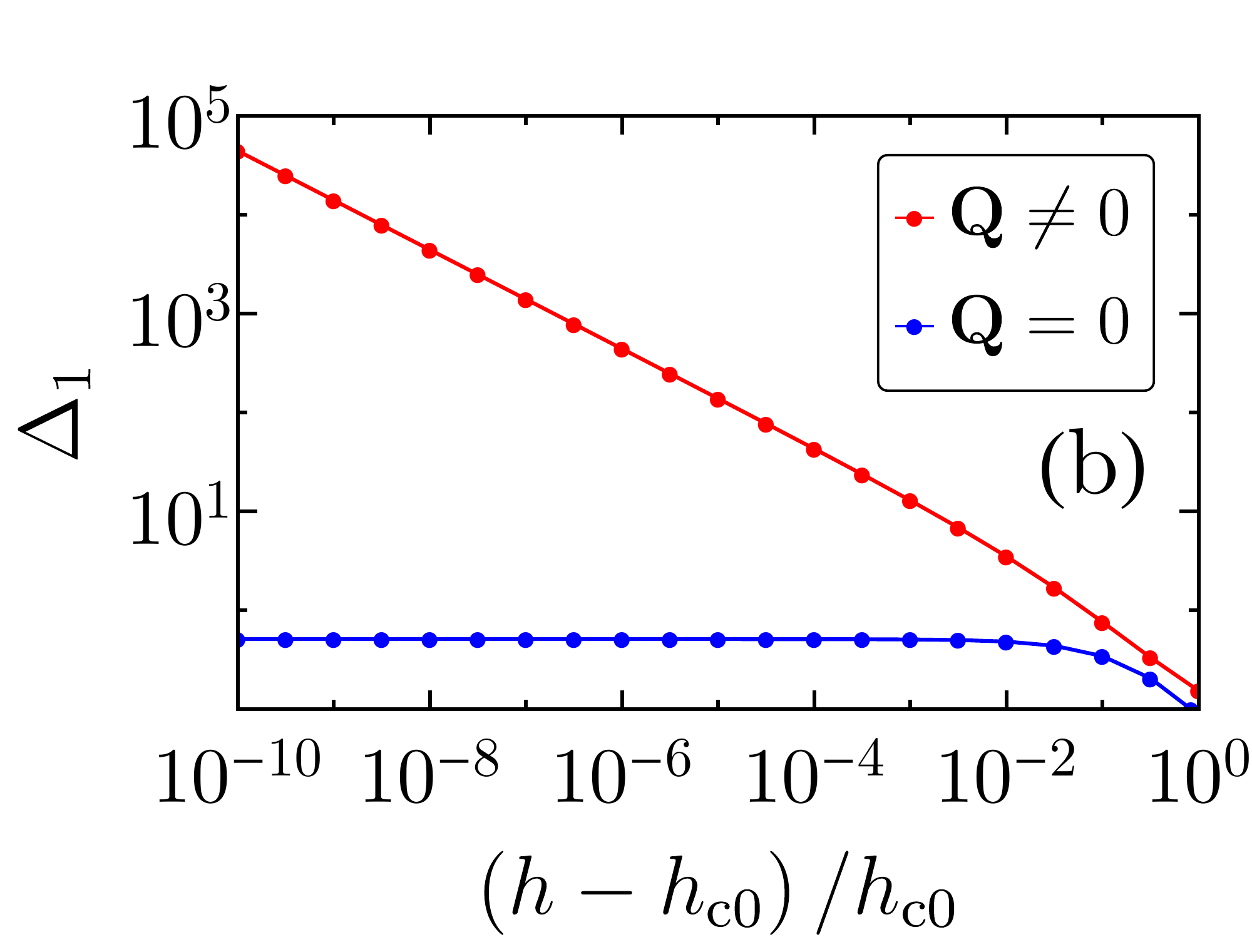}}}$
\caption{(a) Leading-order and (b) NLO contributions to the magnon gap in the polarized phase for $\varphi=0.3\pi$ (blue) and $\varphi=0.7\pi$ (red) with $\textbf{h}\parallel\left[001\right]$. The two values of $\varphi$ are representative for the cases of vanishing (blue) and finite (red) instability wave vectors $\mathbf{Q}$, respectively. These different behaviors justify the need for different conditions to determine the expansion of $1/h_{\mathrm{c}}$.}
\label{fig:gap plots}
\end{figure}

Now we are in the position to construct another $1/S$ expansion for $1/h_{\mathrm{c}}$, based on the criterion $\Delta\to 0$ as $h\to h_{\mathrm{c}}$. There is but one final caveat to bear in mind: The expansion of a physical observable in the vicinity of a quantum phase transition is well defined only if the observable itself is analytic at this transition \citep{joshi2015a,joshi2015b}.
Figure \ref{fig:gap plots}(a) illustrates two different behaviors for the evolution of the gap $\Delta$ as a function of the reduced magnetic field $t\equiv\left(h-h_{\mathrm{c}0}\right)/h_{\mathrm{c}0}$:
Above the N\'eel phase, the gap closes at wave vector $\textbf{Q}=\boldsymbol{0}$ and follows $\Delta_{0}\propto t$. In contrast, in those cases where the gap closes at $\textbf{Q}\ne\boldsymbol{0}$, we have $\Delta_{0}\propto t^{1/2}$, hence $\Delta_0$ is nonanalytic at $h_{\mathrm{c}0}$ whereas $\Delta_0^2$ is analytic.
In the first case, $\Delta_{0}\propto t$, we employ the condition $\Delta(1/h_{\mathrm{c}})=0$ to arrive at
\begin{equation}
\frac{1/h_{\mathrm{c}}}{1/h_{\mathrm{c}0}}=1+\left.\frac{1}{S\,h}\frac{\Delta_{1}}{\left(\partial\Delta_{0}/\partial h\right)}\right|_{h_{\mathrm{c}0}}+\mathcal{O}\!\left(\frac{1}{S^{2}}\right), \quad \text{for $\mathbf Q = 0$.}
\label{eq:1/hc exp quartic Q=0}
\end{equation}
In the second case, $\Delta_{0}\propto t^{1/2}$, we instead expand $\Delta^2$ and use the condition $\Delta^{2}(1/h_{\mathrm{c}})=0$ \citep{joshi2015a,joshi2015b} to find
\begin{equation}
\frac{1/h_{\mathrm{c}}}{1/h_{\mathrm{c}0}}=1+\left.\frac{2}{S\,h}\frac{\Delta_{0}\Delta_{1}}{\left(\partial\Delta_{0}^{2}/\partial h\right)}\right|_{h_{\mathrm{c}0}}+\mathcal{O}\!\left(\frac{1}{S^{2}}\right), \quad \text{for $\mathbf Q \neq 0$}.
\label{eq:1/hc exp quartic Q!=0}
\end{equation}

Interestingly, the NLO contribution to Eq.~\eqref{eq:1/hc exp quartic Q!=0} results from the product of $\Delta_{0}$, which vanishes at $h_{\mathrm{c}0}$, and $\Delta_{1}$. Therefore, $1/h_{\mathrm{c}}$ will only have a correction of order $1/S$ if $\Delta_{1}$ diverges as $t^{-1/2}$ at criticality. As displayed in Fig.~\ref{fig:gap plots}(b), this is precisely what happens for $\textbf{Q}\ne\boldsymbol{0}$. In contrast,  when $\textbf{Q}=\boldsymbol{0}$, Fig.~\ref{fig:gap plots}(b) shows that $\Delta_{1}$ converges at $h_{\mathrm{c}0}$, supporting the need to employ Eq.~\eqref{eq:1/hc exp quartic Q=0} in this case.


\subsection{Quantum corrections to first-order transition lines} \label{subsec:1st-order PTs}

So far, we have tackled the issue of how phase boundaries related to continuous transitions change at NLO in $1/S$. We now aim to do the same for discontinuous transitions. In this case, quantum corrections to the phase boundaries follow from a direct comparison between the ground-state energies of competing phases. By noting that Eq. \eqref{eq:epsilon_q} gives the complete NLO term in Eq. \eqref{eq:ground-state energy} for an arbitrary magnetic order, we thus conclude that LSW theory is sufficient to study the displacement of first-order transition lines, in contrast to the case of continuous transitions.

Consider a point $(\varphi,1/h)=\left(\varphi_{\mathrm{t}0},1/h_{\mathrm{t}0}\right)$ in parameter space, lying on top of a classical first-order transition line. One way to evaluate the shift in the phase boundary is to compute the quantum correction to $1/h_{\mathrm{t}0}$ while keeping $\varphi$ fixed. By demanding the equality of the ground-state energies of the phases above (a) and below (b) the transition, $E_{\mathrm{a}}(\varphi_{\mathrm{t}0},1/h_{\mathrm{t}},1/S) = E_{\mathrm{b}}(\varphi_{\mathrm{t}0},1/h_{\mathrm{t}},1/S)$, and assuming a $1/S$ expansion for $1/h_{\mathrm{t}}$, we find
\begin{equation}
\frac{1/h_{\mathrm{t}}}{1/h_{\mathrm{t}0}}=1+\frac{1}{S\,h_{\mathrm{t}0}}\left.\frac{E_{\mathrm{b}1}-E_{\mathrm{a}1}}{\frac{\partial}{\partial h}( E_{\mathrm{b}0}-E_{\mathrm{a}0})}\right|_{\frac{1}{h_{\mathrm{t}0}}}+\mathcal{O}\!\left(\frac{1}{S^{2}}\right).\label{eq:1/ht 1st order 1/S exp}
\end{equation}

Conversely, one can also study the displacement of a phase boundary by tracking the change in $\varphi_{\mathrm{t}0}$ for a fixed value of $h$. The condition $E_{\mathrm{l}}(\varphi_{\mathrm{t}},1/h,1/S) = E_{\mathrm{r}}(\varphi_{\mathrm{t}},1/h,1/S)$, where the subindices denote the ground states to the left (l) and to the right (r) of the transition line, then yields
\begin{equation}
\frac{\varphi_{\mathrm{t}}}{\varphi_{\mathrm{t}0}} = 1-\frac{1}{S}\left.\frac{E_{\mathrm{l}1}-E_{\mathrm{r}1}}{\frac{\partial}{\partial\varphi} (E_{\mathrm{l}0}-E_{\mathrm{r}0})}\right|_{\varphi_{\mathrm{t}0}}+\mathcal{O}\!\left(\frac{1}{S^{2}}\right).\label{eq:phit 1st order 1/S exp}
\end{equation}

When computing first-order phase boundaries in the $\varphi$-$h$ plane in the next sections, we shall alternate between Eqs.~\eqref{eq:1/ht 1st order 1/S exp} and \eqref{eq:phit 1st order 1/S exp}. In general, both schemes are fully equivalent order by order in the expansion. However, when evaluating the truncated series at particular small values of $S$, the numerical estimates for the phase boundaries can differ. We will use Eq.~\eqref{eq:1/ht 1st order 1/S exp} when we wish to compare the displacement of a certain phase boundary with respect to the critical field above it. Equation \eqref{eq:phit 1st order 1/S exp}, in turn, will prove most useful in studying horizontal shifts in phase boundaries.


\section{Results for $\textbf{h}\parallel\left[001\right]$} \label{sec:h001}

In this section, we apply the theory presented above to extract concrete results for the HK model in a $\left[001\right]$ field. In principle, the fact that we have developed a consistent $1/S$ expansion enables us to evaluate phase diagrams for arbitrary values of $S$. We expect reliable results for large enough $S$ and/or sufficiently away from the Kitaev limits $\varphi = \pm \pi/2$, where the $1/S$ expansion breaks down below $h_{\mathrm{c}0}$ due to a massive degeneracy of classical states \citep{baskaran08,ioannis18}. In the following, we shall focus primarily on the cases $S=1/2$, $1$, $3/2$, and $2$. As discussed in the introduction, the first three cases might be of relevance for current experiments~\citep{janssen19,takagi19,zvereva15,lee20}. The case $S=2$ already turns out to be quite close to the classical limit $S \to \infty$ qualitatively~\citep{janssen16}.

\subsection{Critical field}

\begin{figure}[t]
\centering
\includegraphics[width=8.5cm]{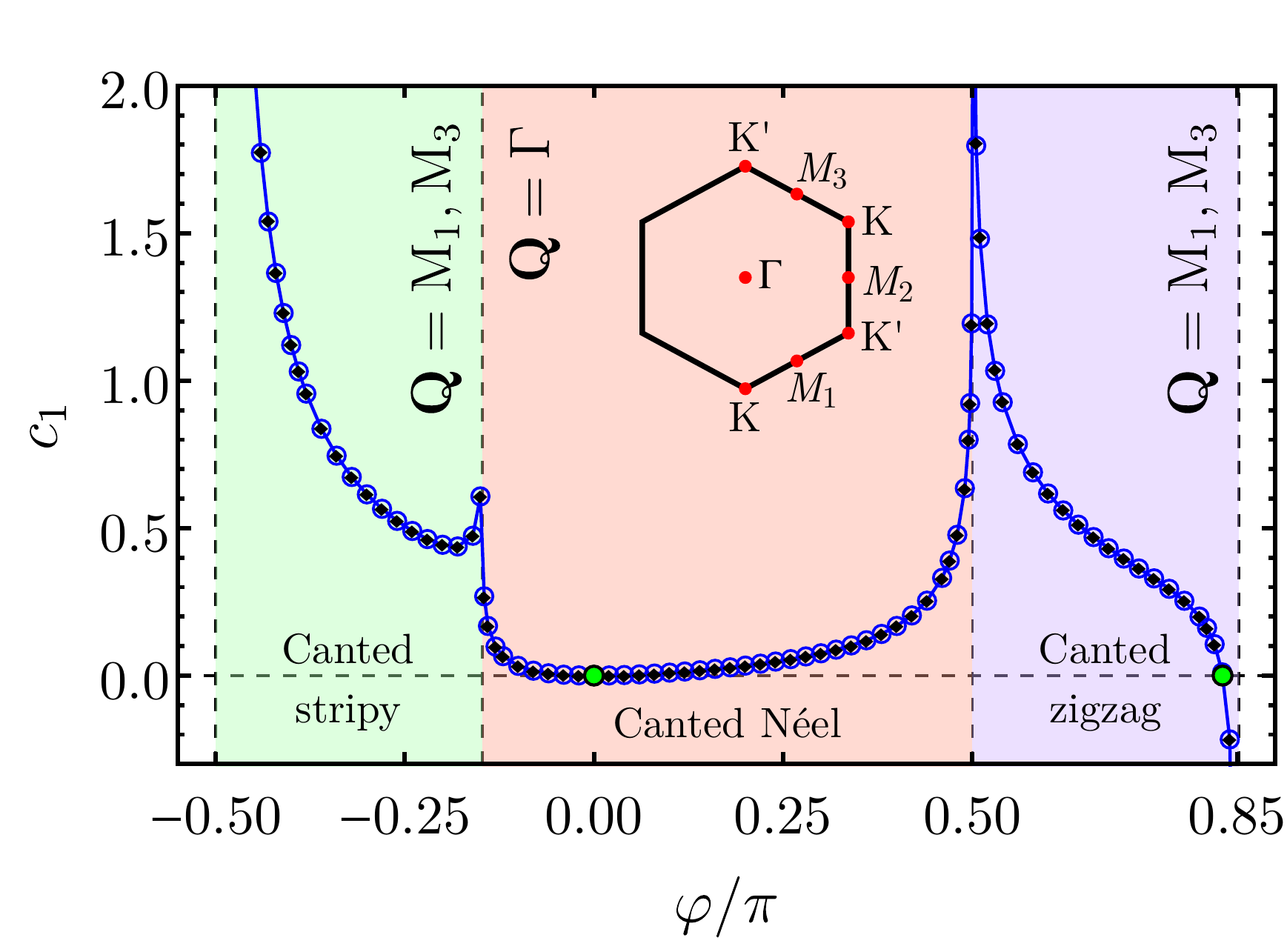}
\caption{$\mathcal O(1/S)$ coefficient $c_1$ in the expansion of the inverse of the critical field, see Eq.~\eqref{eq:1/hc 1/S exp}, as a function of $\varphi$ in the HK model with couplings $J=A\cos\varphi$ and $K=2A\sin\varphi$ in a magnetic field $\textbf{h}\parallel\left[001\right]$. Results using Eq.~\eqref{eq:1/hc exp cos(thtilde)} in the ordered phases (blue open circles) are fully equivalent within our numerical precision with those that follow from applying Eq.~\eqref{eq:1/hc exp quartic Q=0} for instability wave vector $\textbf{Q} = 0$ and Eq.~\eqref{eq:1/hc exp quartic Q!=0} for $\textbf{Q} \neq 0$ in the disordered phase (black diamonds). The inset shows the locations in the first Brillouin zone of the various instability wave vectors corresponding to different intervals of $\varphi$ ($\text{M}_1$, $\text{M}_3$, and $\Gamma$) and different field directions ($\mathrm K$, $\mathrm K'$, and $\mathrm M_2$). The blue line is a guide to the eye. Green dots at $\varphi=0$ and $\varphi\approx0.83\pi$ denote points where the leading-order correction to the critical field vanishes. \label{fig:Correction to 1/hc 001}}
\end{figure}

Let us begin by discussing the changes in the critical field. In Secs.~\ref{subsec:1/hc expansion - 2nd order} and \ref{subsec:HF theory}, we provided two alternatives to evaluate the expansion
\begin{equation}
\frac{1/h_{\mathrm{c}}}{1/h_{\mathrm{c}0}}=1+\sum_{n=1}^{\infty}\left(\frac{1}{S}\right)^{n}c_{n}\label{eq:1/hc 1/S exp}
\end{equation}
up to order $n=1$. As they were based on distinct physical observables and were derived from different classical reference states, the resulting expressions for $c_{1}$ involve apparently unrelated quantities. Yet, after applying both for a range of values of $\varphi$ in the interval with nonzero $h_{\mathrm{c}0}$, we find that Eq.~\eqref{eq:1/hc exp cos(thtilde)} and the combination of Eqs.~\eqref{eq:1/hc exp quartic Q=0} and \eqref{eq:1/hc exp quartic Q!=0} are in fact fully equivalent, see Fig.~\ref{fig:Correction to 1/hc 001}. In addition to confirming the accuracy of our calculations, such an equivalence suggests that one of the methods can be dismissed in favor of the other, even when different field directions are studied. For our purposes, the expansion based on the corrected canting angles turns out to be more efficient, since the application of spin-wave theory to ordered phases is at any rate necessary to analyze first-order phase transitions.

\begin{figure*}
\centering
\includegraphics[width=17cm]{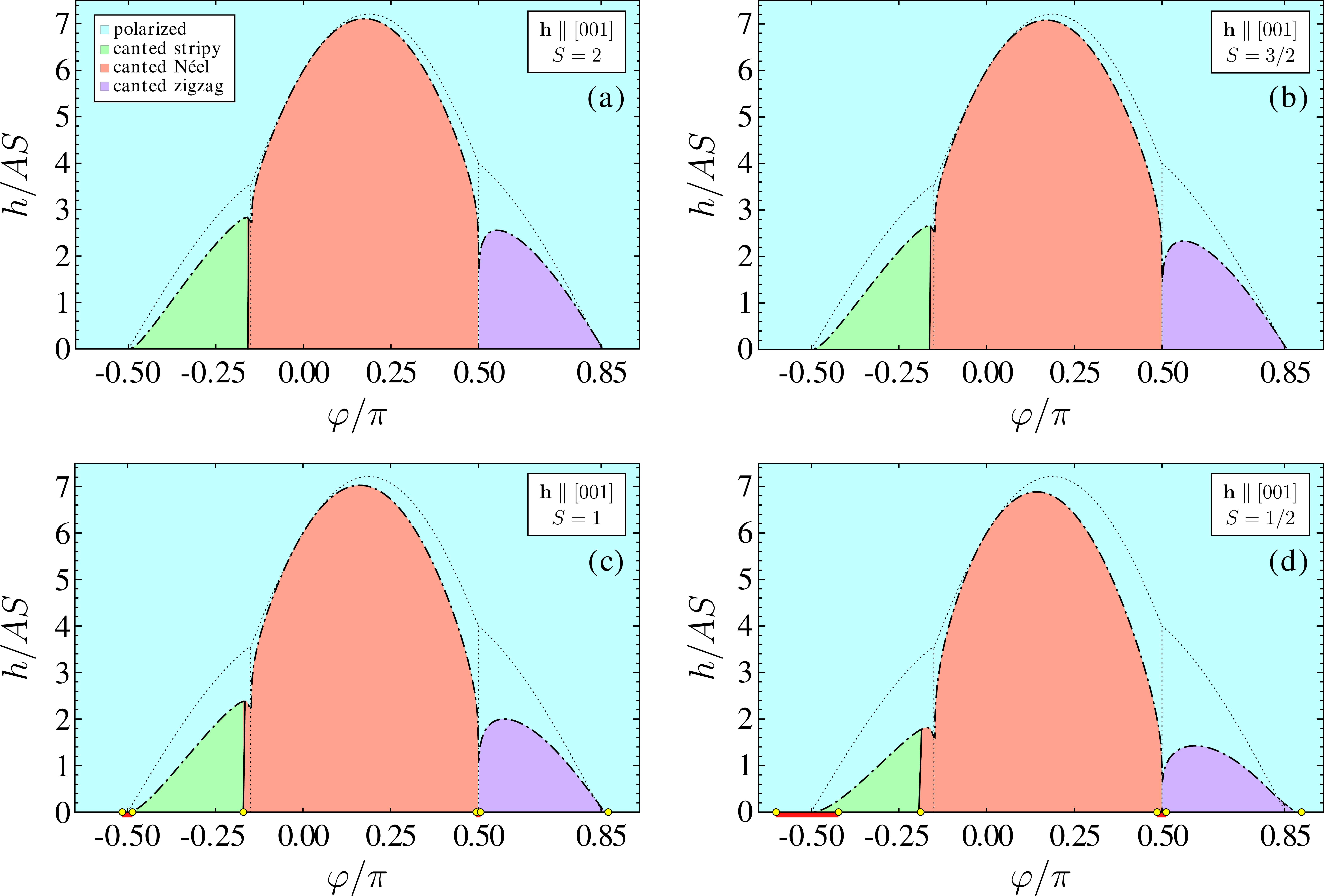}
\caption{\label{fig:pdsh001}%
Phase diagram of the HK model at $T=0$ with $J=A\cos\varphi$ and $K=2A\sin\varphi$ in a magnetic field $\textbf{h}\parallel\left[001\right]$, at next-to-leading order in $1/S$ for (a) $S=2$, (b) $S=3/2$, (c) $S=1$, and (d) $S=1/2$. Dot-dashed and solid lines mark continuous and first-order phase transitions, respectively, whereas the light dotted lines represent the classical phase boundaries, which formally correspond to the limit $S\to\infty$ \citep{janssen16}. The yellow dots added to the $S=1$ and $S=1/2$ diagrams show the $h=0$ phase boundaries according to (c) an infinite density matrix renormalization group study \citep{dong19} and (d) 24-site exact diagonalization results \citep{chaloupka13}. In both cases, the red stripes below the horizontal axis indicate the domains of spin liquid phases. Note that the AF Kitaev spin liquid near $\varphi = \pi/2$ is expected to cover a sizable field range \citep{janssen16,hickey19}, which is not contemplated by our semiclassical expansion~\citep{order}.}
\end{figure*}

Figure \ref{fig:Correction to 1/hc 001} also shows that the corrections to the critical field are finite everywhere except at the Kitaev points $\varphi=\pm\,\pi/2$. Nonetheless, we see that $c_{1}$ is, if not greater than, often comparable to $1$. According to Eq.~\eqref{eq:1/hc exp cos(thtilde)}, this means that the condition $\tan\theta\,\delta\theta(h_{\mathrm{c}0})\ll S$ seldom holds for small values of $S$, and hence that the applicability of a $1/S$ expansion for $h_{\mathrm{c}}$ is limited, as anticipated in Sec.~\ref{subsec:1/hc expansion - 2nd order}.

In fact, a $1/S$ expansion for $h_{\mathrm{c}}$ only becomes reliable in the vicinity of two special values of $\varphi$ for which $c_{1}=0$. One of these is naturally the Heisenberg point, $\varphi=0$, where quantum fluctuations vanish for $h\ge h_{\mathrm{c}0}$. The other occurs near the edge of the canted zigzag phase, at $\varphi\approx0.83\pi$. To the best of our knowledge, there is no special symmetry emerging at this point, so that its precise position should shift as higher orders in $1/S$ are considered. However, it marks a change in the sign of $c_{1}$, which indicates that the critical field \emph{increases} in a small region to the right of $\varphi\approx 0.83\pi$.

Besides the continuous order-to-disorder quantum phase transitions as functions of the field, the classical phase diagram of the HK model in a $[001]$ field has two discontinuous order-to-order transition lines as functions of the interaction parameter $\varphi$. Here, we consider only the transition between the canted N\'eel and stripy states for ferromagnetic $K < 0$. As the classical boundary is a line of constant $\varphi$, we compute the NLO contribution using Eq.~\eqref{eq:phit 1st order 1/S exp}. For the small values of $S$ considered here, we expect the other transition line near the antiferromagnetic Kitaev point at $\varphi =\pi/2$ to be superimposed by a quantum-spin-liquid phase~\cite{gohlke18,zhu20,hickey20,xu20}, which cannot be described within a semiclassical formalism such as spin-wave theory \citep{order}.


\subsection{Phase diagram}

The resulting phase diagrams for a field along the $[001]$ direction are shown for different values of $S$ in Fig.~\ref{fig:pdsh001}. There we can see that NLO contributions (solid and dot-dashed lines) lead to substantial quantitative modifications as compared to the classical phase boundaries (light dotted lines). We find pronounced reductions in the critical field in large parts of the phase diagram, especially in the central portion of the canted zigzag and near the triple point separating the canted stripy, canted Néel and polarized phases. Note that the determination of the corrections to the location of this triple point necessarily involves $1/S$ expansions of different observables, leading to the nonmonotonic behavior of the order-disorder transition line visible near $\varphi \approx -0.15 \pi$.

Furthermore, the phase diagrams reflect the fact that the canted Néel is more stable than the canted stripy by exhibiting a leftward shift in the boundary between both phases. This feature is also observed in numerical studies performed at $h=0$ for both $S=1/2$ \citep{chaloupka13,iregui14,gotfryd17,gohlke17} and $S=1$ \citep{dong19}. For comparison purposes, we reproduce the 24-site exact diagonalization and infinite density matrix renormalization group results from Refs.~\citep{chaloupka13} and \citep{dong19}, respectively, as yellow dots in Figs.~\ref{fig:pdsh001}(d) and \ref{fig:pdsh001}(c). At $h=0$, our spin-wave calculations show the Néel-stripy transition occurring at $\varphi_{\mathrm{t}} \approx -0.193 \pi$ for $S=1/2$, which is in good quantitative agreement with the result $\varphi^\text{ED}_\mathrm{t} \approx -0.189 \pi$ from Ref.~\citep{chaloupka13}. Similar conclusions follow from comparing our estimation to the data obtained in other numerical studies for $S=1/2$ and $S=1$. In the latter case, our result $\varphi_\mathrm{t} \approx -0.170\pi$ coincides with that from Ref.~\cite{dong19} up to the third decimal place.

Finally, we call attention to the rightmost portion of the canted zigzag phase, where the NLO contributions to $1/h_{\mathrm{c}}$ indicate an increase in the critical field. The validity of the $1/h_{\mathrm{c}}$ expansion there ends as soon as the classical domain of the canted zigzag vanishes. However, this does not imply that the phase boundary with the polarized phase drops abruptly to zero. By extrapolating the curve, one can estimate its intercept with the $\varphi$-axis to be $\varphi_\mathrm{t}\approx0.880\pi$ for $S=1/2$, which agrees well with the exact diagonalization result $\varphi^\text{ED}_\mathrm{t} \approx 0.900 \pi$~\citep{chaloupka13}. In the case of $S=1$, our calculations yield $\varphi_\mathrm{t} = 0.862\pi$, which is once more in good quantitative agreement with the numerical result $\varphi_\mathrm{t}^\text{iDMRG} \approx 0.87\pi$ from Ref.~\citep{dong19}.


\subsection{Magnetization curves}

\begin{figure}
\centering
\includegraphics[width=8.5cm]{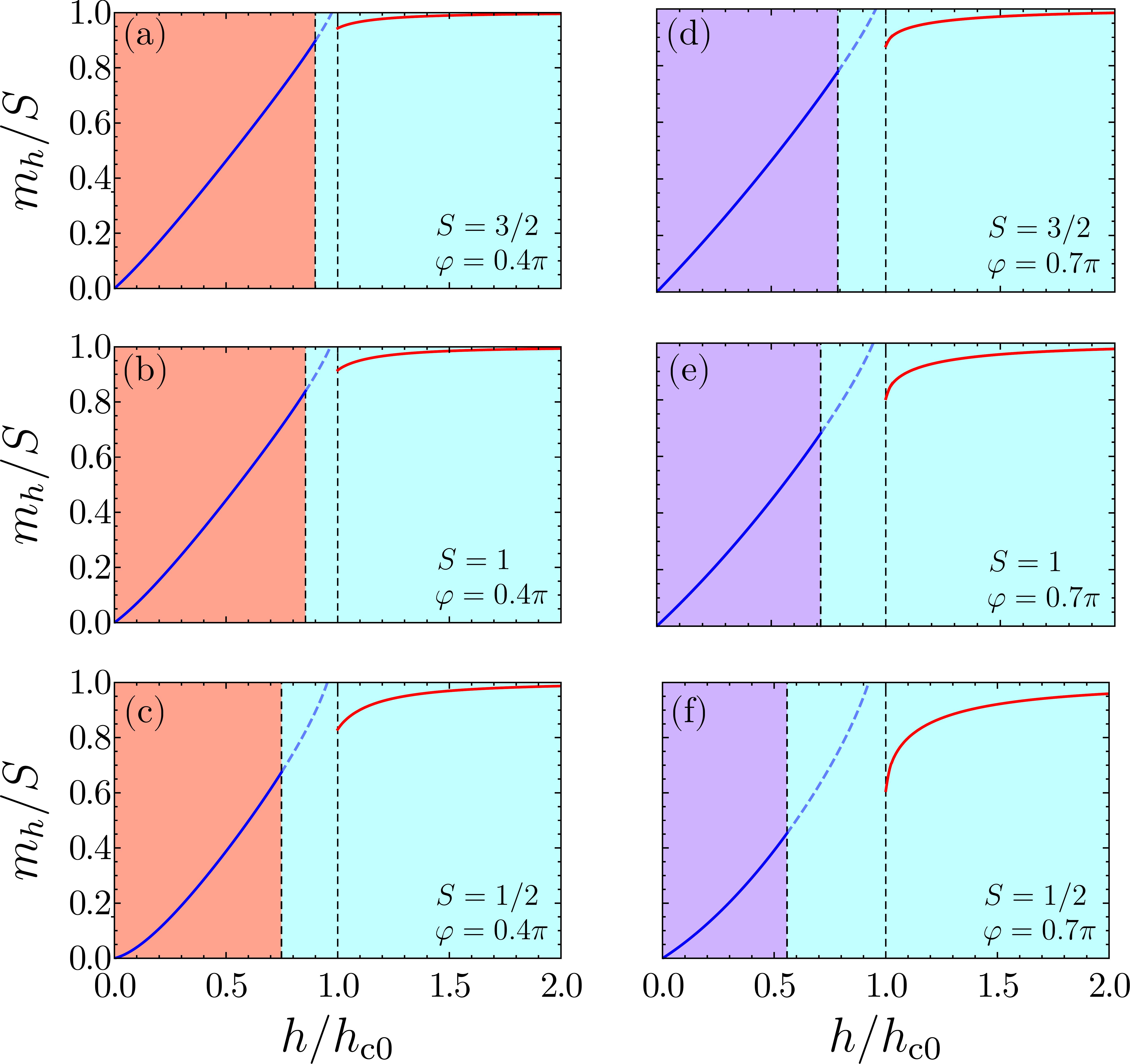}
\caption{Magnetization per site $m_h$ in units of $S$ as a function of the field $h$ in units of $h_{\mathrm{c}0}$ in the HK model with $J=A\cos\varphi$, $K=2A\sin\varphi$, and a magnetic field $\textbf{h}\parallel\left[001\right]$, at NLO in $1/S$. Left panels: $\varphi = 0.4\pi$ above Néel phase for (a) $S=3/2$, (b) $S=1$, (c) $S=1/2$. Right panels: $\varphi = 0.7\pi$ above zigzag phase for (d) $S=3/2$, (e) $S=1$, (f) $S=1/2$. The vertical dashed lines mark the positions of the $1/S$-corrected and classical critical fields, $h_{\mathrm{c}}$ and $h_{\mathrm{c}0}$, respectively. Red curves correspond to the partially polarized phase, whereas blue curves were obtained for the ordered phases below. The dashed portions of the blue curves should therefore be discarded, for they lie in the interval $\left[h_{\mathrm{c}},h_{\mathrm{c}0}\right]$, which is now occupied by the partially polarized phase. Still, one cannot extend the red curve below $h_{\mathrm{c}0}$ because the classical polarized state is unstable in this region.
\label{fig:full mh curves}}
\end{figure}

We further investigate corrections to field-dependent observables at NLO in $1/S$. In Fig.~\ref{fig:full mh curves}, we combine NLO magnetization curves from above and below $h_{\mathrm{c}0}$ with the information on the corrections to $1/h_{\mathrm{c}}$ for $S=1/2$, $1$ and $3/2$.
Figures \ref{fig:full mh curves}(a)--\ref{fig:full mh curves}(c) show that the reduction in $h_{\mathrm{c}}$ at $\varphi=0.4\pi$ eliminates the ill-behaved portion of the magnetization below $h_{\mathrm{c}0}$ (dashed lines) and allows one to smoothly interpolate between the polarized and ordered phase down to the smallest values of $S$. While this tendency remains true for most of the extent of the canted Néel, it breaks down near the Kitaev point, $\varphi=\pi/2$, or for values of $\varphi$ lying within the range of other ordered phases. As an example, consider the case of $\varphi=0.7\pi$, illustrated in Figs.~\ref{fig:full mh curves}(d)--\ref{fig:full mh curves}(f), for which the canted zigzag appears at low fields. Here, the correction to the magnetization in the limit $h\to h_{\mathrm{c}0}^{+}$ is much larger than that observed for $\varphi=0.4\pi$. Thus, a reasonable interpolation between the low and high-field portions is not possible at small $S$, despite the substantial reduction in the critical field. One must therefore go beyond NLO in $1/S$ to obtain magnetization curves which are fully consistent in the vicinity of $h_{\mathrm{c}}$ for small values of $S$. In fact, we can extend this conclusion to all values of $\varphi$ covered by the canted zigzag and canted stripy phases, as previous LSW calculations indicate that $1/S$ corrections reduce the $S=1/2$ magnetization in the limit $h\to h_{\mathrm{c}0}^{+}$ by at least $\sim35\%$ in this entire interval \citep{janssen16}.


\section{Results for $\textbf{h}\parallel\left[111\right]$} \label{sec:h111}

In the previous section, we have seen that our approach provides a consistent way to gauge the stability of the different ordered phases and capture nontrivial changes in the phase boundaries. We can now move on to the more intricate case of $\textbf{h}\parallel\left[111\right]$. As discussed above, we restrict our analysis to ordered phases with at most eight sites per magnetic unit cell. Such a simplification should represent an excellent approximation, though, for it only modifies small slivers of the classical phase diagram \citep{janssen16} and incorporates an overall tendency for magnetic orders with large unit cells to be destroyed by quantum fluctuations. Furthermore, this does \emph{not} affect the classical stability of any region of the phase diagram \citep{kruger20}.

\begin{figure}[t]
\centering
\includegraphics[width=8.4cm]{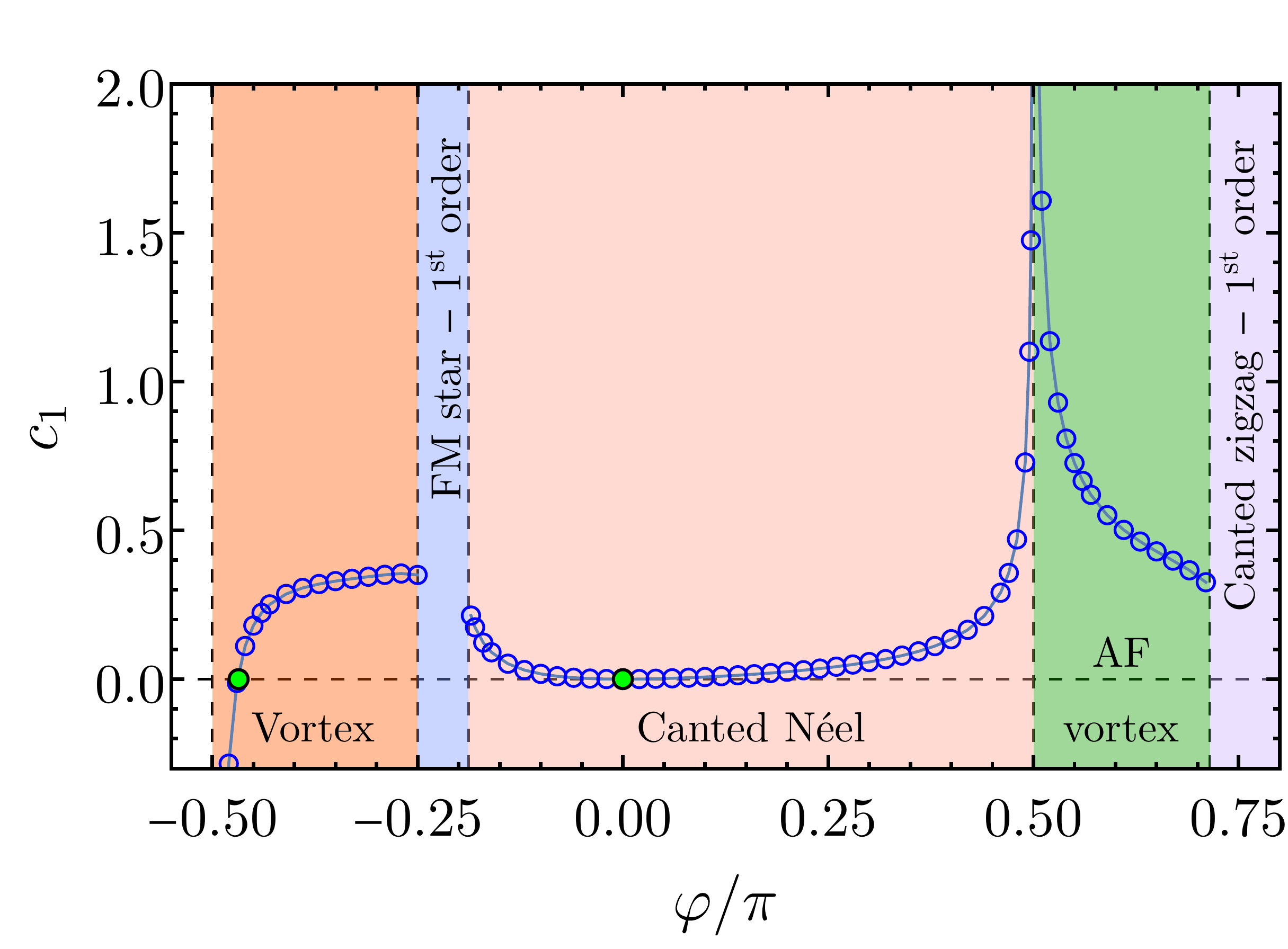}
\caption{$\mathcal O(1/S)$ coefficient $c_1$ in the expansion of the inverse of the critical field, see Eq.~\eqref{eq:1/hc 1/S exp}, as a function of $\varphi$ in the HK model with couplings $J=A\cos\varphi$ and $K=2A\sin\varphi$ in a magnetic field $\textbf{h}\parallel\left[111\right]$, obtained from the spin-wave calculation in the ordered phase [Eq.~\eqref{eq:1/hc exp cos(thtilde)}]. The blue line is a guide to the eye. Green dots at $\varphi=0$ and $\varphi\approx-0.47\pi$ denote points where the leading-order correction to the critical field vanishes. Gaps in the data correspond to intervals of $\varphi$ in which the transition to the polarized phase is discontinuous.
\label{fig:Correction to 1/hc 111}}
\end{figure}

\begin{figure*}
\centering
\includegraphics[width=17cm]{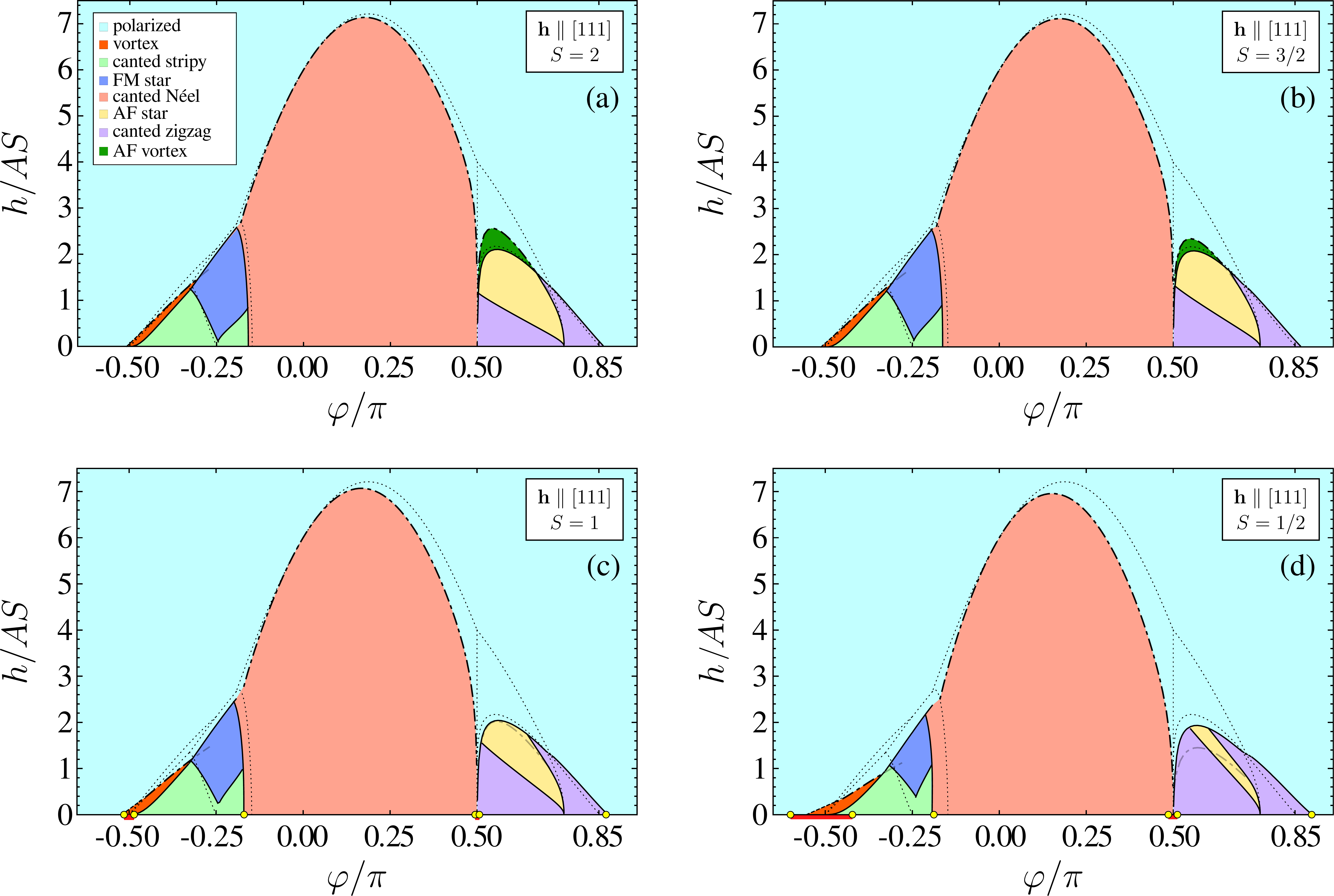}
\caption{Phase diagrams of the HK model as in Fig.~\ref{fig:pdsh001}, but now for a magnetic field $\textbf{h}\parallel\left[111\right]$. Note that the dot-dashed lines representing the critical fields fall below the lower classical boundaries of the vortex and AF vortex phases for small $S$ and an increasing range of $\varphi$ values, leading to a complete disappearance of the AF vortex phase and a strong suppression of the vortex order for $S \le 1$. Note also that the AF Kitaev spin liquid near $\varphi = \pi/2$ is expected to cover a sizable field range \citep{janssen16,hickey19}, which is not contemplated by our semiclassical expansion~\citep{order}.
\label{fig:pdsh111}
}
\end{figure*}

\subsection{Critical field}

In Fig.~\ref{fig:Correction to 1/hc 111}, we present the NLO contributions to $1/h_{\mathrm{c}}$ for all of the continuous phase transitions that appear in the semiclassical limit. As in the case of $\textbf{h} \parallel \left[001\right]$, the corrections to the critical field are finite everywhere except at the Kitaev points. Moreover, the results for the canted Néel are roughly similar in both field directions. None of the remaining continuous transitions, however, have a direct counterpart in a $\left[001\right]$ field; they involve two vortex phases which emerge at intermediate fields for opposite signs of the Kitaev coupling. On the right-hand side of the diagram $\left(K>0\right)$, the AF vortex displays pronounced corrections to $1/h_{\mathrm{c}0}$ even away from $\varphi=\pi/2$. On the left-hand side $\left(K<0\right)$, the corrections inside the vortex change sign at $\varphi\approx-0.47\pi$ before diverging to $-\infty$ at the FM Kitaev point. Hence, much like the behavior uncovered for the canted zigzag when $\textbf{h}\parallel\left[001\right]$, the critical field should increase near the left end of the vortex phase for every $S$, which is qualitatively consistent with the early simulations of Ref.~\citep{jiang11}. However, we note that for $\varphi \approx - 0.5\pi$, we expect the ferromagnetic Kitaev spin liquid to emerge for small values of $S$, which is not captured by our semiclassical calculation.


\subsection{Phase diagram}

We now combine the results presented above with those extracted for first-order phase transitions to assemble phase diagrams for $S=1/2$, $1$, $3/2$, and $2$. Similarly to the previous case, we find a substantial reduction of the critical field between the ordered phases and the partially polarized phase upon the inclusion of $1/S$ corrections in large parts of the phase diagram, Fig. \ref{fig:pdsh111}. Furthermore, we observe a trend whereby phases with large magnetic unit cells tend to be destabilized upon decreasing $S$, in agreement with the general expectation. For $S\leq 1$, the AF vortex phase is completely suppressed and the polarized phase reaches down to the AF star or canted zigzag, depending on $S$ and $\varphi$. On the ferromagnetic-$K$ side of the diagram, the change in the sign of $c_{1}$ at $\varphi\approx-0.47\pi$, see Fig.~\ref{fig:Correction to 1/hc 111}, implies that a finite portion of the vortex phase remains stable at NLO in $1/S$. However, because this phase becomes more concentrated around the ferromagnetic Kitaev point, higher-order corrections in $1/S$ or nonperturbative approaches are necessary to validate its stability for small values of $S$.

By using Eq.~\eqref{eq:1/ht 1st order 1/S exp}, we also verify that the boundary between FM star and the polarized phase is shifted down for decreasing $S$. By employing Eq.~\eqref{eq:phit 1st order 1/S exp} in turn, we find that the FM star phase is suppressed by its neighboring ordered phases as well. From its right side, the whole boundary with the canted Néel undergoes a leftward shift. A similar trend is seen from its left side: Except near the transition to the polarized phase, the boundary with the canted stripy is displaced to the right. Intriguingly, this displacement increases as one follows the classical phase boundary down to the FM Klein point, $(\varphi,h)=(-\pi/4,0)$, where the Hamiltonian exhibits a degenerate (quantum) ground-state manifold in consequence of a hidden SU(2) symmetry \citep{chaloupka10,chaloupka13,chaloupka15,janssen16}. This shifts the FM star phase, which reaches down to $h=0$ at and to the right of the Klein point in the classical limit, to finite fields. Furthermore, by performing LSW calculations at $h=0$, we find that an order-by-disorder mechanism selects the stripy over the FM star everywhere except at the FM Klein point, in agreement with the general expectation \citep{chaloupka10}. Therefore, a finite domain of the canted stripy should exist beneath the FM star for every $\varphi\ne-\pi/4$. The extent of such a domain cannot be determined along the lines of Sec.~\ref{subsec:1st-order PTs}, though, for $1/h_{\mathrm{t}}$ diverges when $S \to \infty$. As an alternative, we estimate the transition line by expanding the equality $E_{\mathrm{a}}(\varphi,h_{\mathrm{t}},1/S)= E_{\mathrm{b}}(\varphi,h_{\mathrm{t}},1/S)$ around $(h_{\mathrm{t}},1/S)=(0,0)$. Here, the indices correspond to the FM star and canted stripy phases above (a) and below (b), respectively, the transition line. Solving for $h_{\mathrm{t}}$, we obtain
\begin{equation}
h_{\mathrm{t}}\left(\varphi\right)=\sqrt{\frac{2}{S}}\sqrt{\left.\frac{E_{\mathrm{a}1}-E_{\mathrm{b}1}}{\frac{\partial^{2}}{\partial h^{2}}\left(E_{\mathrm{b}0}-E_{\mathrm{a}0}\right)}\right|_{h=0}}+\mathcal{O}\!\left(\frac{1}{S}\right),
\label{eq:ht expansion}
\end{equation}
which gives the lower boundary of the FM star for $\varphi > -\pi/4$. As visible in Fig.~\ref{fig:pdsh111}, the FM star turns out to be shifted to finite fields for all values of $\varphi$, even right at the SU(2) symmetric Klein point.

Similarly, NLO contributions in $1/S$ computed via Eq.~\eqref{eq:phit 1st order 1/S exp} favor the canted zigzag over the AF star by moving the boundary between the two to the left. However, the correction to the boundary now vanishes as one approaches the AF Klein point, $(\varphi,h)=(3\pi/4,0)$. On the other hand, by applying the same scheme as in Eq.~\eqref{eq:ht expansion}, we find a large suppression of the AF star from below, which is especially drastic for $S=1/2$. Put together, these results show that the region of stability of the AF star diminishes considerably upon lowering $S$.

Finally, we turn to the transition between the canted zigzag and the polarized phase. Differently from the case of $\textbf{h}\parallel\left[001\right]$, we observe a rightward displacement of the boundary for all finite $h$. This suggests that the canted zigzag order is particularly stable in a $\left[111\right]$ field, as reflected by the large domain it occupies in the diagrams with small values of $S$, see Fig.~\ref{fig:pdsh111}. By inspecting the limit $h \to 0$, we find that the transition between the zigzag and the ferromagnet takes place at $\varphi_{\mathrm{t}} \approx 0.899 \pi$ for $S=1/2$, which is in remarkable agreement with the exact diagonalization result $\varphi^\text{ED}_\mathrm{t} \approx 0.900 \pi$~\citep{chaloupka13}. For $S=1$, our estimation $\varphi_\mathrm{t} \approx 0.877\pi$ also agrees well with the infinite density renormalization group result $\varphi^\text{iDMRG} \approx 0.87\pi$~\citep{dong19}.

In summary, our results indicate a strong suppression at NLO in $1/S$ of the various large-unit-cell and multi-$\textbf{Q}$ classical phases that arise when $\textbf{h}\parallel\left[111\right]$. This generally agrees with the numerical results for $S=1/2$ on small clusters \citep{jiang11,jiang19,hickey19,gohlke20}.


\subsection{Direction of ordered moments: Canted Néel phase} \label{subsec:neel ObD}

As we have seen in the previous subsection, an interesting competition between order-by-disorder and field-selection effects is generally at work at low fields. On one hand, this can lead to shifts in phase boundaries, as in the cases of the transitions between the canted stripy and FM star, and between the canted zigzag and AF star. On the other hand, it can also induce intriguing responses of the direction of the ordered moments to the magnetic field within the same phase. Such a situation occurs in the canted Néel, as we discuss now. First, consider the classical limit, $S \to\infty$, of the HK model. Since all three neighbors of an arbitrary spin have the same configuration in the Néel state, the classical Kitaev term adds up to an effective Heisenberg interaction, and hence preserves SU(2) spin symmetry at zero field. Therefore, when exposed to a small magnetic field, the classical spins initiate uniform canting from the plane perpendicular to the field axis, regardless of which direction this may be.

Quantum corrections, however, lift the SU(2) degeneracy at zero field and are expected to favor states whose ordered moments lie along the cubic axes in spin space by virtue of an order-by-disorder mechanism~\citep{chaloupka16,sizyuk16}. Except for specific field directions, the set of states selected by quantum fluctuations will have no overlap with that selected by the field. This leads, in general, to a competition between the fluctuation effects, most relevant at small fields, and field-selection effects, which dominate at high fields.

\begin{figure}
\centering
\includegraphics[width=6.75cm]{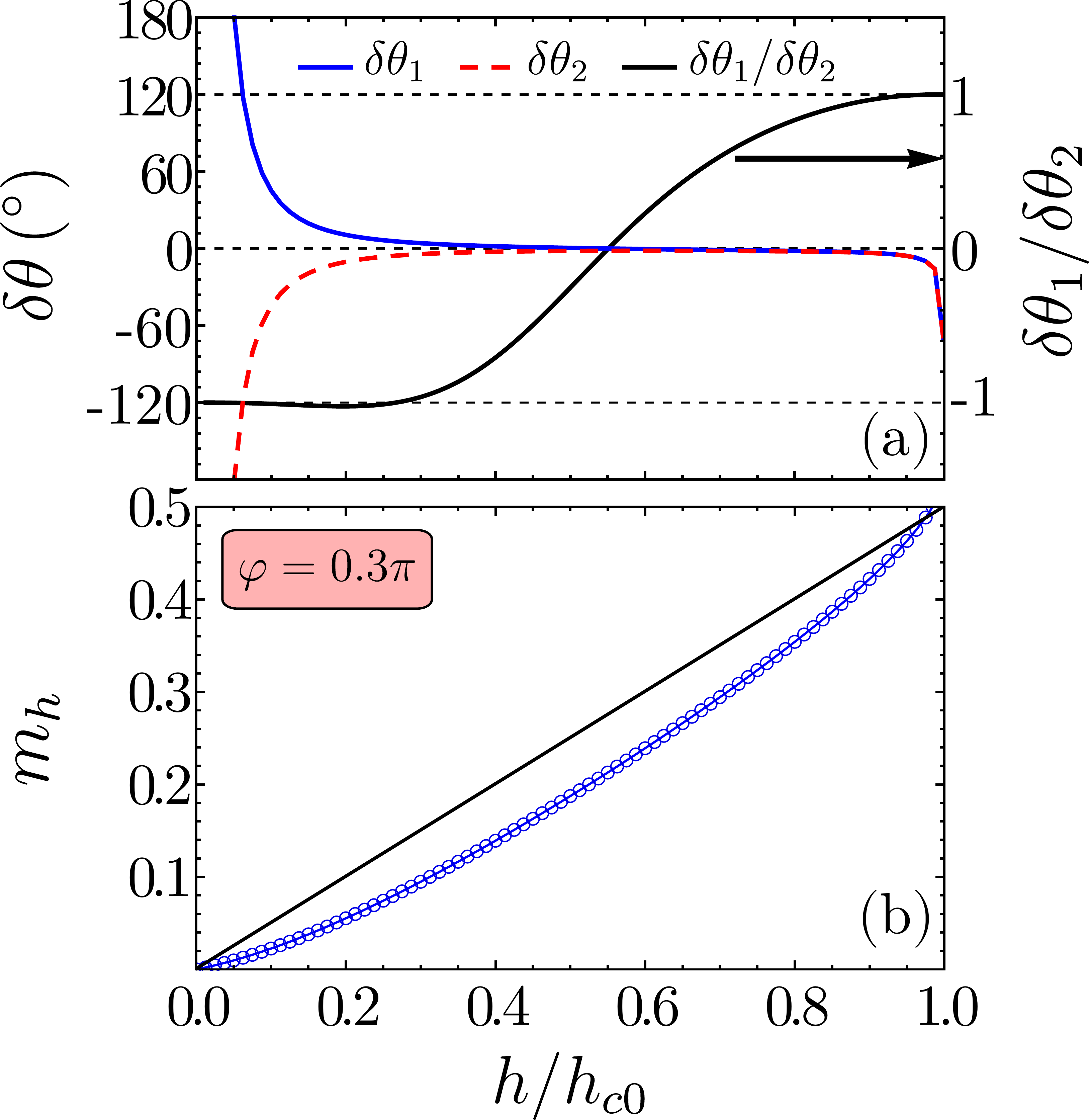}
\caption{Spin-wave-theory results for $\varphi=0.3\pi$ and $\textbf{h}\parallel\left[111\right]$. (a) Corrections to the classical canting angle for spins in the two sublattices of the canted Néel phase. Although the individual angles diverge in the opposing limits of $h\to0$ and $h\to h_{\mathrm{c}0}$, the ratio of $\delta\theta_{1}$ to $\delta\theta_{2}$ shows that the spins tend, respectively, to an antiparallel and a parallel state. (b) Magnetization curves in leading (black) and NLO (blue with markers) order for $S=1/2$. Note that the divergence of $\delta\theta_{1}$ and $\delta\theta_{2}$ as $h\to0$ does not manifest itself in the magnetization.
}
\label{fig:cneel_phi0p3pi}
\end{figure}

In Fig.~\ref{fig:cneel_phi0p3pi}(a), we show the quantum corrections to the canting angles in the canted Néel phase in a $\left[111\right]$ field for a representative value of $\varphi$. On one hand, we see a divergence of the corrections as $h\to h_{\mathrm{c}0}^-$, which we now know is related to the reduction of $h_{\mathrm{c}}$. On the other hand, the plot exhibits two features that distinguish the canted Néel order in a $\left[111\right]$ field from all other magnetic orders considered here, including its counterpart in a $\left[001\right]$ field. First, NLO corrections in $1/S$ impose a fundamental change to the classical parametrization by rendering the canting nonuniform for every $h<h_{\mathrm{c}0}$. Second, both $\delta\theta_{1}$ and $\delta\theta_{2}$ strongly diverge as $h\to0$. No traces of this low-field divergence, however, appear in observables such as the magnetization, see Fig.~\ref{fig:cneel_phi0p3pi}(b).

As hinted above, the key to understanding such an odd behavior lies in the breaking of the classical SU(2) spin symmetry: An order-by-disorder mechanism locks the zero-field Néel order to one of the $xyz$ axes \citep{chaloupka16,sizyuk16}. Since none of the selected states lie on the $ab$ plane, uniform canting in $\left[111\right]$ field cannot be reconciled with the presence of quantum fluctuations. This explains not only the difference between $\delta\theta_{1}$ and $\delta\theta_{2}$ in Fig.~\ref{fig:cneel_phi0p3pi}(a), but also their divergence at low fields. Indeed, if it were not so, the corrections would be suppressed at large but finite $S$. This, however, would be inconsistent with the expectation that the competition between fluctuation and field-selection effects should persist for all finite $S$ and small enough fields.

By tracking the ratio of $\delta\theta_{1}$ to $\delta\theta_{2}$ rather than their individual values, we can find further information hidden in the low-field divergence. As shown by the black curve in Fig.~\ref{fig:cneel_phi0p3pi}(a), $\delta\theta_{1}/\delta\theta_{2}$ converges to $-1$ as $h\to0$. Given that $\theta_{1}=\theta_{2}=\pi/2$ at $h=0$, this implies that the system still approaches an antiparallel state as $h\to0$. Therefore, while the $1/S$ expansion fails to connect high- and low-field parametrizations at NLO, it suggests that, for an infinitesimal field, the system orders in a collinear Néel state lying outside of the plane perpendicular to the field axis, in agreement with the outcome of the order-by-disorder mechanism.

Ultimately, one can interpret these results as a sign of noncommutativity of the limits $h\to0$ and $S\to\infty$ in a $\left[111\right]$ field. After all, the classical parametrizations are obtained by taking $S\to\infty$ before $h\to0$ and are thus completely oblivious to the order-by-disorder mechanism taking place at $h=0$.


\subsection{Order-by-disorder in noncollinear states: Vortex phases} \label{subsec:Vortex ObD}

Finally, we comment on the calculations performed in the vortex and AF vortex phases in further detail to illustrate how an order-by-disorder mechanism acts on noncollinear states at higher orders in $1/S$. As described in Ref.~\cite{janssen16}, both of these phases display an accidental U(1) degeneracy which manifests itself as a free angle $\xi$ in their classical parametrizations: $\phi_\mu = \phi_\mu \left(\xi\right)$. This means that the leading-order term in the $1/S$ expansion of the azimuthal angles, Eq.~\eqref{eq:1/S expansion ph}, is not fully fixed by the minimization of the classical ground-state energy because, unlike higher-order terms in the spin-wave Hamiltonian, $\mathcal{H}_{0} = E_{\mathrm{gs},0}$ does not depend on $\xi$. The appropriate value of $\xi$ is thus determined by minimizing the contribution of quantum fluctuations to the zero-point energy.

\begin{figure}
\centering
\includegraphics[width=8.5cm]{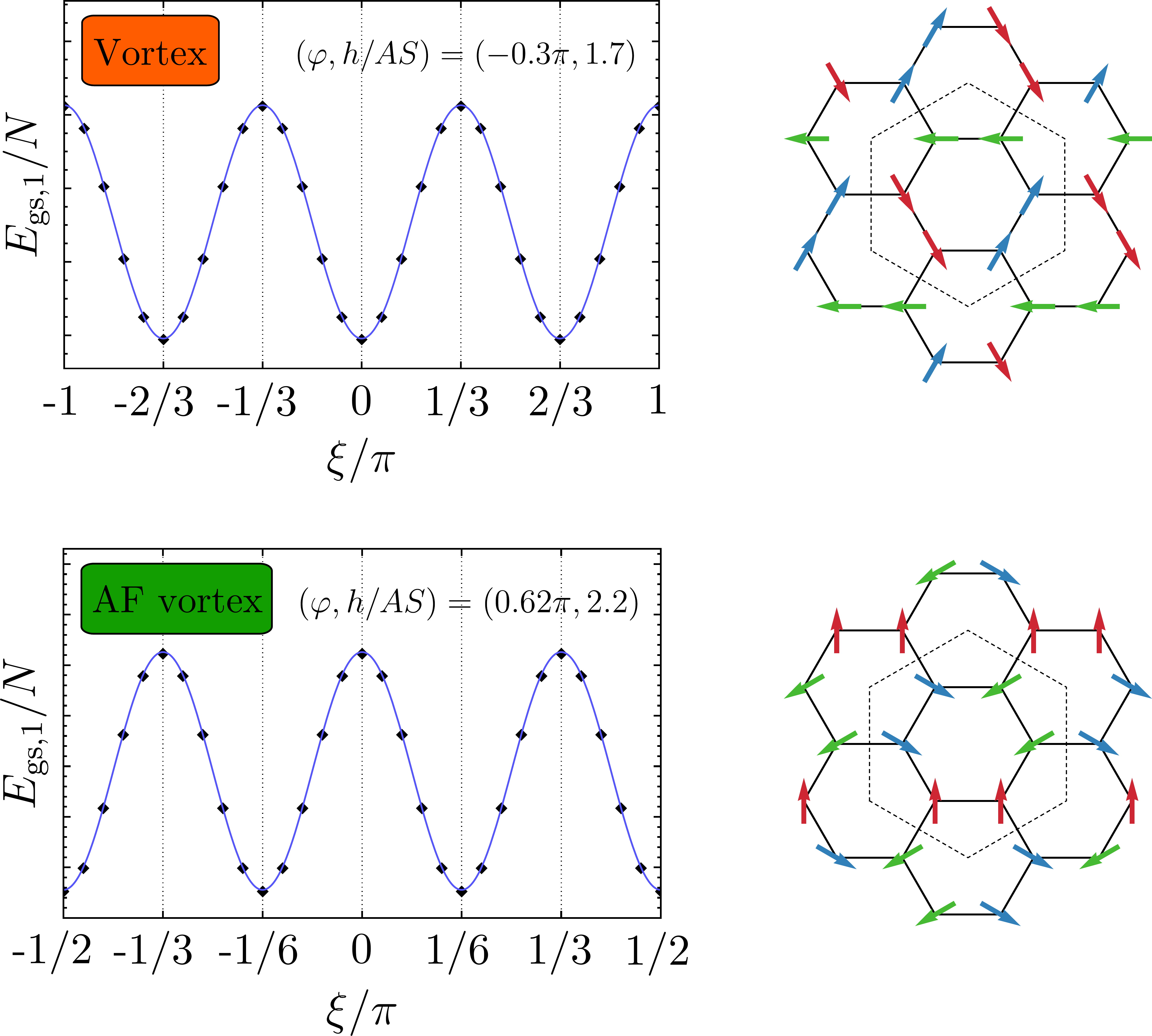}
\caption{Left panels: NLO contribution to the ground-state energies of the vortex and AF vortex phases of the HK model in a $\left[111\right]$ field from LSW theory, illustrating the order-by-disorder mechanism. The blue lines are fits of cosine functions to guide the eyes. Right panels: Projections of the 120° spin configurations selected by quantum fluctuations onto the plane perpendicular to the $\left[111\right]$ direction.\label{fig:ObD plots}}
\end{figure}

Following the usual prescription of order-by-disorder analyses, we have employed LSW theory to compute the NLO contribution to the ground-state energy, Eq.~\eqref{eq:epsilon_q}, as a function of $\xi$. Figure \ref{fig:ObD plots} shows that the results pertaining to the vortex (AF vortex) are well fitted by a cosine function with a period of $2\pi/3$ ($\pi/3$) and a minimum at $\xi^*=0$ ($\xi^*=\pi/6$). When these values of $\xi$ are substituted back into the classical parametrizations, they generate 120° orders whose projections onto the $ab$ plane are parallel or perpendicular to the bonds of the lattice \citep{wu08}.
Therefore, the states selected by the leading-order quantum fluctuations are not only noncollinear, but also noncoplanar. We emphasize that, even in cases such as these, the NLO contribution to the ground-state energy follows entirely from LSW theory. Indeed, suppose we have computed the $1/S$ corrections, $\delta\phi_\mu$ and $\delta\theta_\mu$, to the parametrization angles. The leading-order contributions of such terms to Eq.~\eqref{eq:ground-state energy} are then determined by expanding $E_{\mathrm{gs},0}$ and $E_{\mathrm{gs},1}$ around $\left\{\phi_\mu \left(\xi^*\right), \theta_\mu \right\}$. However, because this set of angles minimizes the classical energy, the NLO term from $E_{\mathrm{gs},0}$ is zero, and Eq.~\eqref{eq:ground-state energy} only receives contributions beyond those given by LSW theory at $\mathcal O(S^0)$~\citep{zhitomirsky98,coletta12}.

Yet, in calculating corrections to the critical field, we have taken the analysis one step further: By using the 120° orders selected within LSW theory as reference states~\citep{xi}, we have implemented the scheme described in Sec.~\ref{subsec:Cubic terms} to compute $\delta\phi_\mu$ and $\delta\theta_\mu$. While the corrections to the polar angles, $\delta\theta_{\mu}$, always turn out to be determinate, we find that the deviations to the azimuthal angles, $\delta\phi_{\mu}$, cannot be expressed independently in any of the two phases. Instead, they are all given in terms of one of the unknowns, say $\xi' \equiv \delta\phi_{1}$.
In the vortex phase, we have $\delta\phi_\mu = \pm \xi'$, where the upper (lower) sign applies to odd (even) $\mu$, corresponding to the two crystallographic sublattices of the honeycomb lattice. Remarkably, the structure of $\delta\phi_\mu$ in this state is completely analogous to the continuous degeneracy appearing in the classical parametrization~\citep{janssen16}, i.e., $\xi$ is simply substituted by $\xi'/S$ in Eq.~\eqref{eq:1/S expansion ph}. By contrast, the angle corrections in the AF vortex phase introduce an asymmetry between the two crystallographic sublattices, since $\delta\phi_\mu = \xi'$ [$\delta\phi_\mu = -(\xi' + \delta\xi')$] for odd (even) $\mu$, with $\delta\xi' = \delta\xi' \left(\varphi,h\right)$. As $\xi'$ appears in a role similar to the one played by $\xi$ at the level of LSW theory, it is to be determined by the minimization of $E_{\mathrm{gs},2}$.

To summarize, for noncollinear states, corrections to the spin angles arising from the cubic terms in the spin-wave Hamiltonian are finite, but do not contribute to the ground-state energy at NLO in the $1/S$ expansion. An accidental continuous degeneracy that occurs at the classical level resurfaces in the $1/S$ corrections to the parametrization angles as a free parameter $\xi'$, which is fixed by minimizing the term of $\mathcal{O}\left(S^0\right)$ of the ground-state energy. The resurgence of such a free parameter is therefore necessary to provide the full angle dependence of the energy at higher orders in the $1/S$ expansion. This guarantees that the energy can be determined consistently order by order and that its minimization fixes the correct values of the spin angles.


\section{Conclusions and outlook} \label{sec:conclusion}

In conclusion, we have studied the effects of quantum fluctuations in the HK model in an external magnetic field. We have applied nonlinear spin-wave theory both to the ordered and the polarized phases to derive a consistent $1/S$ expansion for various observables, allowing us to compute the quantum corrections to the phase diagram at NLO in $1/S$. Our results indicate substantial modifications to the phase boundaries, including an overall tendency of the high-field polarized phase to suppress ordered phases. This effect was found to be especially strong for the several large-unit-cell and multi-$\textbf{Q}$ phases that arise in the classical limit for $\textbf{h}\parallel\left[111\right]$~\citep{janssen16}. In particular, one of the two magnetic vortex states is completely destabilized for $S\le 1$, whereas the other is significantly suppressed. Given that our phase diagrams in Figs.~\ref{fig:pdsh001} and \ref{fig:pdsh111} involve an extrapolation of the $1/S$ expansion to small $S$, more detailed numerical studies are called for, in particular for $S=1/2$ and $S=1$.

We have also computed explicitly the quantum corrections to different observables, such as the direction of the ordered moments, the magnetization, and the spectrum. Our results for the magnetization curves are consistent with the general trend that the transition from an ordered phase to the partially polarized phase is shifted towards lower fields upon increasing $1/S$. The $1/S$ correction to the critical field can be computed either in the ordered phase, by evaluating the angle corrections to the direction of the ordered moments, or in the partially polarized phase, by tracing the spectral gap. We have explicitly demonstrated that these two, seemingly independent, approaches yield the same results.

Our findings may be relevant for higher-spin Kitaev materials. For instance, the antimonates $A_3$Ni$_2$SbO$_6$ ($A = \text{Na}, \text{Li}$) are candidates for $S=1$ Kitaev systems~\citep{stavropoulos19}. Similar to \rucl, they realize a zigzag ground state at low temperatures and zero field~\citep{zvereva15, kurbakov17}. Interestingly, both compounds show metamagnetic transitions towards field-induced intermediate ordered phases. The lower transition has initially been interpreted in terms of a spin-flop mechanism~\citep{zvereva15}; however, recent magnetostriction experiments on Na$_3$Ni$_2$SbO$_6$ appear to be inconsistent with such a simple scenario, and suggest a picture of an anisotropy-governed competition of different antiferromagnetic phases~\citep{werner17}. In contrast to the $S=1/2$ Kitaev materials~\citep{jackeli09,winter16}, the Kitaev interaction in the $S=1$ systems is expected to be \emph{antiferromagnetic}~\citep{stavropoulos19}. This allows a description of the zigzag magnetic order fully within the nearest-neighbor HK model. Our work demonstrates that nontrivial field-induced transitions between different types of antiferromagnetic orders, involving changes in the ordering wave vector and the geometry of the magnetic unit cell, are natural in such a situation. In order to make a more concrete comparison of our predictions for the HK model with the experimental results on the antimonates, in-field neutron diffraction measurements and/or angle-dependent thermodynamic measurements on single crystals would be desirable. This should allow one to elucidate the role of the observed anisotropy~\citep{werner17} and the nature of the field-induced phases.

The Cr-based monolayers that have been proposed as candidates for $S=3/2$ Kitaev systems~\citep{lee20} show a ferromagnetic ground state~\citep{gong17,huang17}, but may potentially be driven to other magnetic or paramagnetic states by epitaxial strain~\citep{xu20}. Our results show that, in such a setup, an external field could also induce nontrivial intermediate phases, and it would be interesting to search for signatures of the corresponding metamagnetic transitions.

In a broader context, the framework developed here can be applied to other spin models with interactions that break SU(2) spin-rotational symmetry as well. This includes extensions of the HK Hamiltonian with additional interactions~\citep{rau14,trebst17,winter17,janssen17} or on other lattices~\citep{kimchi14,lee15,obrien16,kruger20,li20,janssen20}, different classes of compass models~\citep{nussinov15}, and the anisotropic Hamiltonian used to characterize magnetically ordered phases in rare-earth pyrochlores~\citep{ross2011,rau19}. Our approach complements the numerical simulations such as exact diagonalization or density matrix renormalization group that work directly at the desired values of $S$ but are typically constrained to small lattice sizes.

\begin{acknowledgments}
We thank S.\ Koch, W.\ Kr\"uger, and R.\ G.\ Pereira for useful discussions and collaborations on related projects.
P.M.C.\ has been supported by the FAPESP (Brazil) Grant Nos.\ 2017/22133-3 and 2019/02099-0.
The work of L.J.\ is funded by the Deutsche Forschungsgemeinschaft (DFG) through the Emmy Noether Program (JA2306/4-1, project id 411750675).
L.J.\ and M.V.\ acknowledge support by the DFG through SFB 1143 (project id 247310070) and the Würzburg-Dresden Cluster of Excellence on Complexity and Topology in Quantum Matter---\emph{ct.qmat} (EXC 2147, project id 390858490).
E.C.A.\ has been supported by CNPq (Brazil) Grant Nos.\
406399/2018-2 and 302994/2019-0 and FAPESP (Brazil) Grant No.\ 2019/17026-9.
\end{acknowledgments}
\appendix


\begin{figure*}
\centering
\includegraphics[width=6cm]{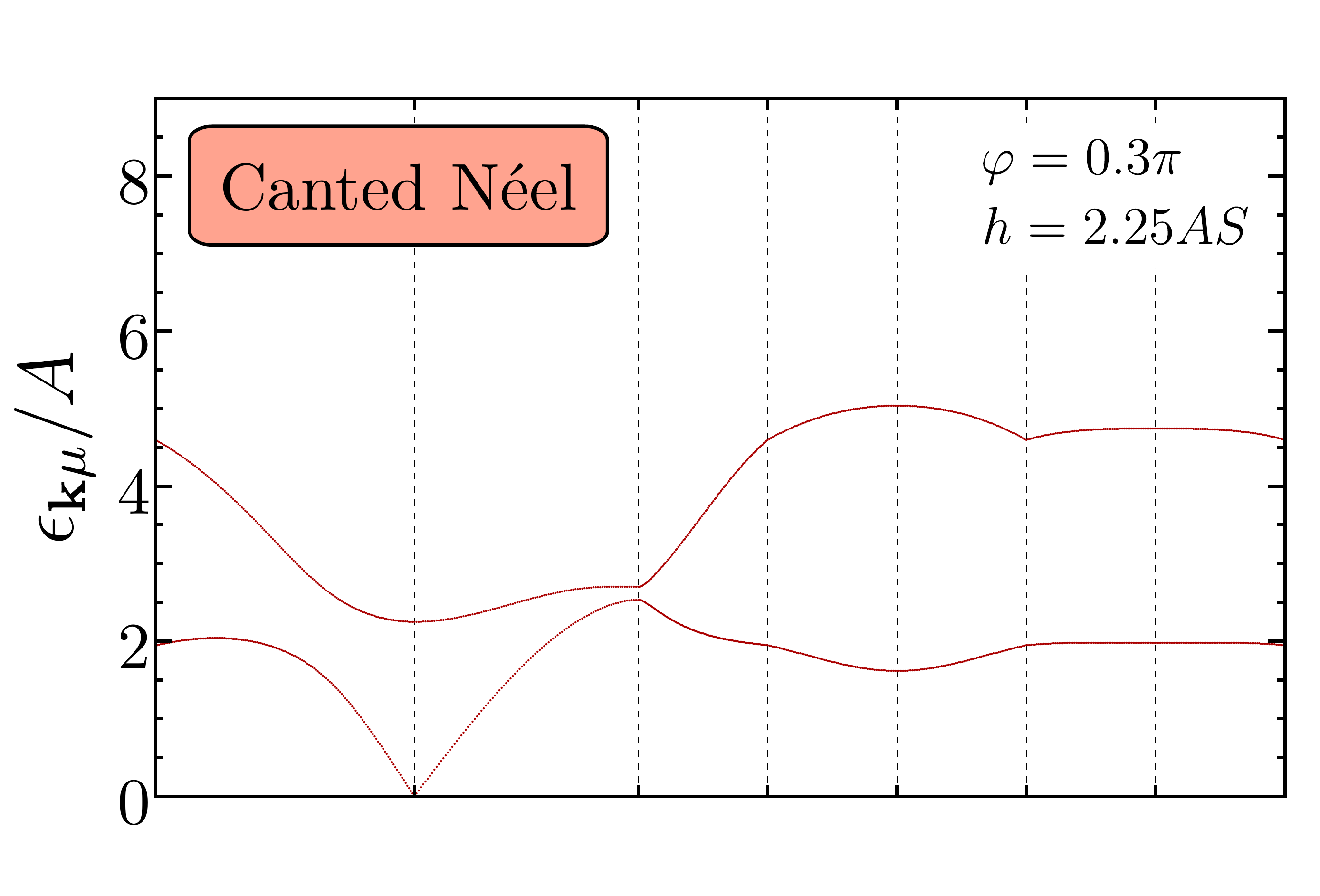}
\hspace{-0.4cm}
\includegraphics[width=6cm]{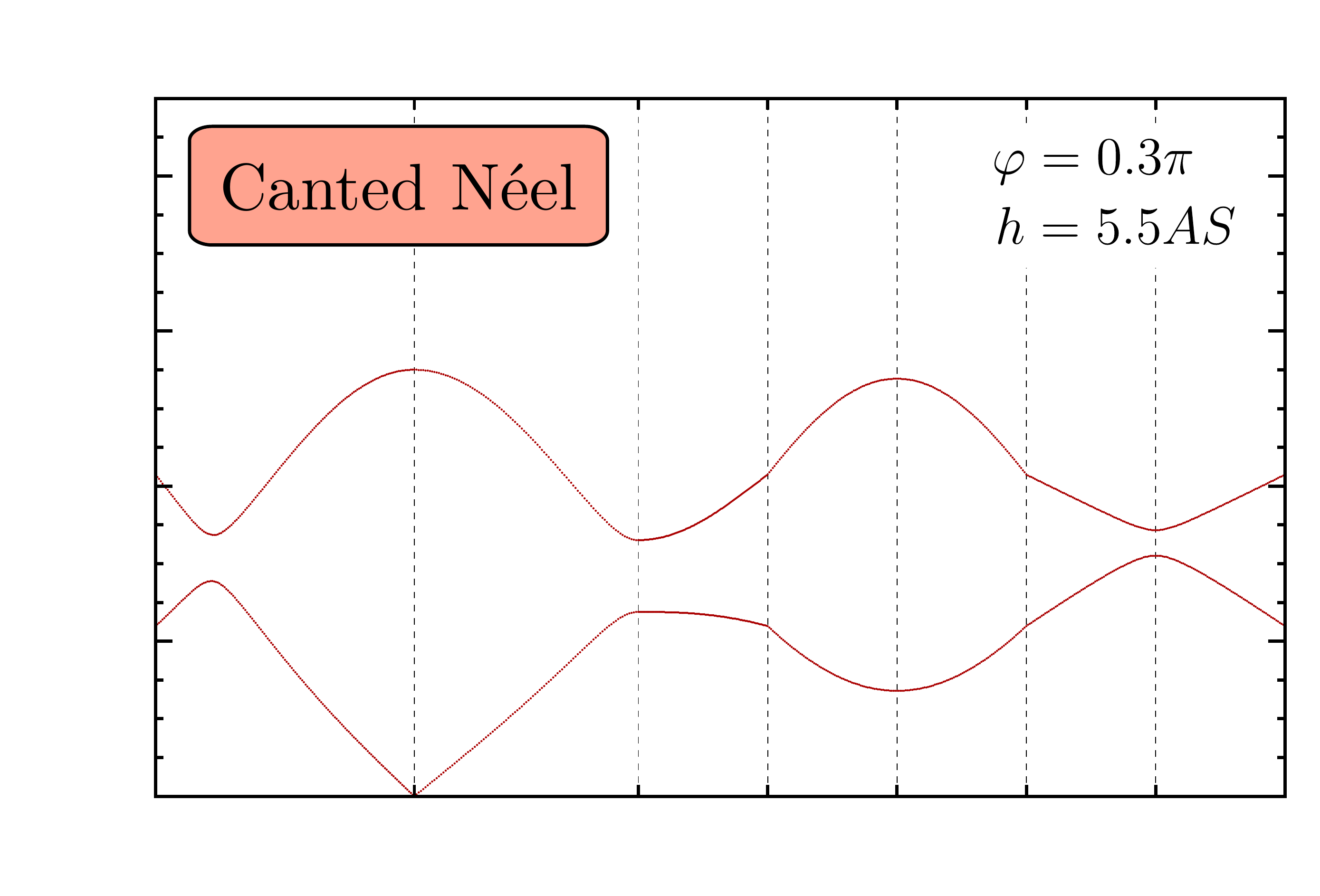}
\hspace{-0.4cm}
\includegraphics[width=6cm]{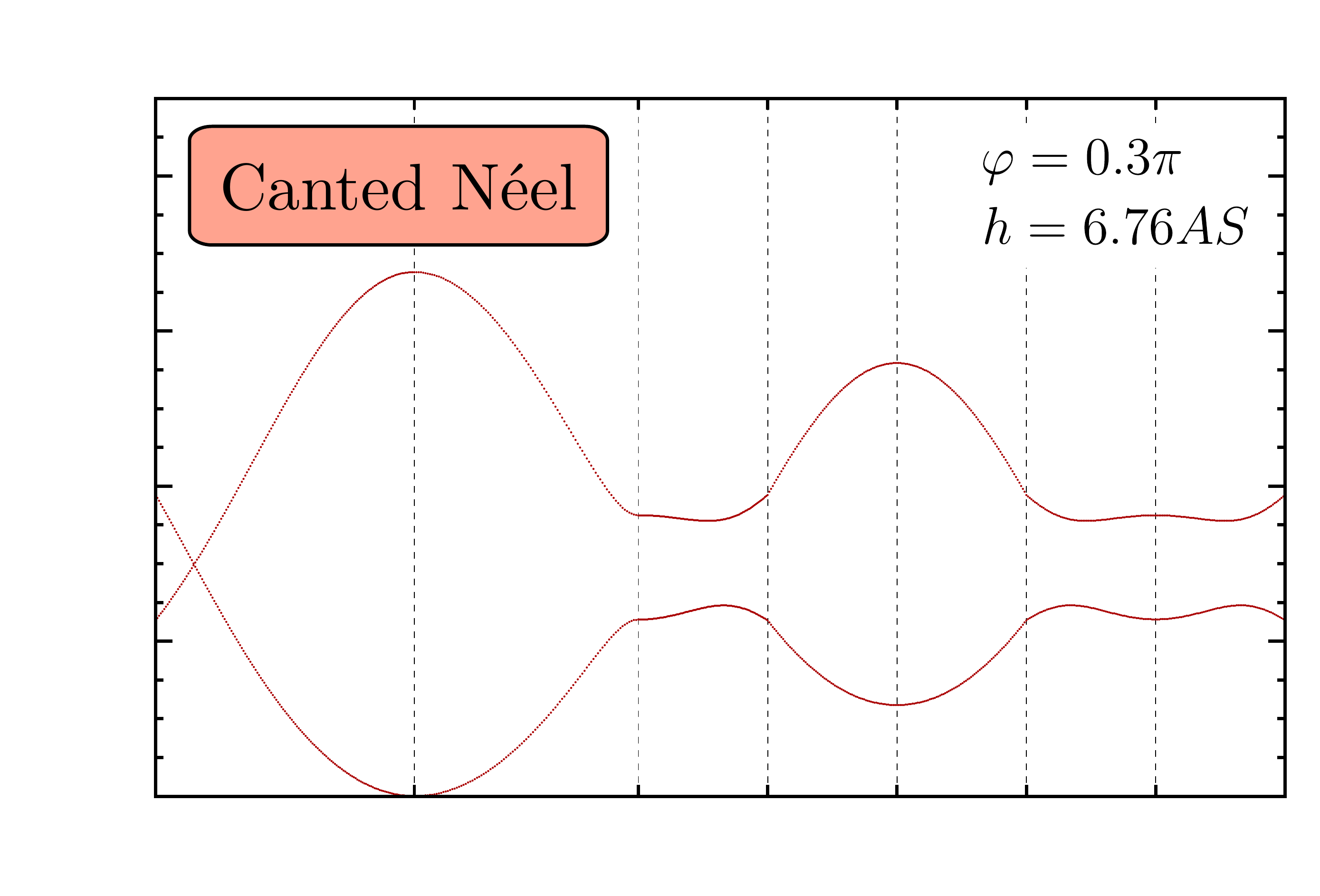}

\vspace{-0.5cm}
\includegraphics[width=6cm]{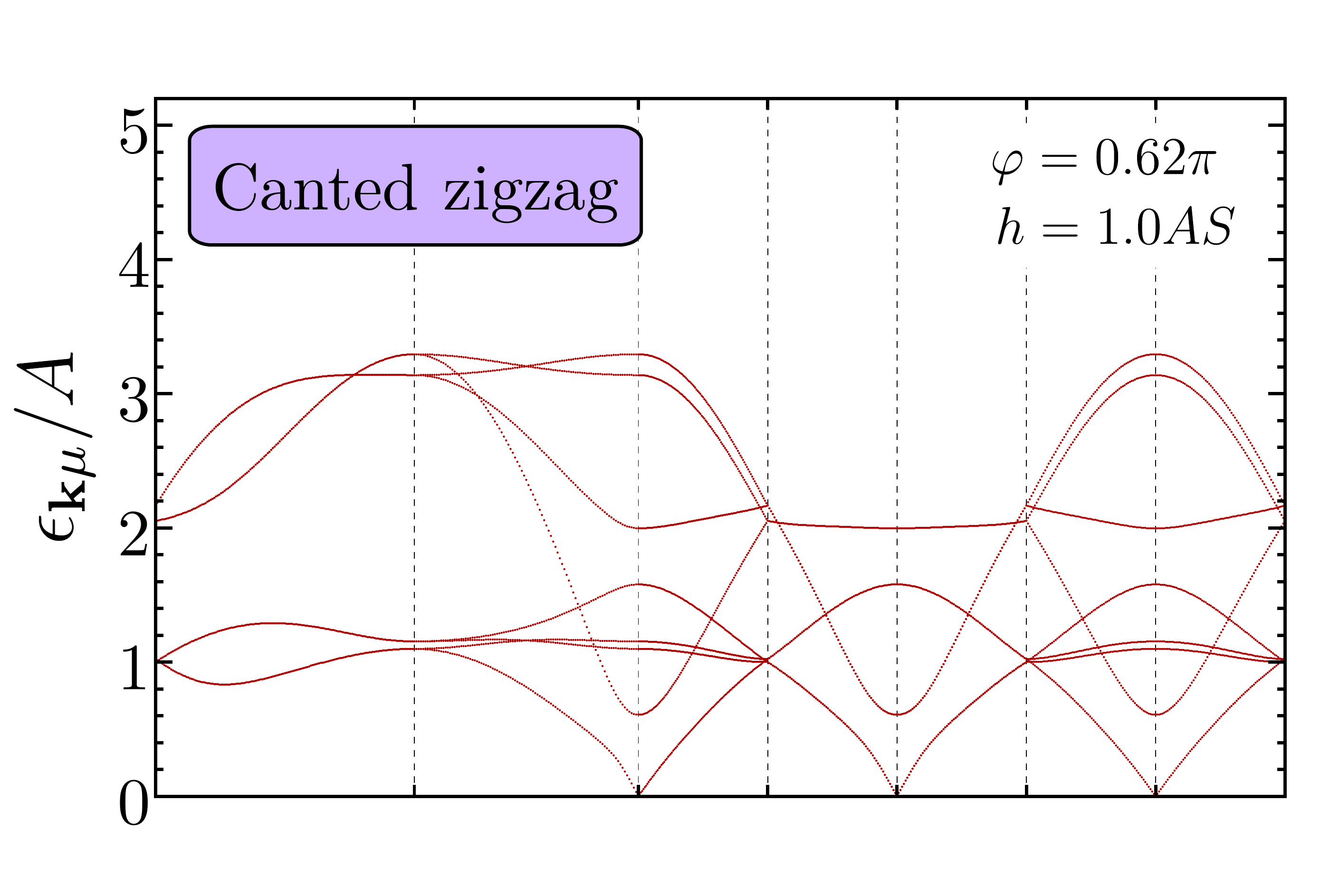}
\hspace{-0.4cm}
\includegraphics[width=6cm]{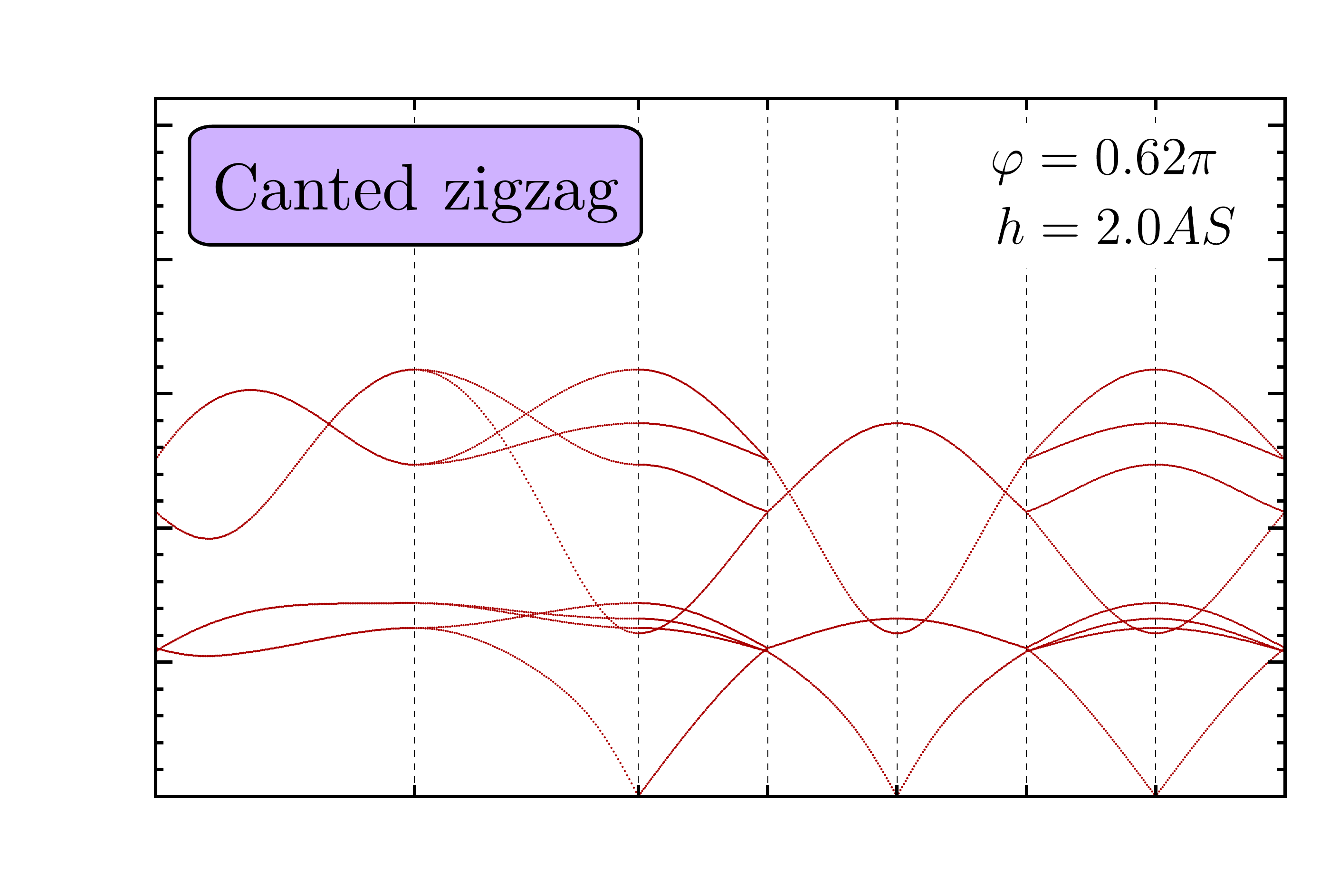}
\hspace{-0.4cm}
\includegraphics[width=6cm]{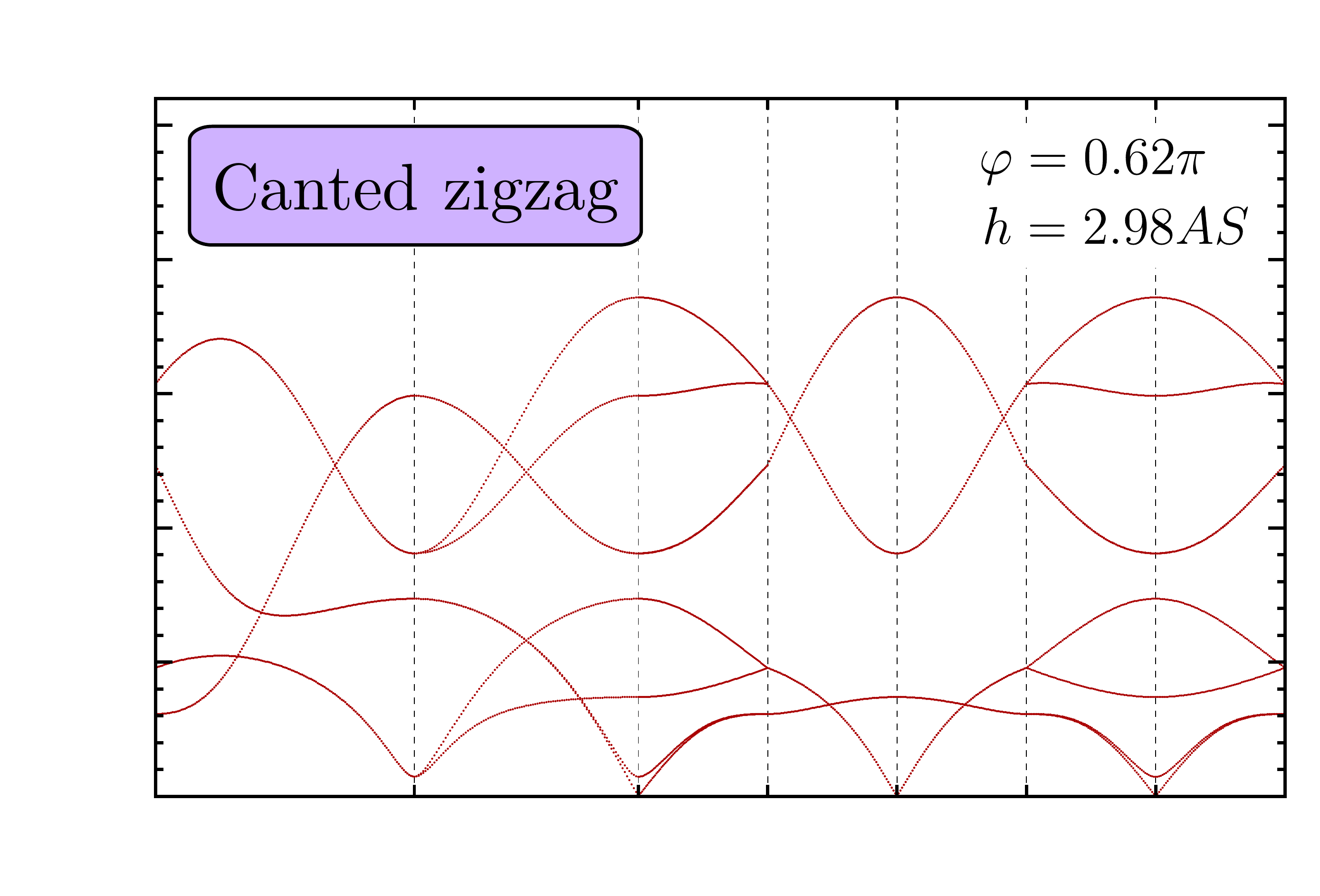}

\vspace{-0.5cm}
\includegraphics[width=6cm]{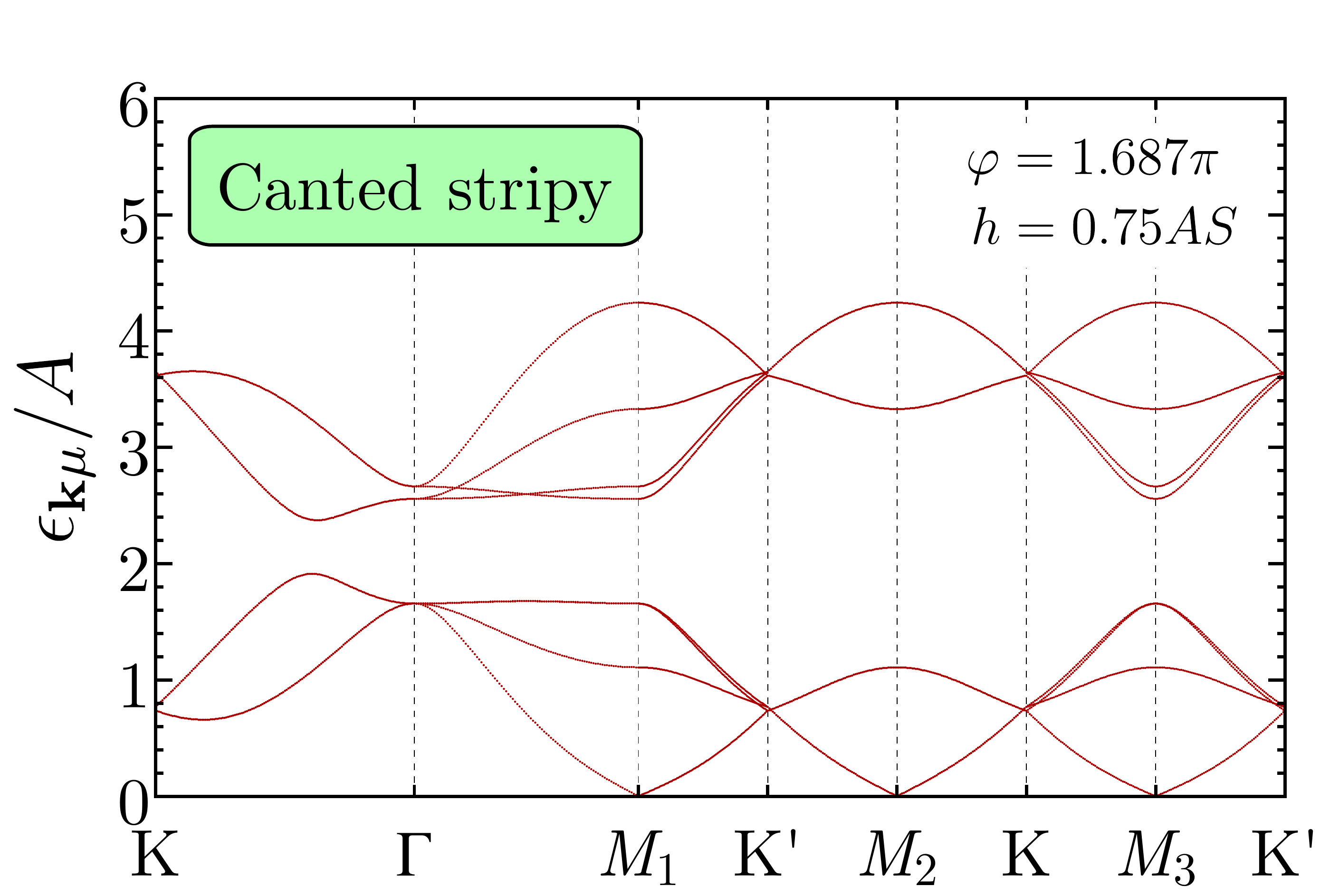}
\hspace{-0.4cm}
\includegraphics[width=6cm]{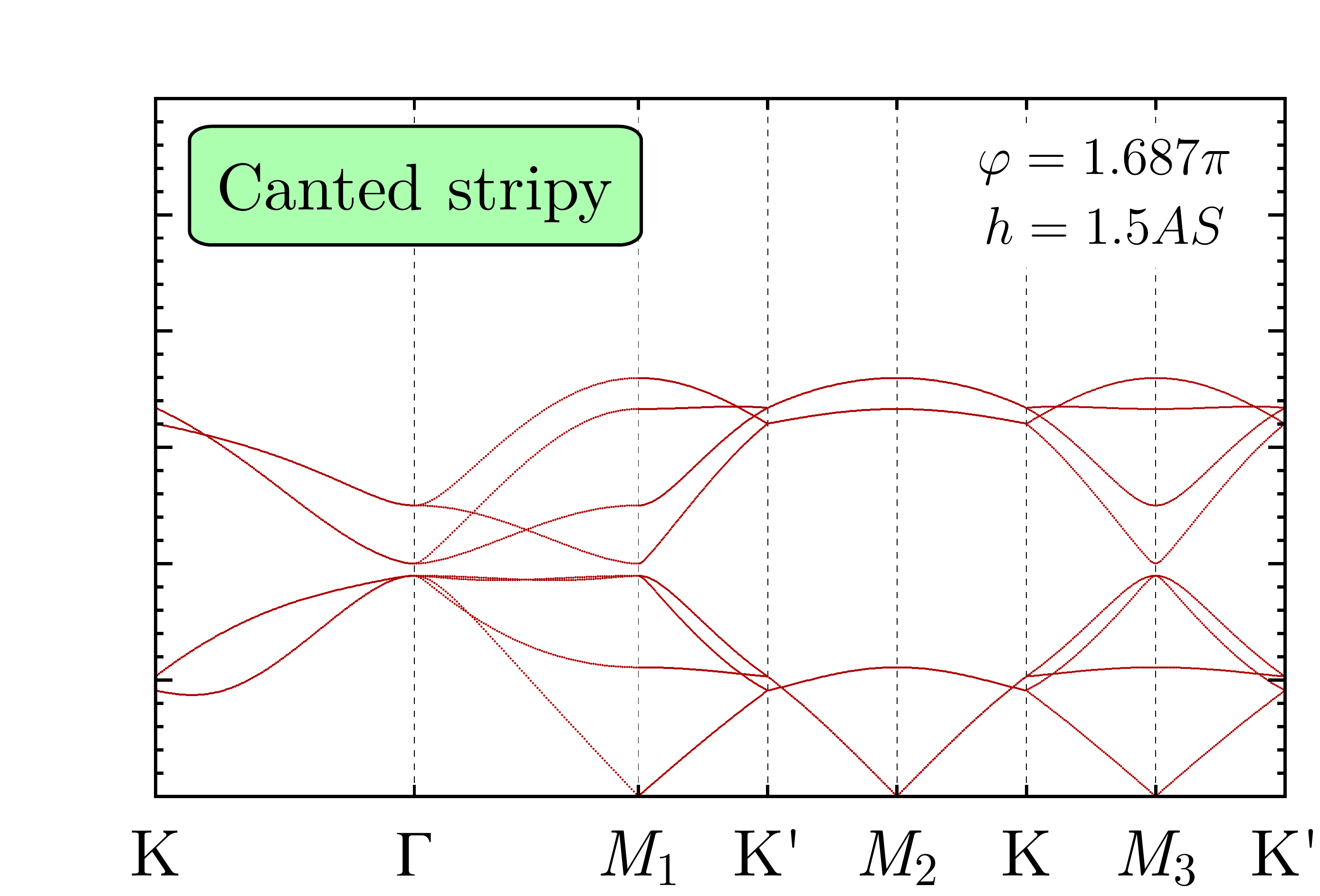}
\hspace{-0.4cm}
\includegraphics[width=6cm]{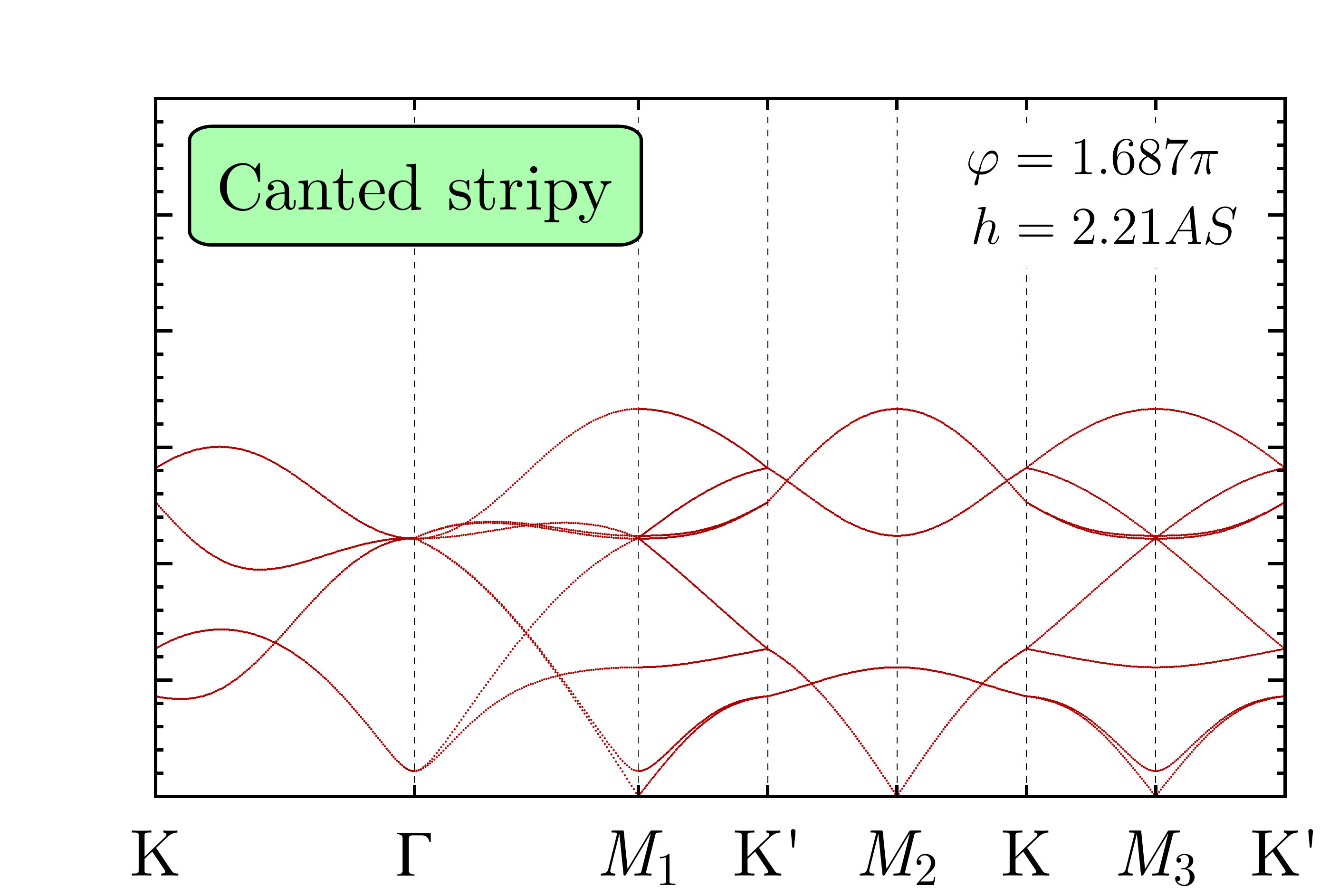}
\caption{Linear spin-wave spectra in the ordered phases in a $\textbf{h}\parallel\left[001\right]$ magnetic field for different values of $h$ and $\varphi$. The right column of the panel represents data immediately below the classical critical field, $h_{\mathrm{c}0}$. The corresponding path along high-symmetry lines of the Brillouin zone is shown in Fig.~\ref{fig:LSW spectra 111}. The plots related to the canted zigzag and canted stripy superimpose the spectra of two degenerate magnetic domains.
\label{fig:LSW spectra 001}}
\end{figure*}

\section{Bogoliubov transformation} \label{sec:app bog transf}

In this appendix, we give some more details on the diagonalization of the LSW Hamiltonian in Sec.~\ref{subsec:LSWT}.
This is accomplished by means of a bosonic Bogoliubov transformation \citep{blaizot1986}, whereby one diagonalizes the modified matrix $\sigma_{3}\mathbb{M}_{\textbf{k}}$, with $\sigma_{3}=\text{diag}\left(\mathds{1}_{N_{\mathrm{s}}},-\mathds{1}_{N_{\mathrm{s}}}\right)$ being a $2N_{\mathrm{s}}\times2N_{\mathrm{s}}$ generalization of the diagonal Pauli matrix. This procedure yields solutions of the form %
\begin{align}
\sigma_{3}\mathbb{M}_{\textbf{k}}V_{\textbf{k}\mu} & =\epsilon_{\textbf{k}\mu}V_{\textbf{k}\mu}\nonumber \\
\sigma_{3}\mathbb{M}_{\textbf{k}}W_{-\textbf{k}\mu} & =-\epsilon_{-\textbf{k}\mu}W_{-\textbf{k}\mu},\label{eq:V and W eigeq}
\end{align}
with $\epsilon_{\textbf{k}\mu}>0$ for all $\textbf{k}, \mu$. Each eigenvector with a negative eigenvalue can be related to an eigenvector with a positive eigenvalue, yet opposite momentum, via the relation $W_{-\textbf{k}\mu}=\sigma_{1}V_{-\textbf{k}\mu}^{*}$, where
\begin{equation}
\sigma_{1}=\begin{pmatrix}0 & \mathds{1}_{N_{\mathrm{s}}}\\
\mathds{1}_{N_{\mathrm{s}}} & 0
\end{pmatrix}.\label{eq:sigma_1}
\end{equation}
Furthermore, one can impose the normalization conditions \citep{blaizot1986}
\begin{gather}
V_{\textbf{k}\mu}^{\dagger}\sigma_{3}V_{\textbf{k}\nu}=-W_{-\textbf{k}\mu}^{\dagger}\sigma_{3}W_{-\textbf{k}\nu}=\delta_{\mu\nu},\nonumber \\
V_{\textbf{k}\mu}^{\dagger}\sigma_{3}W_{-\textbf{k}\nu}=0.\label{eq:V and W normalizations}
\end{gather}
With this, we obtain the Bogoliubov quasiparticles $\left\{b_{\textbf{k}\mu}^{\dagger},b_{\textbf{k}\mu}\right\}$ by means of the transformation $\beta_{\textbf{k}}=\mathbb{T}_{\textbf{k}}\alpha_{\textbf{k}}$, where $\mathbb{T}_{\textbf{k}}^{-1}$ is generally a nonunitary matrix whose first (last) $N_{\mathrm{s}}$ columns correspond to $V_{\textbf{k}\mu}$ ($W_{-\textbf{k}\mu}$).


\section{LSW spectra: Ordered phases} \label{sec:app SW spectra}

\begin{figure*}
\centering
\hspace{1.8cm}
$\vcenter{\hbox{\includegraphics[width=2.4cm]{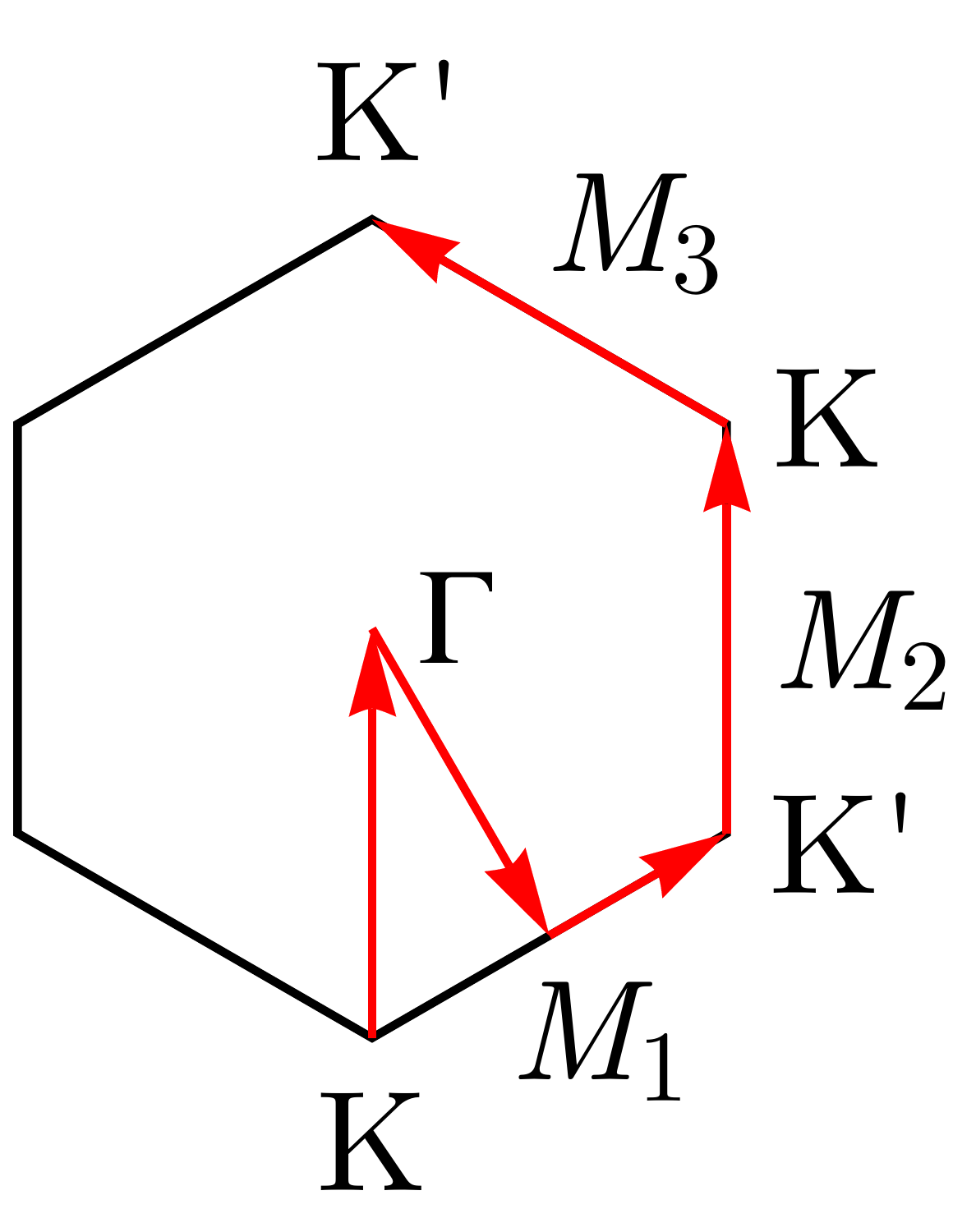}}}$
\hspace{1.8cm}
\hspace{-0.4cm}
$\vcenter{\hbox{\includegraphics[width=6cm]{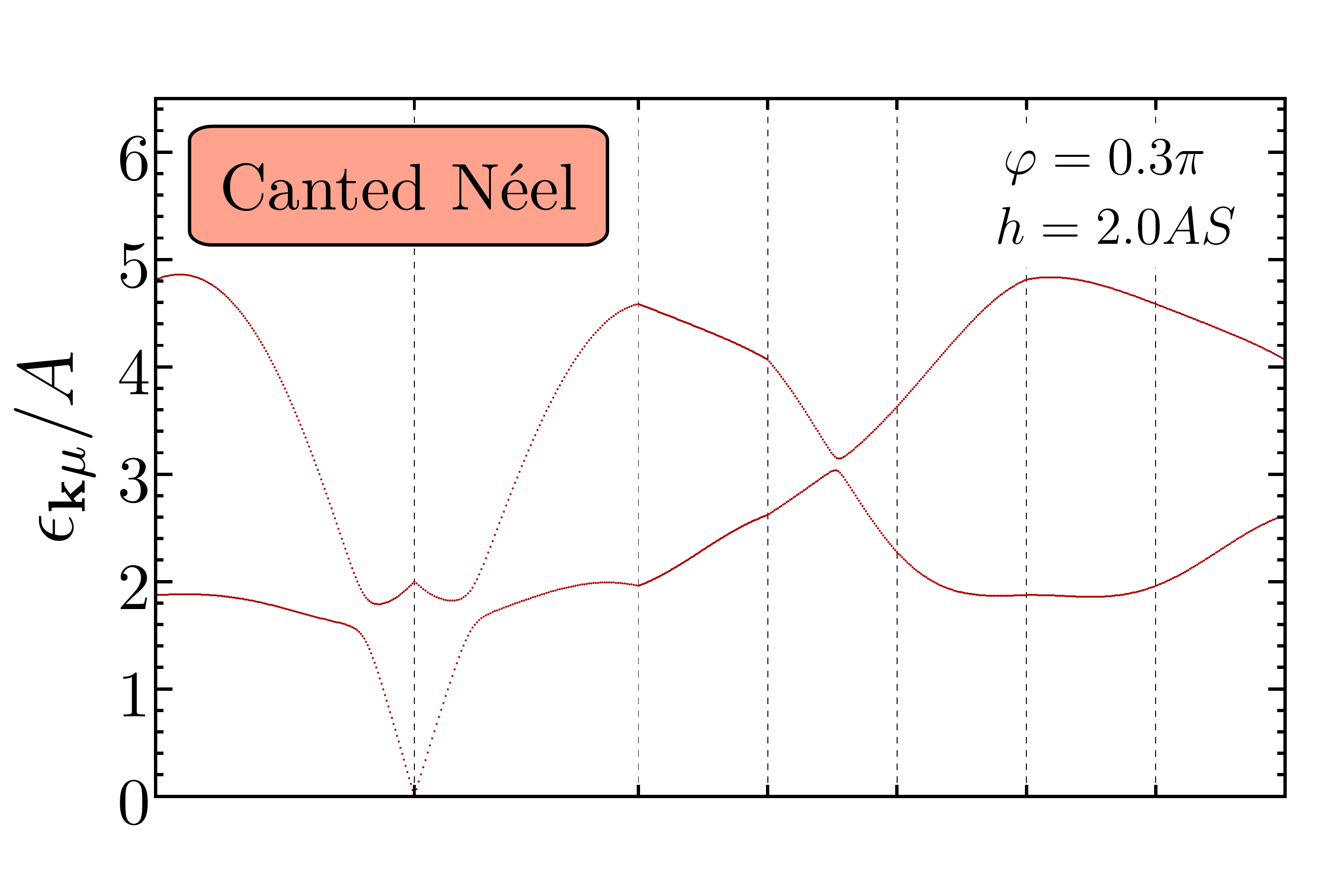}}}$
\hspace{-0.4cm}
$\vcenter{\hbox{\includegraphics[width=6cm]{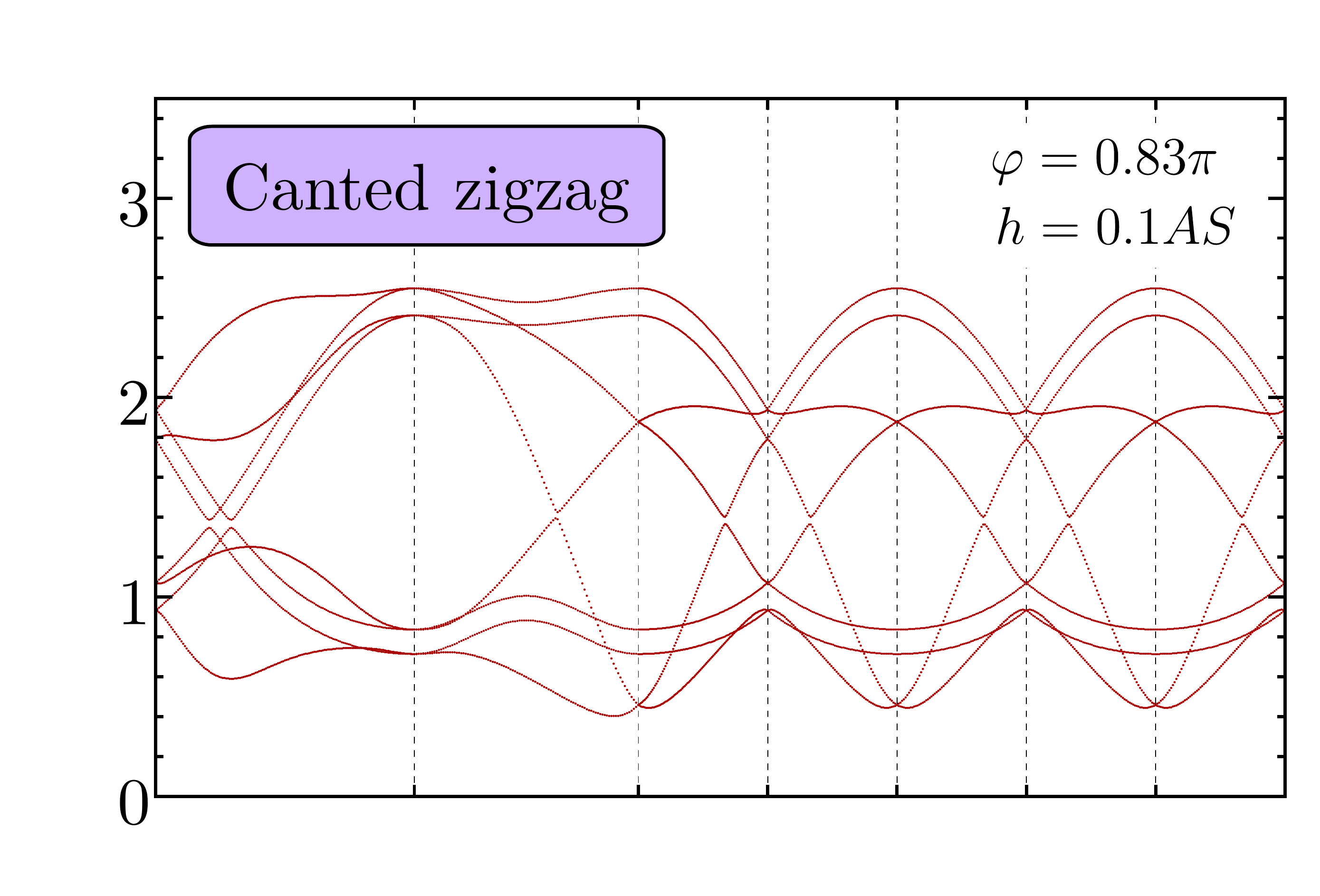}}}$

\vspace{-0.5cm}
\includegraphics[width=6cm]{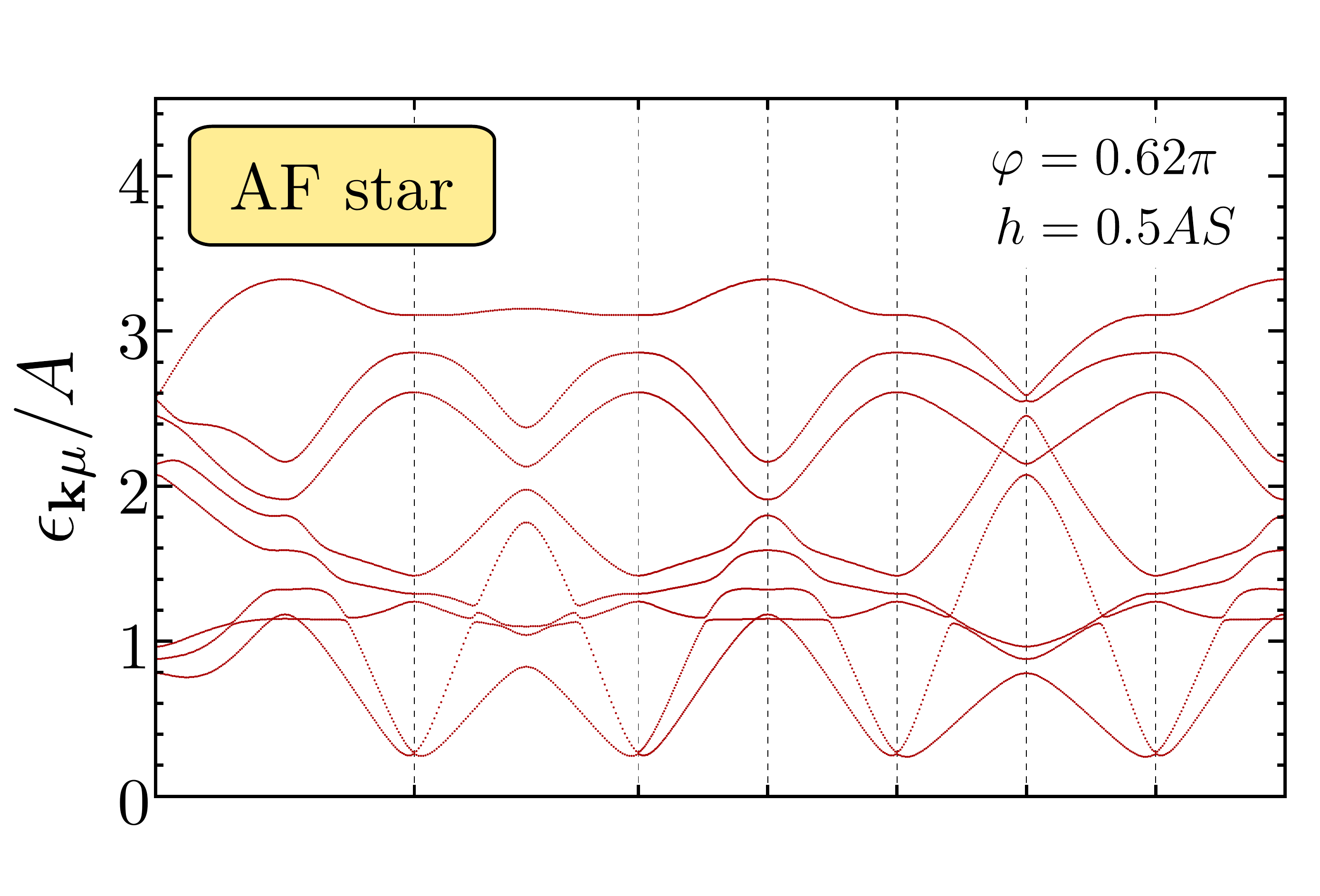}
\hspace{-0.4cm}
\includegraphics[width=6cm]{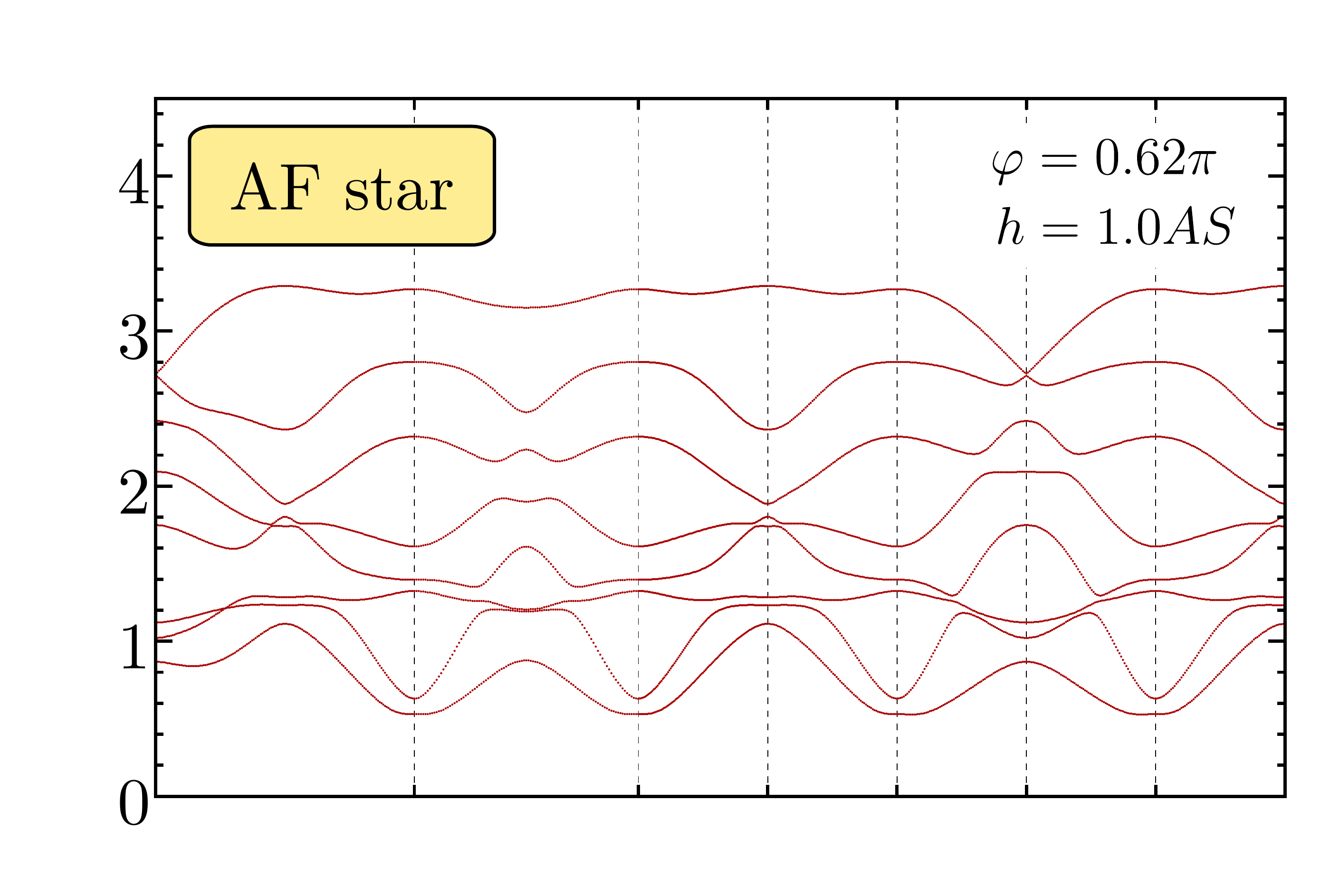}
\hspace{-0.4cm}
\includegraphics[width=6cm]{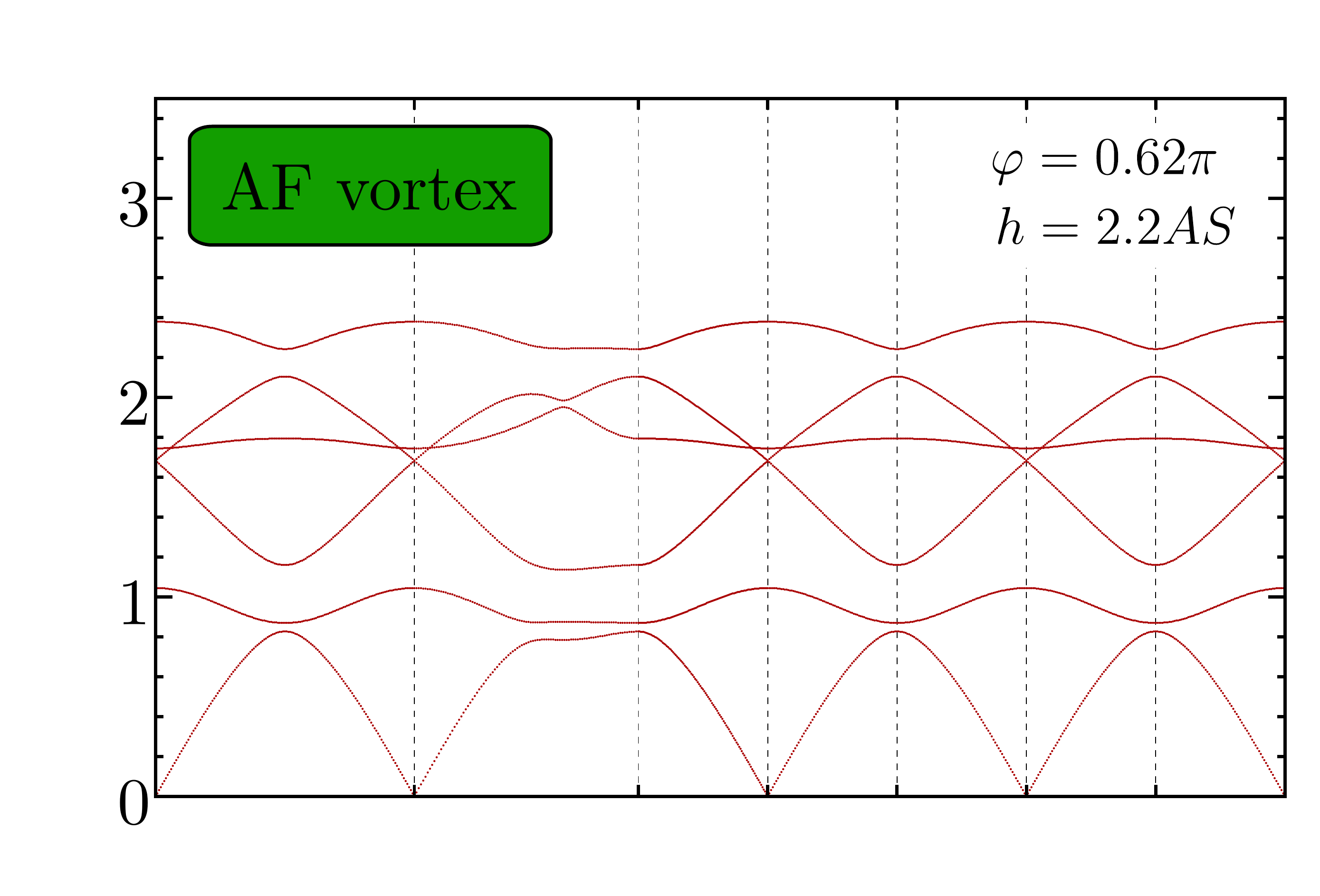}

\vspace{-0.5cm}
\includegraphics[width=6cm]{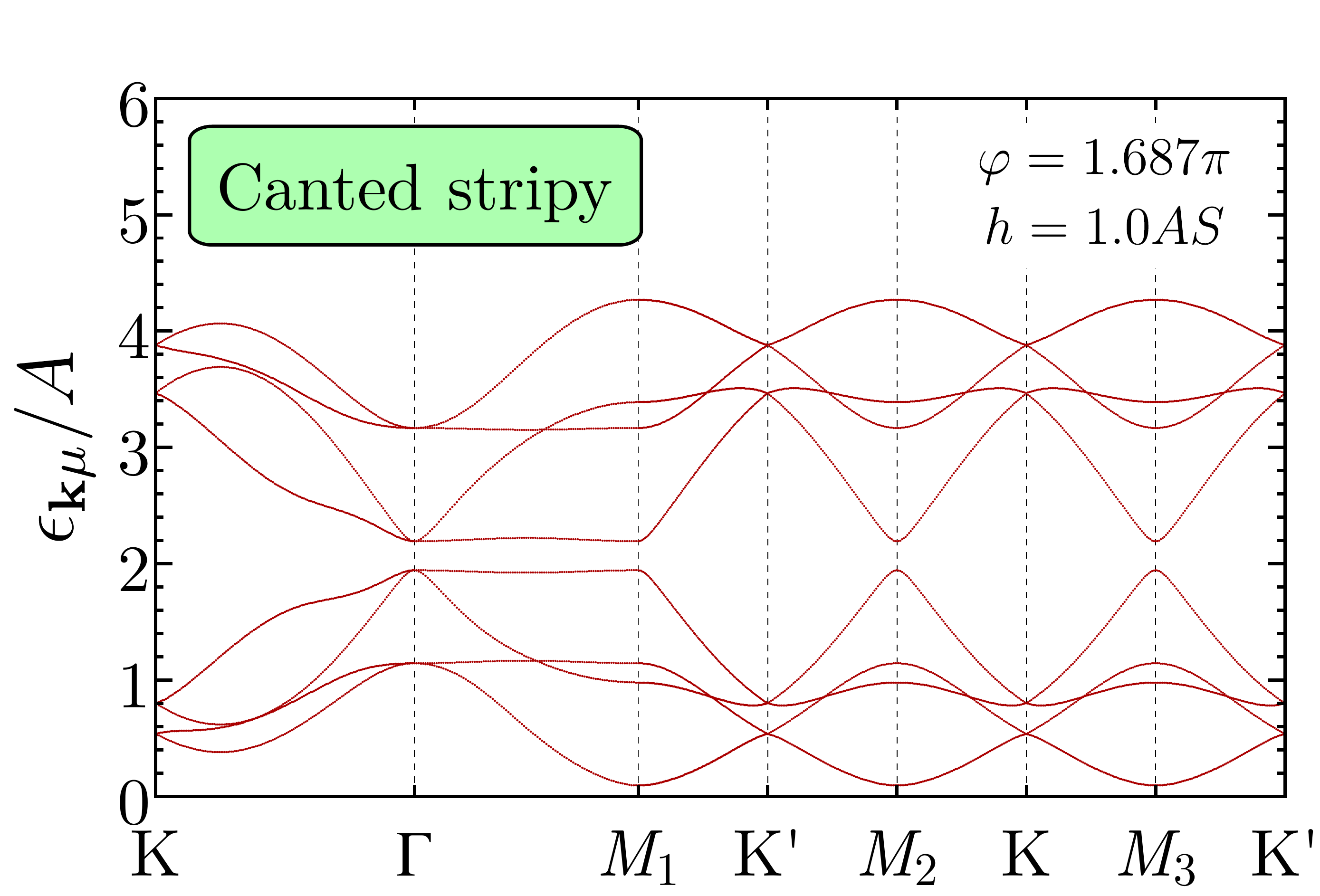}
\hspace{-0.4cm}
\includegraphics[width=6cm]{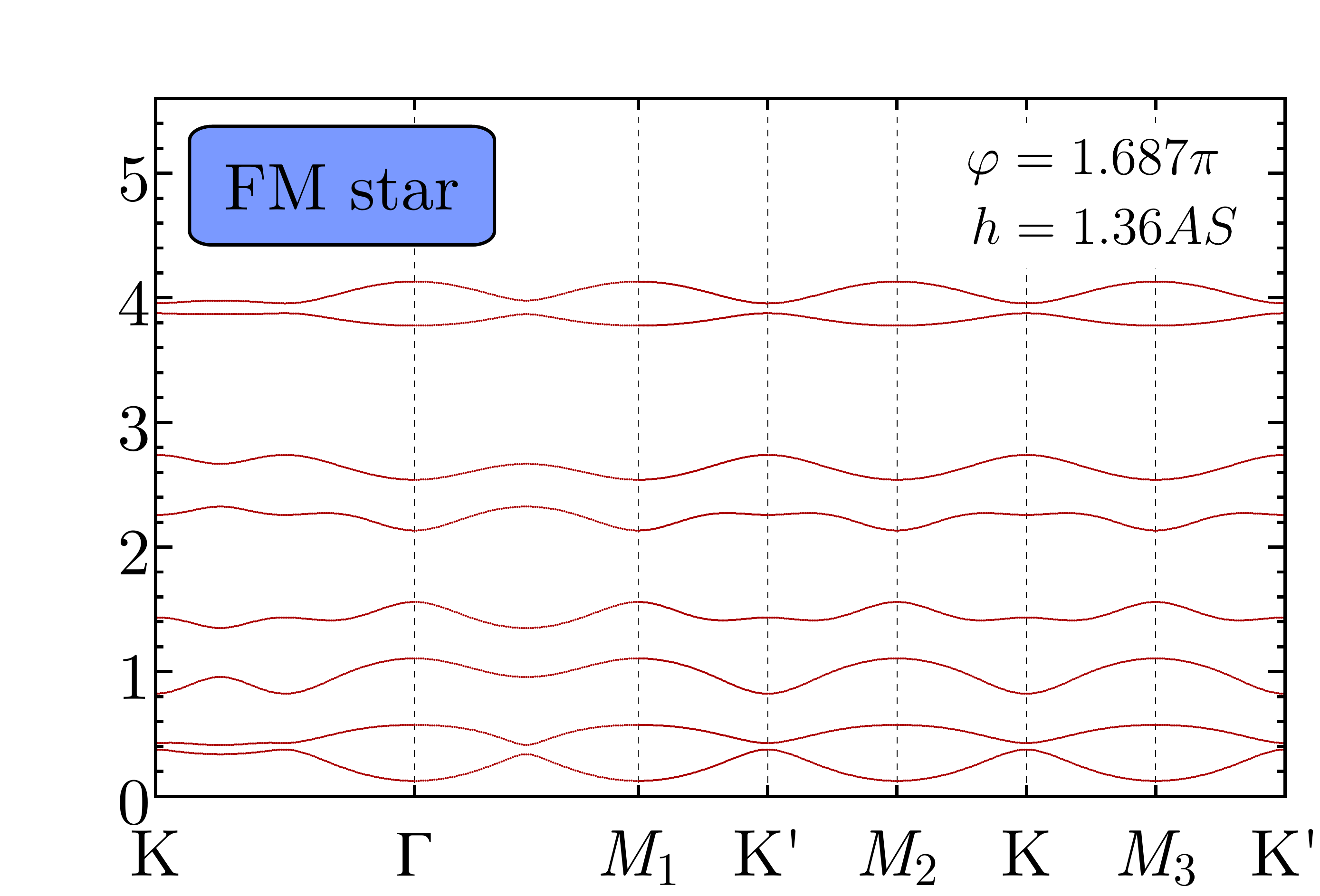}
\hspace{-0.4cm}
\includegraphics[width=6cm]{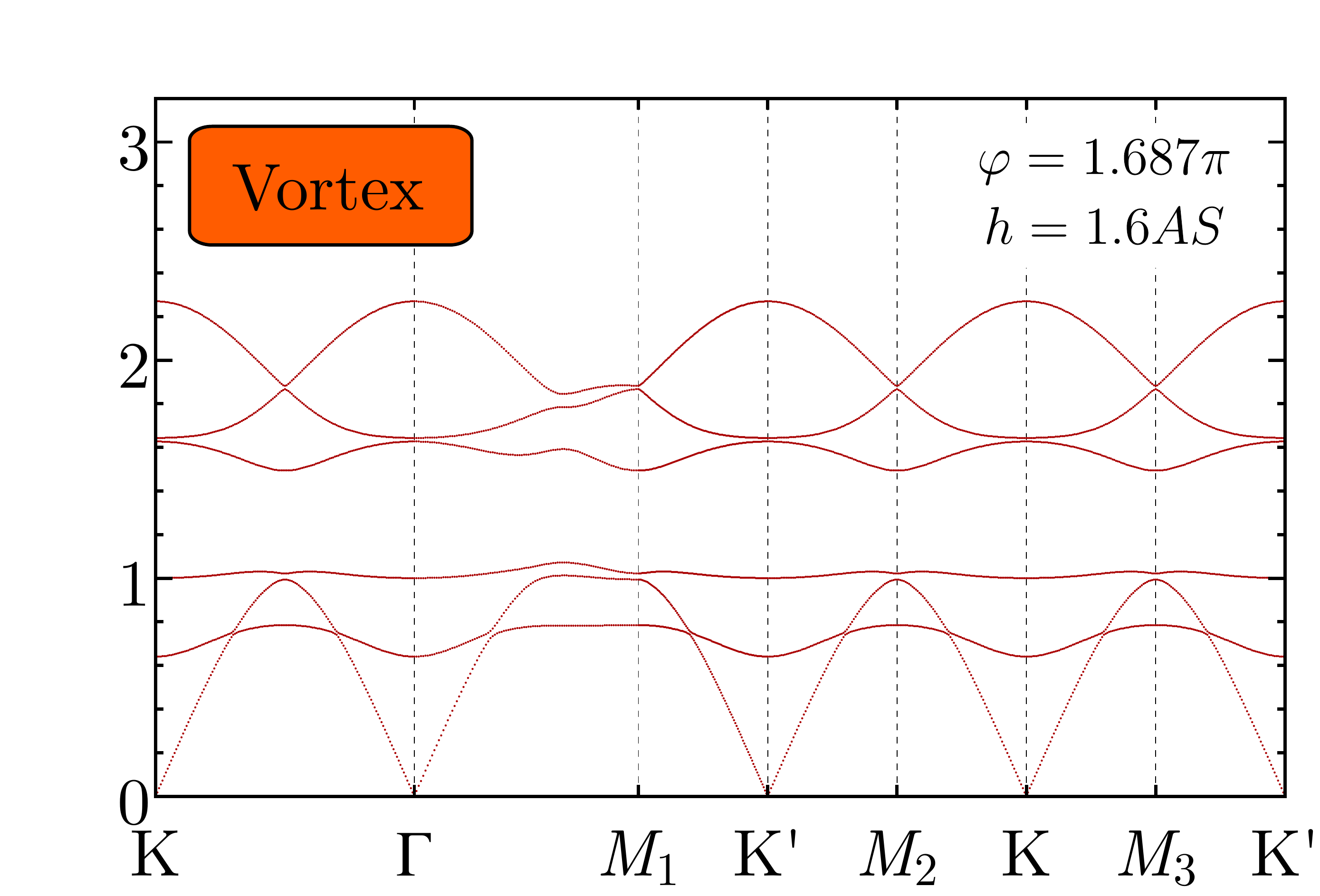}
\caption{Linear spin-wave spectra in several of the ordered phases in a $\left[111\right]$ magnetic field. The corresponding path along high-symmetry lines of the Brillouin zone is shown on the top left. The plots related to the canted zigzag and canted stripy superimpose the spectra of three degenerate magnetic domains. The only spectra that remain gapless under the application of a $\left[111\right]$ magnetic field are those of the canted Néel and vortex phases.\label{fig:LSW spectra 111}}
\end{figure*}

In this appendix, we present a compilation of magnon spectra for the magnetically ordered phases in LSW theory. Spectra at NLO are presented in Appendix~\ref{sec:app NLSWT polarized}.
Figure \ref{fig:LSW spectra 001} illustrates how the LSW spectrum evolves upon increasing the magnitude of $\textbf{h}\parallel\left[001\right]$ in the canted Néel, canted zigzag, and canted stripy phases. The plots shown for the canted zigzag and canted stripy combine the spectra of two degenerate magnetic domains of each phase. The dispersion remains gapless up to $h_{\mathrm{c}0}$ in all three phases, reflecting an accidental continuous degeneracy related to rotations of the magnetic orders around $\textbf{h}$. Such pseudo-Goldstone modes acquire a gap due to quantum fluctuations \citep{rau18}, as an order-by-disorder mechanism selects states which present canting in either the $xz$ or $yz$ plane. In the canted Néel, the low-energy portion of the dispersion gradually changes from a linear to a quadratic shape as the field increases, whereas the opposite trend takes place in the canted stripy. Moreover, as $h\to h_{\mathrm{c}0}^{-}$, one can identify band crossings in each case which also appear in the LSW spectra of the high-field polarized phase. In the canted zigzag and canted stripy, a second band is lowered down to the $M_{1}$ and $M_{3}$ points as we approach $h_{\mathrm{c}0}$, while the gap closes at the $\Gamma$ point as well.

Turning to the case of a $\left[111\right]$ field, Fig.~\ref{fig:LSW spectra 111}, we see that only three of the ordered phases remain gapless for $h>0$. These, however, are but other examples of pseudo-Goldstone modes, as they correspond precisely to the canted Néel, vortex and AF vortex, in which accidental continuous degeneracies may be lifted by order-by-disorder effects, as discussed in Sec.~\ref{subsec:neel ObD}. The spectra of the vortex phases were computed with respect to the classical reference state that minimizes the zero-point energy within LSW theory. The plots corresponding to the canted zigzag and canted stripy now combine the spectra of three degenerate magnetic domains.


\section{Computation of angle corrections} \label{sec:app cubic terms}

In this appendix, we give details on the calculation of angle corrections in the ordered phases by considering the cubic terms in the spin-wave Hamiltonian, cf.\ Sec.~\ref{subsec:Cubic terms}.
It is convenient to locally rotate the spin coordinate system so that the Hamiltonian bears a ferromagnetic ground state in the new reference frame. By using these rotations, we can relate the spin operators in the global $\left\{\hat{\textbf{x}},\hat{\textbf{y}},\hat{\textbf{z}}\right\}$
basis to the ones in the new local $\left\{\hat{\textbf{e}}_{\mu1},\hat{\textbf{e}}_{\mu2},\hat{\textbf{e}}_{\mu3}\right\}$ bases via the rotations
\begin{equation}
\begin{pmatrix}S_{i\mu}^{x}\\
S_{i\mu}^{y}\\
S_{i\mu}^{z}
\end{pmatrix}={R}(\phi_{\mu},\theta_{\mu})\begin{pmatrix}S_{i\mu}^{1}\\
S_{i\mu}^{2}\\
S_{i\mu}^{3}
\end{pmatrix}.
\label{eq:rotation}
\end{equation}
When dealing with noncoplanar states induced by a magnetic field $\textbf{h}$, it is useful to carry out this procedure in three steps represented by the decomposition
\begin{equation}
{R}(\phi_{\mu},\theta_{\mu})={R}_{1}^{\text{T}}{R}_{2}^{\text{T}}(\phi_{\mu}){R}_{3}^{\text{T}}(\theta_{\mu}).\label{eq:R matrix parts}
\end{equation}
The matrix ${R}_{1}$ consists of a global rotation of the original $\left\{\hat{\textbf{x}},\hat{\textbf{y}},\hat{\textbf{z}}\right\}$ basis into a new reference frame $\left\{\hat{\textbf{e}}_{1}^{0},\hat{\textbf{e}}_{2}^{0},\hat{\textbf{e}}_{3}^{0}\right\}$, which is defined so that the unit vector $\hat{\textbf{e}}_{3}^{0} \parallel \textbf{h}$. The two remaining unit vectors may be chosen arbitrarily within the plane perpendicular to $\hat{\textbf{e}}_{3}^{0}$. The second step is encoded in the matrix ${R}_{2}(\phi_{\mu})$, which rotates the $\left\{\hat{\textbf{e}}_{1}^{0},\hat{\textbf{e}}_{2}^{0},\hat{\textbf{e}}_{3}^{0}\right\}$ about the $\hat{\textbf{e}}_{3}^{0}$-axis to give $\left\{\hat{\textbf{e}}_{1\mu}^{0},\hat{\textbf{e}}_{2\mu}^{0},\hat{\textbf{e}}_{3\mu}^{0}\right\}$. The rotation angle $\phi_{\mu}$ is selected in such a way that the orientation of the classical spin on the sublattice $\mu$ lies on the plane generated by the $\hat{\textbf{e}}_{1\mu}^{0}$ and $\hat{\textbf{e}}_{3\mu}^{0}\equiv\hat{\textbf{e}}_{3}^{0}$ vectors. Finally, one maps $\left\{\hat{\textbf{e}}_{1\mu}^{0}, \hat{\textbf{e}}_{2\mu}^{0}, \hat{\textbf{e}}_{3\mu}^{0}\right\}$ onto the target $\left\{\hat{\textbf{e}}_{1\mu},\hat{\textbf{e}}_{2\mu},\hat{\textbf{e}}_{3\mu}\right\}$ basis by performing a rotation ${R}_{3}\left(\theta_{\mu}\right)$ around $\hat{\textbf{e}}_{2\mu}^{0}$.

Now let us gear the formalism to treat the HK Hamiltonian. For convenience, we begin by breaking Eq.~\eqref{eq:hk} into different parts
\begin{equation}
\mathcal{H}=\sum_{\gamma=x,y,z}\mathcal{H}^{(\gamma)}+\mathcal{H}_{h},\label{eq:Hparts}
\end{equation}
where the $\mathcal{H}^{(\gamma)}$ denote spin-spin interaction parts and $\mathcal{H}_{h}$ is the Zeeman term. If we identify the nearest neighbor of site $\mu$ in unit cell $i$ along a $\gamma$ bond by the subindices $j\nu_{\gamma}$, we can write the spin-spin interaction terms as
\begin{align}
\mathcal{H}^{\left(\gamma\right)} & =\sum_{i\mu}\sumprime\left(J\textbf{S}_{i\mu}\cdot\textbf{S}_{j\nu_{\gamma}}+KS_{i\mu}^{\gamma}S_{j\nu_{\gamma}}^{\gamma}\right)\nonumber \\
& =\sum_{i\mu}\sumprime\sum_{m,n=1}^{3}\gamma_{mn}^{\mu}S_{i\mu}^{m}S_{j\nu_{\gamma}}^{n},
\label{eq:H^(gamma)}
\end{align}
with
\begin{align}
\gamma_{mn}^{\mu} & =J\sum_{\ell=1}^{3}{R}_{\ell m}(\phi_{\mu},\theta_{\mu}){R}_{\ell n}(\phi_{\nu_{\gamma}},\theta_{\nu_{\gamma}})\nonumber \\
 &\quad +K{R}_{\gamma m}(\phi_{\mu},\theta_{\mu}){R}_{\gamma n}(\phi_{\nu_{\gamma}},\theta_{\nu_{\gamma}}).
 \label{eq:gamma matrix}
\end{align}
The primed sum in Eq.~\eqref{eq:H^(gamma)} indicates that the sum over $\mu$ only runs through half of the $N_{\mathrm{s}}$ sites in the magnetic unit cell, all of which belong to the same crystallographic sublattice of the honeycomb lattice. The Zeeman term can in turn be expressed as
\begin{equation}
\mathcal{H}_{h}=-h\,\hat{\textbf{e}}_{3}^{0}\cdot\sum_{i\mu}\textbf{S}_{i\mu}=-h\sum_{i\mu\nu}r_{\nu}^{\mu}S_{i\mu}^{\nu}.
\label{eq:H_h}
\end{equation}
Because the coefficients $r_{\nu}^{\mu}(\theta_{\mu})$ are constructed from the rotation matrix ${R}_{3}(\theta_{\mu})$, it follows that $r_{2}^{\mu}=0$.

This construction allows one to write any $n$-boson term of the spin-wave Hamiltonian in a compact manner. Here, however, we are specifically interested in the linear and cubic contributions. The interaction parts of the $n=1$ term read
\begin{align}
\mathcal{H}_{1}^{\left(\gamma\right)} & =\frac{1}{\sqrt{2}}\sum_{i\mu}\sumprime\left[\left(\gamma_{13}^{\mu}+i\gamma_{23}^{\mu}\right)a_{i\mu}^{\dagger}+\left(\gamma_{31}^{\mu}+i\gamma_{32}^{\mu}\right)a_{j\nu_{\gamma}}^{\dagger}+\text{h.c.}\right]\nonumber \\
 & =\sqrt{\frac{N_{\mathrm{c}}}{2}}\sum_{\mu}\left(\gamma_{13}^{\mu}+i\gamma_{23}^{\mu}\right)a_{\boldsymbol{0}\mu}^{\dagger},\label{eq:H_1^(gamma)}
\end{align}
while the field part is
\begin{equation}
\mathcal{H}_{h1}=\sqrt{\frac{N_{\mathrm{c}}}{2}}\frac{h}{S}\sum_{\mu}r_{1}^{\mu}\left(a_{\boldsymbol{0}\mu}^{\dagger}+a_{\boldsymbol{0}\mu}\right).\label{eq:H_h1}
\end{equation}
In both of the expressions above, we have applied a Fourier transform, $a_{\textbf{k}\mu}=N_{\mathrm{c}}^{-1/2}\sum_{\textbf{k}}e^{-i\textbf{k}\cdot\textbf{r}_{i\mu}}\,a_{i\mu}$. Thus, $a_{\boldsymbol{0}\mu}^{\dagger}$ is an operator that creates a boson with momentum $\textbf{k}=\boldsymbol{0}$. Moreover, the last step in Eq.~\eqref{eq:H_1^(gamma)} made use of the fact that $\gamma_{mn}^{\mu}=\gamma_{nm}^{\nu_{\gamma}}$.

As discussed in Sec.~\ref{subsec:Cubic terms}, one must account for $1/S$ corrections to the classical parametrization angles, $\{ \boldsymbol{\phi},\boldsymbol{\theta}\} \to\{ \boldsymbol{\tilde{\phi}},\boldsymbol{\tilde{\theta}}\}$ at NLO in the spin-wave Hamiltonian. If we employ the shorthand notation $\tilde{\gamma}_{mn}^{\mu}=\gamma_{mn}^{\mu}(\boldsymbol{\tilde{\phi}},\boldsymbol{\tilde{\theta}})$ and $\tilde{r}_{1}^{\mu}=r_{1}^{\mu}(\boldsymbol{\tilde{\theta}})$, we get
\begin{align}
\mathcal{H}_{1}(\boldsymbol{\tilde{\phi}},\boldsymbol{\tilde{\theta}}) & =\sqrt{\frac{N_{\mathrm{c}}}{2}}\sum_{\mu}\left[\sum_{\gamma}\left(\tilde{\gamma}_{13}^{\mu}+i\tilde{\gamma}_{23}^{\mu}\right)+\frac{h}{S}\tilde{r}_{1}^{\mu}\right]a_{\textbf{0}\mu}^{\dagger}\nonumber \\
 &\quad +\text{h.c.}\label{eq:H_1 tilde}
\end{align}
As long as we stop the expansion of the spin-wave Hamiltonian at order $n=3$, it is consistent to expand the corrected angles to first order in $S^{-1}$, as in Eqs.~\eqref{eq:1/S expansion ph} and \eqref{eq:1/S expansion th}. This yields $\mathcal{H}_{1}(\boldsymbol{\tilde{\phi}},\boldsymbol{\tilde{\theta}})=S^{-1}\delta\mathcal{H}_{1}+\mathcal{O}(S^{-2})$ with
\begin{equation}
\delta\mathcal{H}_{1}=\sqrt{\frac{N_{\mathrm{c}}}{2}}\sum_{\mu}\left[\left.\textbf{\ensuremath{\boldsymbol{\nabla}}}Z_{\mu}\right|_{\,\boldsymbol{\phi},\boldsymbol{\theta}}\begin{pmatrix}\boldsymbol{\delta\theta}\\
\boldsymbol{\delta\phi}
\end{pmatrix}a_{\textbf{0}\mu}^{\dagger}+\text{h.c.}\right]\label{eq:deltaH_1}
\end{equation}
and
\begin{equation}
Z_{\mu}(\boldsymbol{\tilde{\phi}},\boldsymbol{\tilde{\theta}})=\sum_{\gamma}\left(\tilde{\gamma}_{13}^{\mu}+i\tilde{\gamma}_{23}^{\mu}\right)+\frac{h}{S}\tilde{r}_{1}^{\mu}.\label{eq:Zmu def}
\end{equation}
The gradient of $Z_{\mu}$ is then given by
\begin{equation}
\textbf{\ensuremath{\boldsymbol{\nabla}}}Z_{\mu}=\left(\frac{\partial Z_{\mu}}{\partial\tilde{\phi}_{1}},\ldots,\frac{\partial Z_{\mu}}{\partial\tilde{\phi}_{N_{\mathrm{s}}}},\frac{\partial Z_{\mu}}{\partial\tilde{\theta}_{1}},\ldots,\frac{\partial Z_{\mu}}{\partial\tilde{\theta}_{N_{\mathrm{s}}}}\right).\label{eq:grad Zmu def}
\end{equation}
Each of the partial derivatives above can be written more explicitly as
\begin{align}
\frac{\partial Z_{\mu}}{\partial\tilde{\phi}_{\nu}} & =\sum_{\gamma}\frac{\partial}{\partial\tilde{\phi}_{\nu}}\left(\tilde{\gamma}_{13}^{\mu}+i\tilde{\gamma}_{23}^{\mu}\right),\nonumber \\
\frac{\partial Z_{\mu}}{\partial\tilde{\theta}_{\nu}} & =\sum_{\gamma}\frac{\partial}{\partial\tilde{\theta}_{\nu}}\left(\tilde{\gamma}_{13}^{\mu}+i\tilde{\gamma}_{23}^{\mu}\right)+\delta_{\mu\nu}\frac{h}{S}\frac{\partial\tilde{r}_{1}^{\mu}}{\partial\tilde{\theta}_{\nu}} .
\label{eq:partial devs Z_mu}
\end{align}

With this, we proceed to the cubic term, $n=3$. Since we are only accounting for NLO effects in $1/S$, it suffices to evaluate all $\gamma^{\mu}$ matrices at the classical parametrization angles, $\left(\boldsymbol{\phi},\boldsymbol{\theta}\right)$. We thus obtain
\begin{align}
\mathcal{H}_{3}^{\left(\gamma\right)} & =-\frac{1}{4\sqrt{2}}\sum_{i\mu}\sumprime\left\{ \left(\gamma_{13}^{\mu}-i\gamma_{23}^{\mu}\right)a_{i\mu}^{\dagger}a_{i\mu}a_{i\mu}\right.\nonumber \\
 & +4a_{i\mu}^{\dagger}a_{j\nu_{\gamma}}^{\dagger}\left[\left(\gamma_{13}^{\mu}+i\gamma_{23}^{\mu}\right)a_{j\nu_{\gamma}}+\left(\gamma_{31}^{\mu}+i\gamma_{32}^{\mu}\right)a_{i\mu}\right]\nonumber \\
 & \left.+\left(\gamma_{31}^{\mu}-i\gamma_{32}^{\mu}\right)a_{j\nu_{\gamma}}^{\dagger}a_{j\nu_{\gamma}}a_{j\nu_{\gamma}}\right\} +\text{h.c.}\label{eq:H_3^gamma)}
\end{align}
and
\begin{equation}
\mathcal{H}_{h3}=-\frac{h/S}{4\sqrt{2}}\sum_{i\mu}r_{1}^{\mu}a_{i\mu}^{\dagger}\left(a_{i\mu}^{\dagger}+a_{i\mu}\right)a_{i\mu}.\label{eq:H_h3}
\end{equation}
After combining Eqs.~\eqref{eq:H_3^gamma)} and \eqref{eq:H_h3}, one can use the fact that $\mathcal{H}_{1}(\boldsymbol{\phi},\boldsymbol{\theta})=0$ to simplify $\mathcal{H}_{3}$ considerably. The result is
\begin{align}
\mathcal{H}_{3} & = -\frac{1}{\sqrt{2}} \sum_{i\mu\gamma} \sumprime \left[\left(\gamma_{13}^{\mu} + i\gamma_{23}^{\mu}\right)a_{i\mu}^{\dagger} a_{j\nu_{\gamma}}^{\dagger} a_{j\nu_{\gamma}}\right.\nonumber \\
& \quad \left. + \left(\gamma_{31}^{\mu} + i\gamma_{32}^{\mu}\right)a_{j\nu_{\gamma}}^{\dagger}a_{i\mu}^{\dagger}a_{i\mu}+\text{h.c.}\right].\label{eq:H_3}
\end{align}
According to our discussion in Sec.~\ref{subsec:Cubic terms}, we must now cast Eq.~\eqref{eq:H_3} into normal order, $\mathcal{H}_{3}=\normord{\mathcal{H}_{3}}+\mathcal{H}_{3}^{\left(1\right)}$. By using Wick's theorem, one finds that the residual linear term, $\mathcal{H}_{3}^{\left(1\right)}$, depends on the averages
\begin{align}
m_{\mu\nu,\gamma} & = \left\langle a_{i\mu}^\dagger a_{j\nu_\gamma}\right\rangle,  &
\Delta_{\mu\nu,\gamma} & =\left\langle a_{i\mu}a_{j\nu_\gamma}\right\rangle, \nonumber \\
n_\mu & =\left\langle a_{i\mu}^\dagger a_{i\mu} \right\rangle,  &
\delta_\mu & =\left\langle a_{i\mu} a_{i\mu} \right\rangle .
\label{eq:HF averages}
\end{align}
To avoid ambiguity, we have explicitly indicated the bond type $\gamma$ involved in the parameters $\Delta_{\mu\nu,\gamma}$ and $m_{\mu\nu,\gamma}$. In fact, this distinction is essential here due to the anisotropy introduced by the Kitaev exchange. Our considerations from Appendix \ref{sec:app bog transf} allow us to express all of the quantities above in terms of the eigenvectors of $\sigma_{3}\mathbb{M}_{\textbf{k}}$ with positive eigenvalues. If we denote by $\boldsymbol{\delta}_{\gamma}$ the vector that connects a site $\mu$ to its nearest neighbor $\nu$ along a $\gamma$ bond, we obtain
\begin{align}
m_{\mu\nu,\gamma} & =\frac{1}{N_{\mathrm{c}}}\sum_{\textbf{k}\lambda}e^{-i\textbf{k}\cdot\boldsymbol{\delta}_{\gamma}}V_{\textbf{k}\lambda,N_{\mathrm{s}}+\nu_{\gamma}}^{*}V_{\textbf{k}\lambda,N_{\mathrm{s}}+\mu},\nonumber \\
\Delta_{\mu\nu,\gamma} & =\frac{1}{N_{\mathrm{c}}}\sum_{\textbf{k}\lambda}e^{-i\textbf{k}\cdot\boldsymbol{\delta}_{\gamma}}V_{\textbf{k}\lambda,N_{\mathrm{s}}+\nu_{\gamma}}^{*}V_{\textbf{k}\lambda,\mu},\nonumber \\
n_{\mu} & =\frac{1}{N_{\mathrm{c}}}\sum_{\textbf{k}\lambda}\left|V_{\textbf{k}\lambda,N_{\mathrm{s}}+\mu}\right|^{2},\nonumber \\
\delta_{\mu} & =\frac{1}{N_{\mathrm{c}}}\sum_{\textbf{k}\lambda}V_{\textbf{k}\lambda,N_{\mathrm{s}}+\mu}^{*}V_{\textbf{k}\lambda,\mu}.\label{eq:HF averages eigvecs}
\end{align}
The single-boson term $\mathcal{H}_{3}^{\left(1\right)}$then reads
\begin{align}
\mathcal{H}_{3}^{\left(1\right)} & =-\sqrt{\frac{N_{\mathrm{c}}}{2}}\sum_{\mu\gamma}\sumprime\left[m_{\mu\gamma}^{*}\left(\gamma_{31}^{\mu}+i\gamma_{32}^{\mu}\right)\right.\nonumber \\
 &\quad \left.+\Delta_{\mu\gamma}\left(\gamma_{31}^{\mu}-i\gamma_{32}^{\mu}\right)+n_{\nu_{\gamma}}\left(\gamma_{13}^{\mu}+i\gamma_{23}^{\mu}\right)\right]a_{\textbf{0}\mu}^{\dagger}+\text{h.c.}\label{eq:H_3^(1)}
\end{align}

The corrected reference state is determined by demanding the additional linear term to be zero, $\mathcal{H}_{3}^{\left(1\right)}+\delta\mathcal{H}_{1}=0$. From Eqs.~\eqref{eq:deltaH_1} and \eqref{eq:H_3^(1)}, one can see that this leads to a system of linear equations
\begin{equation}
\boldsymbol{\nabla} Z_{\mu} \bigr|_{\;\boldsymbol{\phi},\boldsymbol{\theta}}
\begin{pmatrix}
\boldsymbol{\delta\theta} \\
\boldsymbol{\delta\phi}
\end{pmatrix}
= x_{\mu},
\quad\mu = 1, \ldots, N_{\mathrm{s}},
\label{eq:linear eq}
\end{equation}
with coefficients
\begin{align}
x_{\mu} & =\sum_{\gamma}\bigl[n_{\nu_{\gamma}}\left(\gamma_{13}^{\mu}+i\gamma_{23}^{\mu}\right)+m_{\mu\gamma}^{*}\left(\gamma_{31}^{\mu}+i\gamma_{32}^{\mu}\right) \nonumber \\
 &\quad + \Delta_{\mu\gamma}\left(\gamma_{31}^{\mu}-i\gamma_{32}^{\mu}\right)\bigr].\label{eq:x_mu def}
\end{align}

Although we have developed the results with reference to the HK Hamiltonian, it is worth noting that the formalism remains valid for other spin models under a suitable adaptation of the $\gamma^{\mu}$ matrices, Eq.~\eqref{eq:gamma matrix}.

To illustrate the procedure, we present some explicit results for $\textbf{h}\parallel\left[001\right]$. In this case, all ordered phases are coplanar, so that the azimuthal angles $\phi_{\mu}$ are exempt from $1/S$ corrections. Moreover, the fact that order-by-disorder mechanisms do not interfere with the uniform canting allows us to compute a single quantity, $\delta\theta=\delta\theta_{\mu}$ for all $\mu\in\left\{ 1,\ldots,N_{\mathrm{s}}\right\} $, in each phase.

In the canted Néel phase, we have
\begin{equation}
\delta\theta=\frac{\cot\theta}{3J+K}\left[\left(J\sum_{\gamma=x,y,z}+\frac{K}{2}\sum_{\gamma=x,y}\right)\left(\Delta_{\gamma}+m_{\gamma}-n_{1}\right)\right].\label{eq:dtheta Neel}
\end{equation}
Interestingly, the expression above singles out the source of the divergence of $\delta\theta$ at $h=h_{\mathrm{c}0}$. In the presence of a nonzero Kitaev interaction, the LSW Hamiltonian becomes nondiagonal in the Holstein-Primakoff bosons $\{a_{\textbf{k}\mu}^{\dagger},a_{\textbf{k}\mu}\}$ at $h=h_{\mathrm{c}0}$. This fact, which basically follows from the polarized state not being an eigenstate of the Hamiltonian, causes the mean-field averages $\Delta_{\gamma}\equiv\Delta_{12,\gamma}$, $m_{12,\gamma}\equiv m_{\gamma}$, and $n_{1}$, as well as the entire term in square brackets in Eq.~\eqref{eq:dtheta Neel}, to have nonzero values. Therefore, we conclude that $\delta\theta$ diverges as $\cot\theta$ when $h\to h_{\mathrm{c}0}$, whereas the product $\tan\theta\,\delta\theta$ is generally nonzero and finite away from the Kitaev point $\varphi=\pi/2$.

\begin{figure}
\centering
\includegraphics[width=4cm]{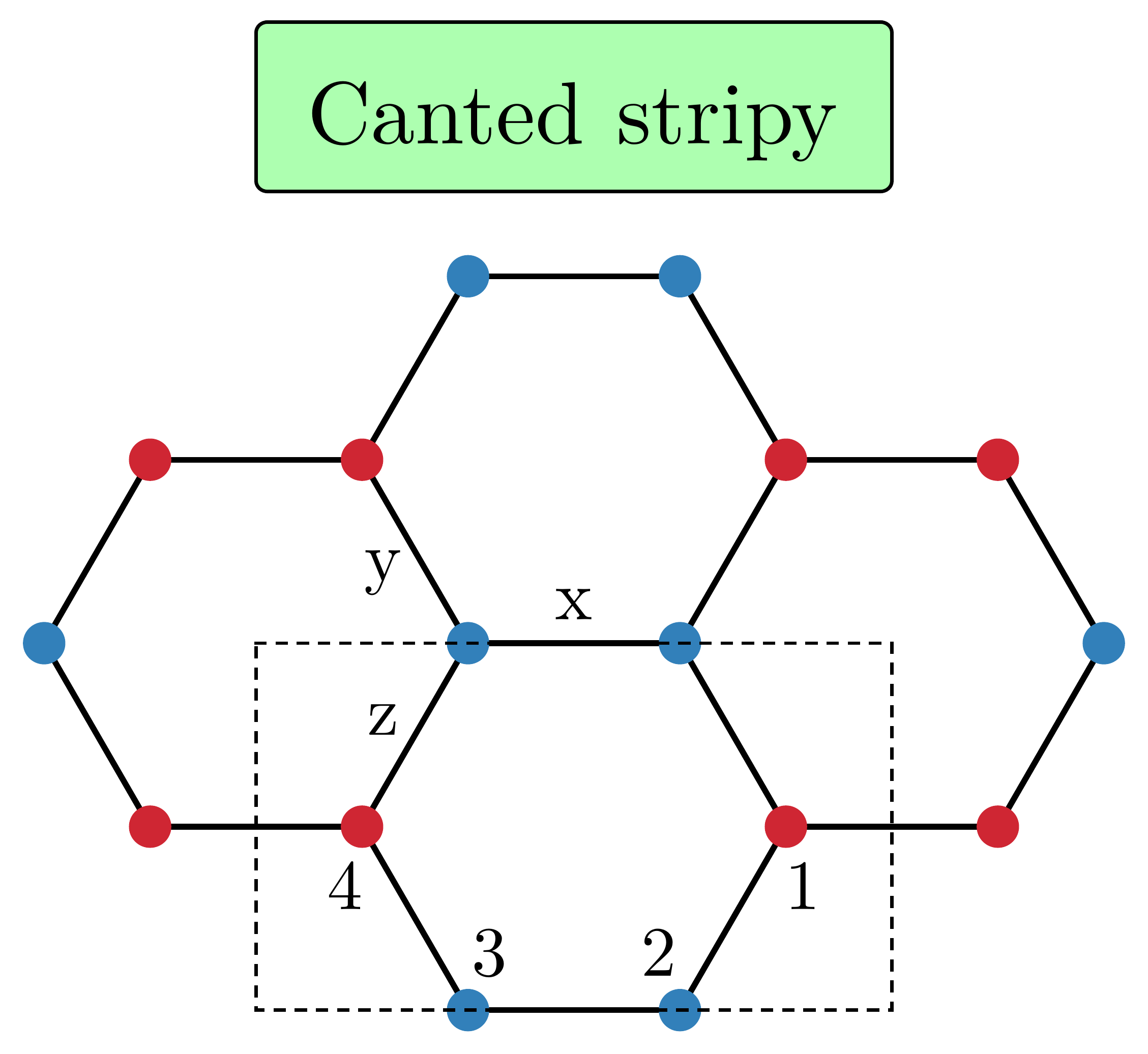}
\hspace{0.4cm}
\includegraphics[width=4cm]{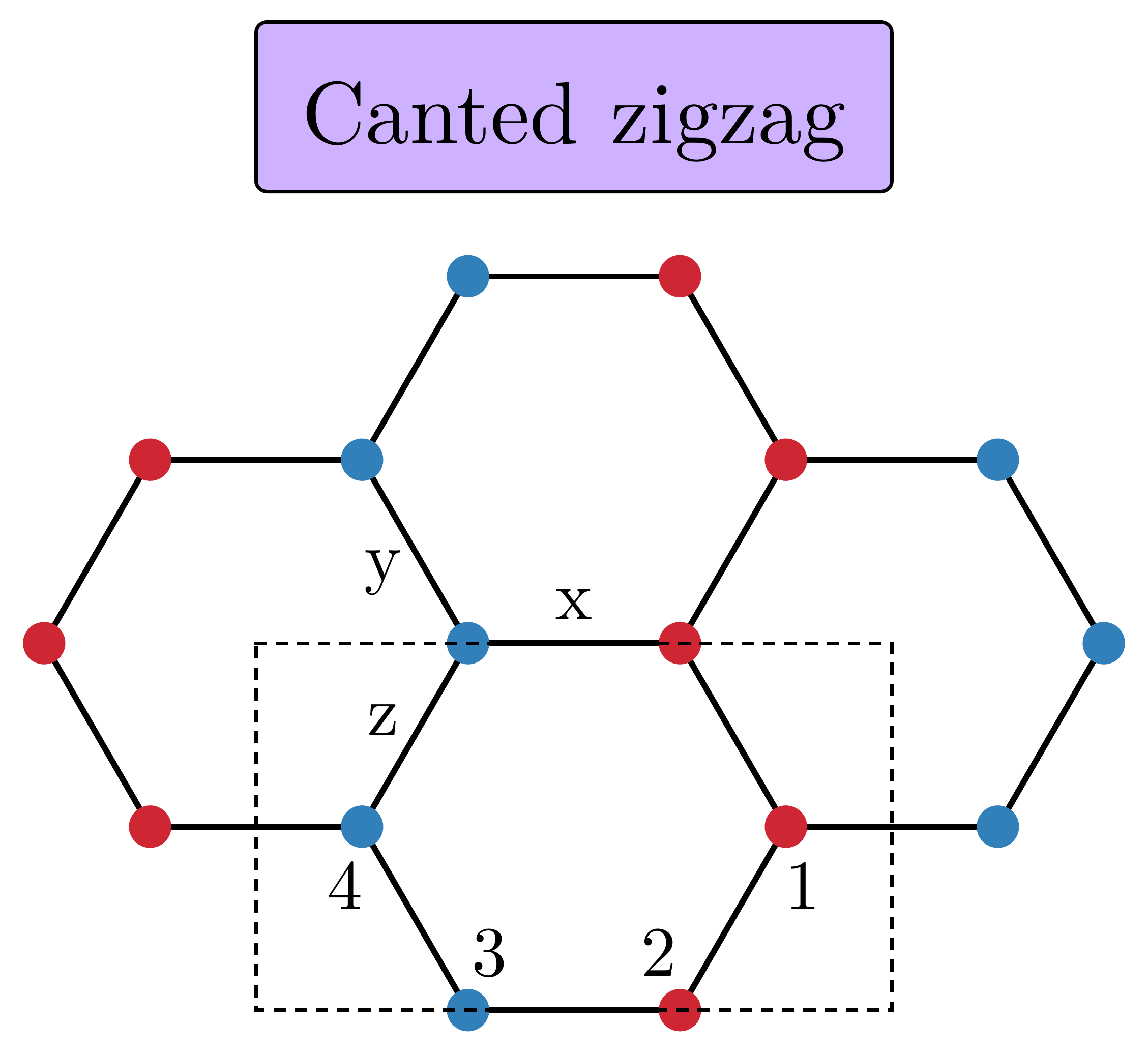}
\caption{Representation of the domains of the canted stripy and canted zigzag phases used to obtain the results in Eqs. \eqref{eq:dtheta stripy001} and \eqref{eq:dtheta zigzag001}. Each of the four magnetic sublattices is labeled by a number from $1$ to $N_{\mathrm{s}}=4$. \label{fig:patterns001}}
\end{figure}

In the treatment of the canted stripy and canted zigzag phases, one must bear in mind that a magnetic field along the $\left[001\right]$ direction partially lifts the degeneracy between the three magnetic domains~\cite{janssen19}. In the case of the canted stripy (zigzag), the pattern with stripes (zigzag chains) running parallel (perpendicularly) to the $z$ bonds becomes unfavorable. By using the configurations represented in Fig.~\ref{fig:patterns001}, one finds
\begin{align}
\delta\theta & =\frac{\cot\theta}{J+K}\left[\left(J+\frac{K}{2}\right)\left(\Delta_{32,x}+m_{32,x}^{*}-n_{2}\right)\right.\nonumber \\
 & \left.-\frac{K}{2}\left(\Delta_{34,y}+m_{34,y}^{*}+n_{4}\right)\right]\label{eq:dtheta zigzag001}
\end{align}
for the canted zigzag and
\begin{align}
\delta\theta & =\frac{\cot\theta}{2J}\left[J\sum_{\gamma=y,z}\left(\Delta_{34,\gamma}+m_{34,\gamma}^{*}-n_{4}\right)\right.\nonumber \\
 & \left.+\frac{K}{2}\left(\Delta_{32,x}+\Delta_{34,y}+m_{32,x}^{*}+m_{34,y}^{*}+n_{2}-n_{4}\right)\right]\label{eq:dtheta stripy001}
\end{align}
for the canted stripy. Note that Eqs.~\eqref{eq:dtheta zigzag001} and \eqref{eq:dtheta stripy001} are also proportional to $\cot\theta$, so that the argument presented below Eq.~\eqref{eq:dtheta Neel} applies for all ordered phases in a $\left[001\right]$ field.


\section{Quantum corrections to the spectrum: Partially polarized phase} \label{sec:app NLSWT polarized}

\begin{figure*}
\centering
\includegraphics[width=18cm]{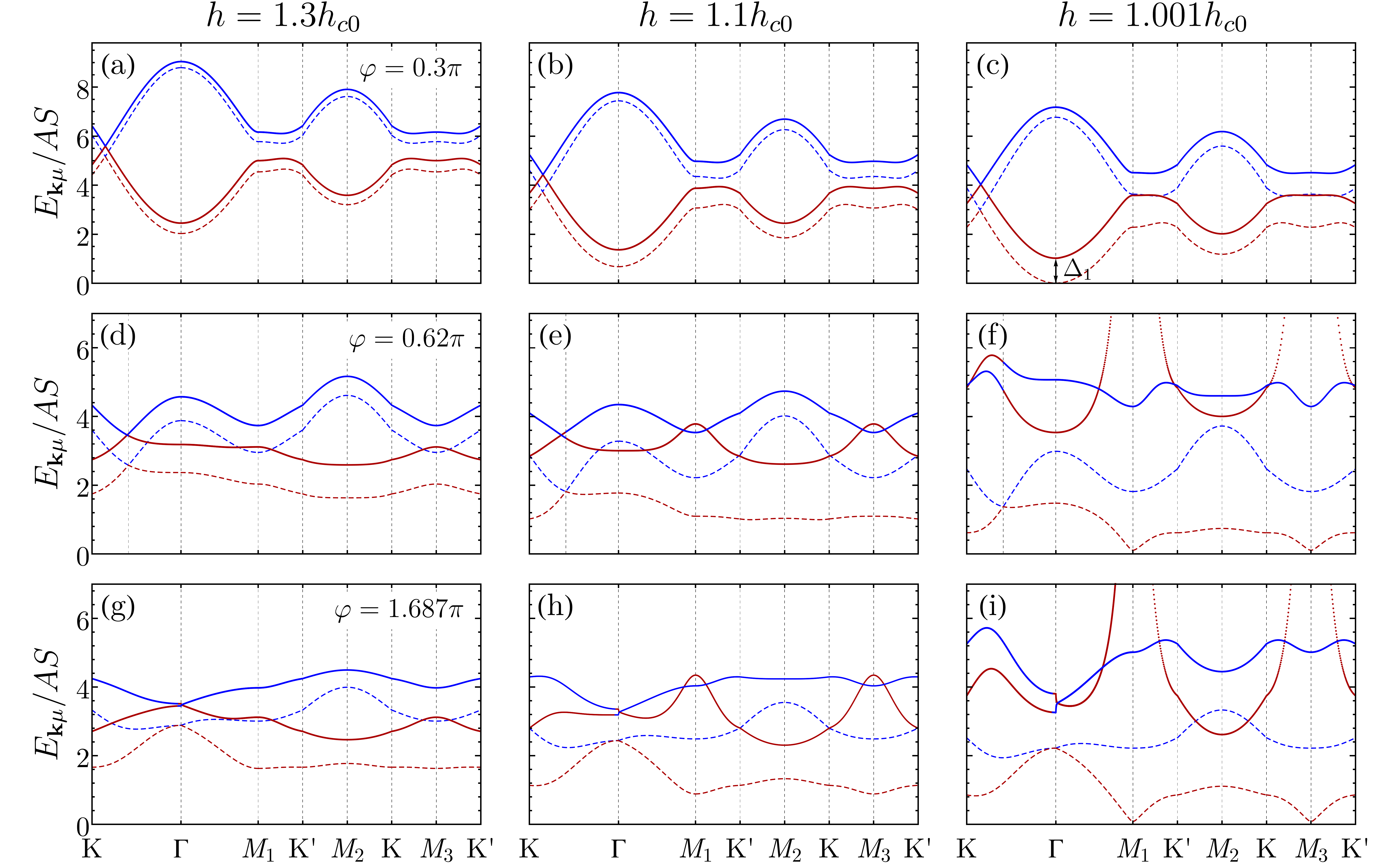}
\caption{Nonlinear spin-wave spectra (dots) in the $\left[001\right]$ high-field polarized phase including NLO contributions in $1/S$ for $S=1/2$. The dashed lines correspond to the LSW results. Each row illustrates the effect of lowering the magnetic field from 130\% to 100.1\% of the classical critical field, $h_{\mathrm{c}0}$, at a constant value of $\varphi$. Plots (a)-(c) show data for $\varphi=0.3\pi$, whereas (d)-(f) and (g)-(i) correspond to $\varphi=0.62\pi$ and $\varphi=1.687\pi$, respectively. The spectrum acquires a finite and nonzero gap as $h\to h_{\mathrm{c}0}^{+}$ above the canted Néel phase, whereas the gap diverges as one approaches the transition to the canted zigzag or canted stripy phases. This is consistent with the discussion regarding the reduction of the critical field in Sec.~\ref{subsec:HF theory}. \label{fig:NLSW spectra 001}}
\end{figure*}

Finally, we discuss details concerning the computation of the magnon spectrum at NLO in the $1/S$ expansion. In general, the NLO contributions are generated by the cubic and quartic terms of the spin-wave Hamiltonian. For simplicity, consider only the partially polarized phase. In this case, the classical reference state is collinear, such that combinations of the type $S_{i\mu}^{\pm}S_{j\nu}^{3}$ do not appear after one performs the required rotations to the spin coordinate system. Consequently, no contributions with an odd number of bosons are produced by the Holstein-Primakoff transformation. For this reason, we shall focus solely on the quartic terms of the spin-wave Hamiltonian. The decoupling $\mathcal{H}_{4}=\normord{\mathcal{H}_{4}}+\normord{\mathcal{H}_{4}^{\left(2\right)}}+\mathcal{H}_{4}^{\left(0\right)}$ leads to a quadratic term with a general form
\begin{align}
\normord{\mathcal{H}_{4}^{\left(2\right)}} & =f_{1}\sum_{\textbf{k}\mu}\normord{a_{\textbf{k}\mu}^{\dagger}a_{\textbf{k}\mu}}\nonumber \\
 & +\sum_{\textbf{k}}\left[f_{2}\left(\textbf{k}\right)\normord{a_{\textbf{k}1}^{\dagger}a_{-\textbf{k}2}^{\dagger}}+f_{3}\left(\textbf{k}\right)\normord{a_{-\textbf{k}1}a_{-\textbf{k}2}^{\dagger}}\right]\nonumber \\
 & +f_{4}\sum_{\textbf{k}\mu}\normord{a_{\textbf{k}\mu}^{\dagger}a_{-\textbf{k}\mu}^{\dagger}}+\text{h.c.},\label{eq:H_4(2) generic}
\end{align}
with functions $f_{1}, \dots, f_{4}$, which generically depend on the wave vector $\textbf{k}$ and the parameters in the Hamiltonian. Note, however, that $f_{1}$ and $f_{4}$ are independent of $\textbf{k}$ because they multiply pairs of bosons related to the same sublattice. Now one uses the Bogoliubov transformation to rewrite Eq.~\eqref{eq:H_4(2) generic} in terms of the Bogoliubov quasiparticles, reading
\begin{align}
\mathcal{H}_{4}^{\left(2\right)} & =\frac{1}{2}\sum_{\textbf{k}}\beta_{\textbf{k}}^{\dagger}\left(\sum_{n=1}^{n}\mathbb{S}_{\textbf{k}n}\right)\beta_{\textbf{k}}+\text{h.c.}\label{eq:H_4(2) post bogoliubov}
\end{align}
with
\begin{align}
\mathbb{S}_{\textbf{k}1} & =f_{1}\sum_{\mu=1}^{4}\Ket{p_{\mu}}\Bra{p_{\mu}}\nonumber \\
\mathbb{S}_{\textbf{k}2} & =f_{2}\left(\textbf{k}\right)\Ket{p_{1}}\Bra{p_{4}}+f_{2}\left(-\textbf{k}\right)\Ket{p_{2}}\Bra{p_{3}}\nonumber \\
\mathbb{S}_{\textbf{k}3} & =f_{3}\left(\textbf{k}\right)\Ket{p_{3}}\Bra{p_{4}}+f_{3}\left(-\textbf{k}\right)\Ket{p_{2}}\Bra{p_{1}}\nonumber \\
\mathbb{S}_{\textbf{k}4} & =f_{4}\sum_{\mu=1}^{2}\left(\Ket{p_{\mu}}\Bra{p_{\mu+2}}+\Ket{p_{\mu+2}}\Bra{p_{\mu}}\right)\label{eq:Sks}
\end{align}
and
\begin{equation}
\Bra{p_{\mu}}=\begin{pmatrix}V_{\textbf{k}1,\mu} & V_{\textbf{k}2,\mu} & W_{-\textbf{k}1,\mu} & W_{-\textbf{k}2,\mu}\end{pmatrix}\label{eq:p_mu def}
\end{equation}
for $\mu = 1,\dots,4$. Thus, we find the self-energy
\begin{equation}
\Sigma_{\textbf{k}}=\sum_{n=1}^{4}\mathbb{S}_{\textbf{k}n}+\text{h.c.},\label{eq:=00005Csigmak}
\end{equation}
which is evidently Hermitian.

As in Appendix \ref{sec:app cubic terms}, we can use Wick's theorem to compute the coefficients $f_{n}$ in terms of the averages from Eq.~\eqref{eq:HF averages}. While laborious, this procedure is straightforward. In the case of $\textbf{h}\parallel\left[001\right]$, the results simplify considerably due to the fact that all averages from Eq.~\eqref{eq:HF averages} are real and obey the relations
\begin{equation}
\begin{cases}
n_{1}=n_{2}=n,\\
\delta_{1}=\delta_{2}=0,\\
m_{x}=m_{y},\\
\Delta_{x}=-\Delta_{y}\text{ and }\Delta_{z}=0.
\end{cases}\label{eq:symmetries averages 001}
\end{equation}
Taking all of this into account, we arrive at
\begin{equation}
\begin{cases}
f_{1}=\frac{J}{2}\left(3n-\sum_{\gamma}m_{\gamma}\right)+K\left(n-m_{x}-\Delta_{x}\right)\\
f_{2}\left(\textbf{k}\right)=\left(J\Delta_{x}-Kn\right)\left(e^{i\textbf{k}\cdot\boldsymbol{\delta}_{x}}-e^{i\textbf{k}\cdot\boldsymbol{\delta}_{y}}\right)\\
f_{3}\left(\textbf{k}\right)=J\sum_{\gamma}\left(m_{\gamma}-n\right)e^{i\textbf{k}\cdot\boldsymbol{\delta}_{\gamma}}\\
\phantom{f_{3}\left(\textbf{k}\right)}+K\left[2m_{z}e^{i\textbf{k}\cdot\boldsymbol{\delta}_{z}}-n\left(e^{i\textbf{k}\cdot\boldsymbol{\delta}_{x}}+e^{i\textbf{k}\cdot\boldsymbol{\delta}_{y}}\right)\right]\\
f_{4}=0.
\end{cases}\label{eq:f's 001}
\end{equation}
Here, $\boldsymbol{\delta}_{\gamma}$ denotes the nearest-neighbor vector along the $\gamma$ bond. In units of the lattice constant, a possible set of choices is $\boldsymbol{\delta}_{x}=\left(-1/2,\sqrt{3}/2\right)$, $\boldsymbol{\delta}_{y}=\left(-1/2,-\sqrt{3}/2\right)$, and $\boldsymbol{\delta}_{z}=\left(1,0\right)$. Along with the eigenvectors of the Bogoliubov transformation, Eq.~\eqref{eq:V and W eigeq}, the expressions above complete the information necessary to compute the spectrum to NLO in $1/S$.

Examples of the resulting spectra are shown in Fig.~\ref{fig:NLSW spectra 001}. Overall, this panel is a good illustration of the key concepts discussed in Sec.~\ref{subsec:HF theory}. First, note how magnon interactions at $\varphi=0.3\pi$ lead to a finite gap, $\Delta_{1}$, as $h\to h_{\mathrm{c}0}^{+}$. In contrast, the spectra immediately above the transitions to the canted zigzag and canted stripy are shown to diverge at the corresponding instability wave vectors, $\textbf{Q}=M_{1},M_{3}$, as $h\to h_{\mathrm{c}0}^{+}$. Both of these observations are consistent with the considerations from Sec.~\ref{subsec:HF theory}. A common feature of all dispersions is that interactions cause the energy of the excitations to increase. Finally, we note that the nonlinear spin-wave spectra in the bottom row of Fig.~\ref{fig:NLSW spectra 001} display a kink at the $\Gamma$ point, which becomes more pronounced as one approaches $h_{\mathrm{c}0}$. This may be understood as an enhancement of the asymmetry between the $k_x$ and $k_y$ directions already seen in the LSW spectrum, possibly due to three-magnon decay processes. Further clarification on this point is left for future work.



\begin{thebibliography}{95}
\bibitem [{ord()}]{order}%
	\BibitemOpen
	\href@noop {} {}\bibinfo {note} {It is possible to estimate the boundary
		between the ordered phases and the Kitaev spin-liquid within LSWT studying
		the vanishing of the ordered moment due to NLO contribution \cite{gotfryd17}.
		In this study, however, we focus solely on the boundary between ordered
		phases.}\BibitemShut {Stop}%
\end{thebibliography}

\end{document}